%% file: main.tex
\RequirePackage{fix-cm}
\documentclass[twocolumn]{article}                     % onecolumn (standard format)
\usepackage{graphicx}
\usepackage{type1cm}        % activate if the above 3 fonts are
\usepackage{makeidx}         % allows index generation
\usepackage{graphicx}        % standard LaTeX graphics tool
\usepackage{multicol}        % used for the two-column index
\usepackage[bottom,hang]{footmisc}% places footnotes at page bottom

\usepackage{newtxtext}       % 
\usepackage[utf8]{inputenx}
\usepackage{amsmath,amssymb,mathrsfs}
\usepackage{commath}
\usepackage{calc}
\usepackage{mathptmx}
\usepackage[superscript]{cite}
\usepackage{color}
\usepackage{empheq}
\usepackage{multirow}

\usepackage{zref-savepos}
\usepackage[ruled,vlined]{algorithm2e}
\usepackage{setspace}
\input{./definitions/blackboard}

\begin{document}
\title{Aerostructural Wing Shape Optimization assisted by Algorithmic Differentiation}
%\title{\inred{Algorithmic Differentiation-based} Aerostructural Wing Shape Optimization using Open-source Software}%\thanks{Grants or other notes



\author{
  Rocco Bombardieri \\ Bioengineering and Aerospace Engineering Department, \\ University Carlos III of Madrid.  \and
   Rauno Cavallaro \\ Bioengineering and Aerospace Engineering Department, \\ University Carlos III, Madrid. AIAA Member. \and
  Ruben Sanchez \\ SU2 Foundation, California, USA. \and
  Nicolas R. Gauger \\ Chair for Scientific Computing, TU Kaiserslautern, Germany.
  }


\maketitle

\input{./Chapters/1_Abstract}
\input{./Chapters/2_Introductionv2}
\input{./Chapters/3_Theoretical}
\input{./Chapters/4_Testcases}
\input{./Chapters/4b_Testcases}
\input{./Chapters/5_Application}
\input{./Chapters/5b_Application}

\input{./Chapters/5c_Application}
\input{./Chapters/5d_Application}
\input{./Chapters/6_Conclusions}
\input{./Chapters/7_Replication}
\input{./Chapters/1b_Statements}

%\input{./Chapters/Templates}
%\begin{acknowledgements}
%\end{acknowledgements}


\medskip

\bibliographystyle{unsrt}      % basic style, author-year citations
\bibliography{./bibliography/Bibliography}   % name your BibTeX data base

%\begin{thebibliography}{}
%\end{thebibliography}

\end{document}

%% file: definitions/blackboard.tex
\usepackage{soul}

%% file: Chapters/1_Abstract.tex
\begin{abstract}
With more efficient structures, last trends in aeronautics have witnessed an increased flexibility of wings, calling for adequate design and optimization approaches. 
To correctly model the coupled physics, aerostructural optimization has progressively become more important, being nowadays {performed} also considering higher-fidelity discipline methods, i.e., CFD for aerodynamics and FEM for structures. 
%
%not only more sophisticated discipline solvers are needed earlier, but, also the couplings needs to be reproduced.
%needs to be included. 
%
%In particular, aerostructural optimization has became a 
%
In this paper a methodology for high-fidelity gradient-based aerostructural optimization of wings, including aerodynamic and structural nonlinearities, is presented. 
%
%One of its attractive feature is modularity, exploited interfacing the discipline solvers, each provided with its own adjoints evaluated by means of algorithmic differentiation, to obtain both solution of the FSI primal problem and the exact gradient of the coupled problem. 
%
The main key feature of the method is its modularity: each discipline solver, independently employing algorithmic differentiation for the evaluation of adjoint-based sensitivities, is interfaced at high-level by means of a wrapper to both solve the aerostructural primal problem and evaluate exact discrete gradients of the coupled problem. 
The implemented capability, ad-hoc created to demonstrate the methodology, and freely available within the open-source SU2 multiphysics suite, is applied to perform aerostructural optimization of aeroelastic test cases based on the ONERA M6 and NASA CRM wings.
%of industrial interest based, on the ONERA M6 and NASA CRM wings.
%of well-known test cases (\emph{ONERA M6} and \emph{NASA CRM}). % wing shape optimization test cases.
%
%The implemented capability, ad-hoc created to demonstrate the methodology, is available within the open-source \emph{SU2} multiphysics suite.
%
%Such test cases are based on well-known wings (ONERA M6 and NASA CRM), and optimizations are carried out to find wing's optimal outer mold line in terms of aerodynamic efficiency for a nominal flight condition, and considering the wing able to deflect (aerostructural coupling).
%
%Validation is pursued by application on aeroelastic test cases of industrial interest, based on the ONERA M6 and NASA CRM wings. 
Single-point optimizations, employing Euler or RANS flow models,  are carried out to find wing optimal outer mold line in terms of aerodynamic efficiency. 
%, for a nominal flight condition.
%, and considering the wing able to deflect (aerostructural coupling)

%Both Euler and RANS flow models are employed. Aerostructural optimization is carried out to find wing's optimal outer mold line in terms of aerodynamic efficiency for a nominal flight condition.
%, and considering the wing able to deflect (aerostructural coupling)
%
%
%ding the wing's optimal outer mold line considering the  
%\inred{Results show the importance of taking into account the aerostructural coupling and warn against potential disadvantages associated to the use of aerodynamic shape optimization carried out on rigid wings as opposed to aerostructural optimization.}
%performing aerodynamic shape optimizations on fixed wing flying shapes.
%rigid wings. %
%potential large disadvantages associated to the use of aerodynamic shape optimizations carried out on rigid wings. % as opposed to aerostructural optimization.  
%
%Also RANS modeling of the flow is considered, and, thanks to algorithmic differentiation providing exact gradients even for the full-turbulence case, good quality of optimization convergence is found. \inred{[TOOCATA DI MARONI]}     

%selected aeroelastic test cases to show the 
%potentials of such solution for the wing shape optimization 
%
%\inred{Results show the importance of considering aerostructural coupling when performing wing shape optimization.}
%
Results remark the importance of taking into account the aerostructural coupling when performing wing shape optimization.

\end{abstract}

%% file: Chapters/2_Introductionv2.tex
\section{Introduction}
\label{sec:intro}
%------------------------------------------%
%
Multi-Disciplinary Optimization (MDO) applied to aerostructural wing design problems has been attracting the interest of both research and industry for the last fifty years and is still today a widely researched topic. 
%
%The term \emph{aerostructural} clearly indicates the simultaneous action of aerodynamics and solid mechanics, nevertheless, an aerostructural optimization can have several different forms. 
%Typically, it considers as  state (or discipline) equations the ones relative to the steady aerodynamics and solid mechanics. 
Typically, when referring to aerostructural optimization, a system is considered, governed by two state (or discipline) equations: one set relative to steady aerodynamics and the other to solid mechanics.
Commonly, Design Variables (DVs) consist in aerodynamic shape and/or structural parameters (e.g., thickness of the skin), while constraints consider the integrity of the structure and/or the trim of the aircraft (mainly in terms of lift to be generated). 
Very often, constraints and/or objective functions regard aircraft performances, such as, range or fuel burn, which in turn, depend on aircraft weight and aerodynamic efficiency. 
\par
A critical aspect is the coupling between the physics of the two considered disciplines: for a given flight condition, the deflected shape of the wing is tightly coupled with the generated aerodynamic forces in a two-way relation. 
%\inred{Current trends to design more efficient and light structures end up having wings operating in flight conditions with highly deflected configurations}.
Current trends to design more efficient and light structures end up having more flexible wings, showing significant deflections while in operation.
This, in terms of modeling, exacerbates both the need to include such aeroelastic tight coupling and to employ nonlinear and higher-fidelity solvers. 
%\inmag{for} an accurate design direction.
%With the current trends to design more efficient structures, the deformed wing shape in flight conditions are remarkable. 'Nuovo periodo' exacerbating both  the need for including such coupling into the optimization and use nonlinear and higher-fidelity solvers to provide an accurate design direction. 
%
%Different optimization strategies in which other aeroelastic phenomena are included have also been carried out (e.g., considering flutter speed), however, efforts featuring mainly aerostructural optimization problems will be reviewed in the following.  
%
\par
%
% Haftka
One of the pioneers of aerostructural optimization was Haftka~\cite{HAFTKA-1977} who proposed a method for the optimization of flexible wings subjected to stress, strain  and drag constraints. The adopted %fidelity level of the 
discipline tools reflected the computatuonal availability of the time, and consisted in a Lifting Line Method (LLM) for aerodynamics and a Finite Element Method (FEM) for the structure. 
A penalty method formulation converted the constrained problem into an unconstrained one. 
% Grossman
\par
The work of Grossman \textit{et al.}~\cite{GROSSMAN_1986} applied an integrated aircraft design analysis and optimization on a sailplane, employing a LLM and a beam-based FEM. 
Two optimization strategies were explored: in the first one, aerodynamics and structure were optimized sequentially, i.e., isolated aerodynamic and structural optimizations were repeatedly performed, while in the second, an integrated (concurrent) procedure was used. 
The integrated approach demonstrated to be superior, {capitalizing on} the interaction between disciplines and finding better results with respect to sub-optimal ones given by sequential optimization. 
%A physical interpretation of the different outcomes was also provided. \inblue{[Beh, conviene citarla in una riga altrimenti meglio togliere]}.
%
%Grossman 2
Following this last work, a higher-fidelity aerostructural model was later used for the optimization of the weight of a forward-swept subsonic aircraft wing subjected to range constraint~\cite{GROSSMAN_1989}. 
%For the aerodynamics, a vortex lattice method was used and viscous drag estimation was computed using section lift coefficient and drag polar, based on provided data; various design variables were employed, based on geometry, structure and performance requirements. 
%
Regarding the performances of sequential/integrated approaches, Martins~\textit{et al.}\cite{Martins2004HighFidelityAD} recently confirmed the observed trends also with higher-fidelity approaches.
%The advantage of coupled aerostructural optimization with respect to the sequential one was also recently confirmed by Martins \textit{et al.}\cite{Martins2004HighFidelityAD} employing an Euler flow model within the context of their aerostructural solver.
%
\par
In the last twenty years, researchers have started to employ higher-fidelity tools to perform aerostructural optimization. 
Such increased level of fidelity came with several consequences, being the first one an increased computational demand for each of the discipline solver. 
%When considering CFD, a further consequence is the need to consider a volume mesh that deforms when wing shape changes, both as consequence of the jig-shape change and the deflection in flight. 
When dealing with CFD, it is necessary to consider a volume mesh that deforms when wing shape changes, both as consequence of the jig-shape variation and the deflection in flight. 
Furthermore, given the high sensitivity of transonic flows with respect to small geometric features~\cite{UC3M_ASO1Pustina}, a large number of DVs may be needed to fully exploit the potentials of the optimization process. 
%\inred{vera una, vera l'altra ma siamo sicuri che la prima implichi la seconda?}
For such driving reasons, gradient-based optimization\cite{Peter_etal_2010} is preferred, as it generally converges faster to a minimum, and comes with a reduced computational cost if adjoint-technique is used for sensitivity evaluation; 
however, chance of hitting a local minimum exists due to the possible non-convexity of the problem~\cite{Lyu2015AerodynamicSO}.
%As further measure to increase efficiency, adjoint methods to evaluate the sensitivity is adopted as it is not time-intensive (although depending on the strategy, large memory peaks can be reached).
%
\par
%  Maute
One of the pioneering works in this direction is the one of Maute \textit{et al.}~\cite{maute_2001}, featuring a gradient-based optimization framework with a three-field approach (structures, aerodynamics, mesh) to model the nonlinear coupled problem. 
A staggered (also known as partitioned or segregated) procedure was set up to solve the coupled fields; aerodynamics was modeled with CFD-Euler, structure with a linear FE model and fluid mesh deformation with a spring-analogy method. 
Among the several proposed applications, the aerodynamic efficiency of the ARW2 wing was optimized, employing lift, stress and displacement constraints while using as DVs wing sweep angle and twist, and structure thicknesses.
%, employing a staggered procedure for the solution (also known as partitioned or segregated). Aerodynamics consisted in an Euler flow model, structure in a linear FEM model and the motion of the fluid grid was computed by means of a spring analogy method. 
For the evaluation of sensitivity a direct approach was chosen, in which partial derivatives where analytically calculated and a staggered scheme was used to solve the system of equations. Direct approach is, anyway, only convenient when then number of design variables is small.
The authors stated that three-field formulations are on average 25\% slower than two-field ones (i.e., strategies that bypass fluid mesh deformation problem) but allow for a more robust handling of large structural displacements, and are more general and reliable for a large variety of cases. 
\par
%%%  Martins 1 & 2
Martins \textit{et al.}~\cite{martins2001ab} proposed a framework for the calculation of coupled aerostructural sensitivities for cases in which aeroelastic interactions were significant. 
The method employed high-fidelity models for both aerodynamics (CFD-Euler) and structure (linear FEM with two element types).  
This work is considered, to the best of the authors' knowledge, one of the first efforts to employ in aerostrucutral optimization an adjoint method, which makes the time for calculating sensitivities almost independent of
%makes the sensitivities calculation time almost on-dependent of 
the number of DVs~\cite{Martins_Review_Der2013}.
%(although, depending on the implementation, large memory peaks can be reached). 
%
%for the evaluation of high-fidelity aerostructural sensitivities~\cite{Martins_Review_Der2013}, hence dropping the dependence of computational time on the  number of DVs 
%
%
%\inred{[X RAUNO:Occhio da controllare se non fosse gia' stato usato su low fidelity, ci sono papers negli anni 90 che controllero']}.
%Such method allows a faster evaluation of sensitivities in case of a high number of design variables, making the computational time and cost to be almost independent of their number and, for such reason, the ideal to be applied for industrial cases. 
The proposed sensitivity evaluation framework was a lagged-coupled adjoint, in which the single discipline adjoint equations were lagged in a similar fashion to the primal solver solution strategy, and partial derivatives were evaluated analytically or by finite differences (FD). 
The proposed implementation was able to calculate aerostructural sensitivity of drag coefficient $C_D$ with respect to bump shape functions applied to the wing jig-shape. 
The accuracy of the method was compared both to FD and complex step (CP)-based ones.
\par
%
%%%%% Martins 2004
In a following effort\cite{Martins2004HighFidelityAD} a similar framework was applied to the design of a supersonic business jet; the selected objective function was a weighted sum of structural weight and drag coefficient evaluated for a design lift coefficient; Kreisselmeier-Steinhauser function lumped individual stresses of the structure in a single structural constraint. 
A total number of 97 design variables was used representing both structural thicknesses and aerodynamic shapes. 
Gradient evaluation time was observed to be almost independent of the number of DVs.
\par
%
%%%%  Maute
Following their previous effort~\cite{maute_2001}, Maute \textit{et al.}~\cite{maute2003911} proposed a different method for aeroelastic optimization. For the same aerostructural problem, gradient calculation was achieved by analytically deriving the adjoint sensitivity equations of the primal problem: a staggered solution algorithm was implemented, where partial derivatives could be calculated analytically or by automatic differentiation. The paper highlighted the computational problems relative to storing partial derivative matrices of large-scale problems for adjoint-based sensitivity calculations. 
%
%\par
%
%In a following effort\cite{Martins2004HighFidelityAD}, Martins \textit{et al.} used a similar optimization framework  to optimize the design of a supersonic business jet. In this framework the flow was modeled with Euler equations while the structure consisted in a detailed linear FEM model. A comparison of adjoint-based sensitivities with FD and CP-based ones showed an average error of 2\%. The optimization was performed minimizing a combination of the aircraft $C_D$  and weight, constraining its $C_L$ and the Kreisselmeier-Steinhauser function to lump individual stresses of the structure. A total number of 97 design variables was used, divided into structural and aerodynamic surface ones, showing the almost independence of the gradient evaluation time from their number.
%
\par
%%%%  BARCELOS 
The work of Barcelos \textit{et al.}~\cite{Barcelos20081813} presented an optimization methodology for fluid-structure interaction (FSI) problems in which aerodynamics took into account turbulence by means of RANS with Spalart-Allmaras (SA) turbulence model~\cite{SpalartAllmaras}. 
The formulation was based on the three-field strategy used in Maute \textit{et al.}~\cite{maute_2001,maute2003911}, whereas the structure was modeled with by  geometrically nonlinear FEM. 
This counts, to the best of the authors' knowledge, as the highest level of fidelity adopted so far for aerostructural optimization of wings. 
Calculation of sensitivities followed the direct approach; all partial derivative contributions were evaluated analytically or via FD. 
With the direct approach, evaluation of total derivatives of the state variables for the coupled aerostructural problem needs to be repeated for each design variable.
Nevertheless it was considered an advantageous approach compared to the adjoint one in case of large-scale problems, for which the Jacobian of the flow problem is difficult to transpose and keep in memory to be used at need.
%; nevertheless, the authors preferred this approach to the adjoint one to overcome issues related to storage of the Jacobian of the flow problem.
%in case of large-scale problems, for which the Jacobian of the flow problem is difficult to transpose and keep in memory to be used at need. 
%
This is one of the common drawbacks due to memory requirements when using adjoint-based sensitivity evaluation strategies, largely mentioned in literature~\cite{maute2003911,SanchezThesis} which, in {our} effort, is conveniently by-passed with the use of the library \textit{CoDiPack}~\cite{sagebaum2017high} for Algorithmic Differentiation (AD).
\par
%
%%% Brezillon
Other literature works presented gradient-based aerostructural optimizations using adjoint methods. Brezillion \textit{et al.}~\cite{Brezillon2012} proposed an articulated high-fidelity optimization framework wrapping DLR's TAU code for CFD (RANS-SA) which includes a discrete-adjoint model with non-frozen turbulence and ANSYS for linear structural FEM. 
%It is claimed, here, that relying only on high-fidelity models, besides being very costly, is unnecessary, and that multi-fidelity optimization strategies should be considered.
%
\par
%%% Ghazlane 
A similar approach was pursued by Ghazlane \textit{et al.}~\cite{ghazlane2012}. 
In the presented aerostructural framework, aerodynamics (Euler flow model) was simulated by ONERA's \emph{elsA} code~\cite{Cambier2002}, which used an iterative fixed-point scheme for the solution of the aerodynamic adjoint; for the structure, a linear structural FEM module was analytically differentiated. 
%\inred{[Si, ma il coupling come?]}
%
\par
%%% Kenway 1 e 2
In effort~\cite{kenway2014} Kenway \textit{et al.} introduced a high-fidelity framework that could perform aerostructural optimization with respect to thousands of multidisciplinary DVs, thanks to an improved parallel scalability of the method; a fully-coupled Newton-Krylov approach was employed for the solution of the aerostructural and the relative adjoint systems.  %%%%%%%%%
The aerostructural solver was based, for the aerodynamic part,  on \emph{SUmb}~\cite{Weide2006} code, featuring  automatic differentiation (\emph{ADjoint}), and, for the structural part, on \emph{TACS}\cite{TACS2010}, also able to evaluate adjoint-based sensitivities. 
In the cited effort, Euler flow model was used together with a linear detailed FEM model of the structure. 
To solve the aerostructural adjoint equations a combination of analytic, forward, and reverse AD methods was adopted.
The method was demonstrated on an aerostructural CRM test case~\cite{Vassberg-2008} using a CFD mesh with over than $16$~millions cells, a FEM grid with over 1 million degrees of freedom (DOFs) and more than $8$~millions state variables, using 512 processors. 
One of the purposes of the use of such large number of DOFs and DVs was that, even though adjoint-based gradient evaluation techniques should be almost independent of the number of considered design variables, in all previous efforts in the literature their number was limited. 
%\inmag{Real scalability of the method, with respect to the number of DVs can, in fact, only be achieved with a correct implementation.} 
%This is due to the fact that real scalability of the method can only be achieved with a careful parallelization of its implementation. 
%
%\inmag{One of the purpose of such large number of DOFs and DVs was to demonstrate, at the practical level,  the independence of adjoint-based gradient evaluation of the number of DVs
%
%overcoming the state of art at the time, which didn't really investigate the practical independence of adjoint-based gradient evaluation techniques on the number of DVs: in fact, real scalability of the method can only be achieved with a careful parallelization of its implementation. 
%
\par
Similarly, in another study~\cite{KenwayMultipoint2014} Kenway \textit{et al.} presented a multipoint high-fidelity aerostructural optimization of a CRM aeroelastic model, alternatively minimizing the takeoff gross weight and the fuel burn. 
%The CRM detailed CSM is built based on its aerodynamic mold line at 1 \textit{g} load factor. 
Flow model was based on Euler equation augmented with a low fidelity viscous drag estimate; linearly behaving structures were considered. A massive parallel supercomputer (more than 400 processors) performed the calculations. 
%\inred{[Dare occhiata a Krylov vs Block Gauss]}.
%
\par
%% Kennedy
A similar work was carried on by Kennedy \textit{et al.}\cite{Kennedy2014e}, optimizing a Quasi-CRM wing model introducing also composite materials for the structure. 
After an aero-structural optimization based on lower-fidelity tools, a final RANS-based (SUmb) aerodynamic shape optimization was performed over the winner configuration.
Further studies were conducted by the authors~\cite{kennedyScitech2014}, in which RANS-based aerostructural optimizations were performed on the same model.
%
%Gradient-based aerostructural optimization is carried out using medium-fidelity tools: a combination of a panel method enhanced with induced and viscous drag evaluation (Tripan) together with a linear structural model (TACS). For coupled aerostructural adjoint analytic derivatives are implemented in the framework. After an aerostructural optimization which aims at minimizing a linear combination of takeoff gross weight and fuel burn a final RANS-based (SUmb) aerodynamic shape optimization is performed over the winner configuration aiming at minimizing $C_D$. Add on to the study is done by the authors in a follwing effort~\cite{kennedyScitech2014}, in which RANS-based aerostructural optimizations are performed on the same aeroelastic model.
%
\par
%%%% Kennedy undeflected 
Works of Kenway \textit{et al.}~\cite{KenwayScitech2014} and Brooks \textit{et al.}~\cite{Brooks2018} performed a similar RANS-based aerostructural optimization on variations of the NASA CRM (undeflected and higher aspect ratio versions) using a similar computational infrastructure as the one mentioned above.  %by the same authors.  
%
%are dedicated to define, with an inverse design procedure and for aerostructural optimization purposes, undeflected versions with different Aspect Ratios (AR) of the CRM (uCRM). This because the CRM is born as aerodynamic benchmark only and its outer mold line represents the aerodynamic shape for 1 \textit{g} load factor. 
%For the found configurations gradient-based aerostructural optimizations were performed, employing RANS-based CFD and linear FEM model with a similar computational infrastructure as the one mentioned above by the same authors. 
%
\par
%%%%% Hoogervorst
The study of Hoogervorst \textit{et al.}\cite{Hoogervorst2017WingAO} proposed an unconventional approach in the landscape of gradient-based aerostructural optimization. The selected MDO architecture was based on an Individual Discipline Feasible (IDF) approach\cite{CramerIDF_1994,MartinsIDF}: the disciplines of the aerostructural problem were decoupled and convergence was ensured imposing additional equality constraints on the interdisciplinary state variables, which became additional surrogate design variables. 
This allowed to lower the computational cost of the problem, especially in cases of strong nonlinearities, but only when the number of DVs was kept small.
In the study, flow was modeled with the Euler equation using the open-source code \emph{SU2}~\cite{SU2_AIAAJ2016} while the structural solver FEMWET was used to solve the linear structural equations. A combination of continuous adjoint approach and FD was used for gradient evaluation. The method was applied to reduce the fuel weight of an Airbus A320 aircraft for a fixed nominal range. %
DVs were both relative to external aerodynamic shape and structural thicknesses, together with the additional surrogate variables requested by the approach.
%geometric shapes Design variables are defined as the displacements of a Free-Form Deformation (FFD) box control points and selected structural model panels thicknesses, together with the additional surrogate variables requested by the approach.
%
\par
Finally, an example of genetic-based aerostructural optimization algorithm was proposed by Nikbay \textit{et al.}~\cite{Nickbay2009} who opted for a more industry-oriented strategy, in which the commercial software \emph{modeFRONTIER} was used to wrap various commercial codes (FLUENT for Euler-based aerodynamics, ABAQUS for linear FEM-based structure and CATIA for parametric solid geometry handling) to perform a loosely coupled analysis. The framework was built to handle the interfaces between such codes and the approach was validated on two aeroelastic test cases: the AGARD 445.6 wing and the ARW-2 wing. 
The effort was driven by industrial practices mainly, which rely (partially or entirely) on assessed commercial codes; hence, the winning strategy had to rely on modular optimization frameworks to which each unit (conveniently interfaced) could connect.
\subsection{Contributions of the present study}
%%%%%%%%%%%%%%
%
%The studies mentioned above represent, mostly in chronological order,  ahttps://it.overleaf.com/project/5efb1207fd58d9000184ea1f part of the attempt of the aerospace community to perform high-fidelity, i.e., employing CFD for fluid and FEM for structures, coupled aerostructural optimization of aircraft wings.
The studies mentioned above highlight the effort of the aerospace community to perform high-fidelity aerostructural optimization of aircraft wings: i.e., employing CFD for fluid and FEM for structures.
%, with particular emphasis on frameworks relying on high-fidelity CFD and CSM models. 
This investigation falls into such high-fidelity aerostructural optimization class, and brings several original contributions on several levels.
\par
Main contribution of this work is to demonstrate a methodology for high-fidelity aerostructural design and optimization of wings, including aerodynamic and structural nonlinearities. 
The proposed gradient-based optimization method relies on a novel strategy that pursues modularity and uses AD within each modular discipline to evaluate the coupled aerostructural gradients with the adjoint method. 
%Main contribution of this work is to introduce an open-source, high-fidelity framework for aerostructural design and optimization of wings, including aerodynamic and structural nonlinearities. The proposed gradient-based optimization  relies on an innovative and, to the best of the author's knowledge, never used strategy, based on coupled AD for gradient evaluation, which includes nonlinearities both in the CFD and CSM problems. 
%
The framework implementing the proposed methodology has been released as \emph{open-source} software within the SU2 suite~\cite{Palacios2015, Pini2017, Gori2017, Molina2017, Zhou2017, albring2016};
%\emph{open-source}, delivered within the SU2 suite~\cite{Palacios2015, Pini2017, Gori2017, Molina2017, Zhou2017, albring2016}; 
it relies on SU2 for the CFD part and employs a geometrically nonlinear beam FEM solver, ad-hoc coded for this investigation. 
Differently than the capability already available in SU2 suite~\cite{sanchez2018}, in which FSI optimization problems were tackled using AD at native level, and in which aerostructural problems were solved for matching fluid and structural meshes (i.e., no interface or spline strategies) the here-presented approach is modular and based on a Python-wrapped interface between the different solvers, namely CFD, FEM and Interface/Spline modules. 
Such methodology provides extreme flexibility: different solvers, each one with the sought level of fidelity, can be interfaced to perform aerostructural analyses and/or optimization, once provided with a standard interface. 
Addition of the interface/spline module and its integration in the sensitivities evaluation workflow provide further flexibility, allowing to employ the framework on cases with non-matching structural/fluid interfaces, typically found in aeronautical applications.
%provide extra flexibility 
%Furthermore, with respect to the previous effort~\cite{sanchez2018}, the introduction of the interface/spline module and its integration in the sensitivities evaluation workflow are attractive features as allow application of the framework to non-matching structural/fluid interfaces, typically found in aeronautical applications.} 
%
%for applications to on-matching structural/fluid interface
%provide a more general approach, addressed to aeroelastic applications,
%for which non-matching structural/fluid interfaces are common. }
%The work also introduces a dedicated Interface module for non-matching structural/fluid interfaces providing a more general approach for applications 
%This investigation takes inspiration from the work of Sanchez \textit{et al.}~\cite{sanchez2018} where FSI problems and coupled sensitivities evaluation were performed at native level within the SU2 suite by means of AD.
%
\par
A geometrically nonlinear beam FEM solver (\emph{PyBeam}), has been developed as a model code to show how simple the integration of the chosen AD library (CoDiPack) would be on existing codes, avoiding manual implementations of adjoint algorithms which are usually complex and time-consuming to perform, despite recent advantages in this direction~\cite{DAFoam2019}.
Being open-source, this framework provides an easy access to an aerostructural optimization tool tailored for aircraft applications to a potentially large user audience. 
%
%The framework, to be part of the open-source SU2 suite~\cite{Palacios2015, Pini2017, Gori2017, Molina2017, Zhou2017, albring2016}, relies on SU2 for high-fidelity CFD (Euler, RANS) and on a geometrically nonlinear beam FEM solver, namely pyBeam. Differently than previous efforts, in which FSI optimization problems were tackled by SU2 at native level, based on an AD strategy for gradient evaluation~\cite{sanchez2018}, the proposed approach is modular and based on a Python-wrapped interface between different solvers (i.e. CFD, FEM and interpolation). It is considered, in fact, that such approach is an ideal solution for a wider infrastructure in which more tools of various fidelity levels, provided with a standard interface, can be incorporated to perform both analysis and optimization at the different stages of the design process. It will also facilitate an easy access to high-fidelity multidisciplinary optimization tools for aircraft design to a potentially large user audience. PyBeam has also been developed as a model code to show how simple the integration of the chosen AD library (CoDiPack) would be on existing codes, avoiding manual implementations of adjoint algorithms which are usually complex at code level and time-consuming to perform, despite recent advantages in this direction~\cite{DAFoam2019}.
%
\par
Application of the method is carried out on aeroelastic test cases of potential industrial interest, based on ONERA M6 and NASA CRM wings and featuring relevant structural deflection. 
Several levels of fidelity are employed in the analyses: together with a geometrically nonlinear structural model, both Euler and RANS-SA flow models are used for aerodynamics. 
For RANS-SA based applications, particular attention has been payed to solve the fluid primal and relative adjoint problems following the approach of the full-turbulence (or non-frozen-turbulence)~\cite{Lyu2015AerodynamicSO,Barcelos20081813}.
Relevance of the aerostructural coupling on the optimization results is highlighted, showing how neglecting it can lead to a less performing design with respect to the initial nonoptimized configuration.
%Further contributions of this paper come from the application of the present framework to optimize the outer mold line of wings. Relevance of  aerostructural coupling on the optimization results is highlighted showing how, neglecting it can lead less performant design with respect to the initial nonoptimized configuration.  
%not having performed an optimization might have been the best option. 
%
%neglecting it, can lead to  a degradation of performances with respect to the ones relative to the 
%Especially for highly deformable wings, aerostructural coupling needs to be taken into account 
%a%re in showing, thanks to this framework, effects of aerostructural coupling when performing aerodynamic shape optimization, especially for highly deformable wings. 
%
%
%
%%%%%%%%%
\subsection{Organization of the paper}
%%%%%%%%%%%
%
%\inred{Quello che scrivi sotto e' un'arama a doppio taglio. Va scritto in maniera diversa non citando esplicitamente quello che e' stato fatto precedentemente]}
%This work follows up a previous effort by Bombardieri \textit{et al.} \inred{cite book chapter}, in which pyBeam module was presented and the primal aeroelastic solver was shown. 
%Later on, the novel AD-based gradient evaluation strategy was introduced and the calculation of selected aerostructural cross sensitivities was validated. 
%
%Following the previous effort by Bombardieri \textit{et al.}~\cite{Bombardieri_uc3m_EUROGEN}, in the current work, pyBeam capabilities are extended with nonlinear rigid elements, the optimization framework is set up and applied to cases of industrial interest. 
%
The remainder of this paper is organized as follows: in Section~\ref{sec:theory} the theoretical background for the solution of the aerostructural problem and for sensitivities evaluation is  detailed; in Section~\ref{sec:testcases} the aeroelastic test cases are introduced (i.e., aeroelastic models based on the ONERA M6 wing and on the CRM) whereas in Section~\ref{sec:results} aerostructural sensitivities validation is presented and results of the optimization are shown and discussed. 
Section~\ref{sec:concl} concludes the paper and provides recommendations for future works.

%% file: Chapters/3_Theoretical.tex
%------------------------------
\section{Theoretical background}
\label{sec:theory}
%------------------------------
A more in-depth overview of the theoretical background and solvers is presented in this section. 
First, the primal problem, i.e., the static aeroelastic equilibrium of a flexible wing subjected to a given flow, is formulated; each of the discipline solvers is briefly reviewed and their coupling and interfacing within the framework is discussed.
%interfaced within the framework architecture. 
Later on, the same primal problem is reformulated in the form of fixed-point iterations, which is the most suitable one for the implementation of the used AD-based adjoint method. 
State and design variables are introduced and the complete set of equations of the adjoint problem is shown, together with the reverse computational path followed by the algorithm for the evaluation of sensitivities. 
It is relevant to stress out that structural, aerodynamic and mesh solvers are implemented in  independent modules, each featuring its own AD-based sensitivity capability. 
Hence, the approach to be described is suitable for different combinations of solvers provided with the adequate primal/adjoint interfaces. 
\par
Last topic covered in this section is the aerostructural wing shape optimization formulation.
%
%
%------------------------------
\subsection{Primal problem}
\label{sec:primal}
%------------------------------
%------------------------------
\paragraph{Structural FEM solver}
\label{sec:fem_solver}
%------------------------------
%
The structural in-house solver pyBeam relies on a 6-dof geometrically nonlinear beam model following the work of Levy~\cite{LevyBook}. 
%The formulation is a Total Lagrange \inred{[Vedere Belytchko se effettivamente e' Total Lagrangian]} with a corotational frame to extract the strain part from the large displacements; small strain and linear material are assumed. 
The Euler-Bernoulli beam kinematic assumption is considered; the formulation follows an Updated Lagrangian approach with a corotational~\cite{Belytschko_book1} frame to extract the strains from  large displacements.
 %
 %and small strains are identified from the large displacement field using a corotational strategy~\cite{Belytschko_book1}. 
%
Nonlinear rigid elements are employed for a correct transfer of displacements and forces between the structural and fluid meshes (see Section~\ref{sec:testcases}). The implementation is based on the penalty method proposed by Belytschko~\cite{Belytschko_book1}.
%
%The description of continuous mechanics is based on the classic solid mechanics theory, for which:
%
% ---------------------------------
%       Solid mechanics
% ---------------------------------
%\begin{eqnarray} \label{e:Structural_problem}  
%\mathcal{S} (\mathbf{U}) = \mathbf{0} \Leftrightarrow
%\begin{cases}
%\mathbf{\nabla} \cdot \mathbf{\sigma} + \mathbf{F_b} = \mathbf{0} \\
%\mathbf{\epsilon} = \mathbf{\epsilon}(\mathbf{U}) \\
%\mathbf{\sigma} = \mathcal{C} : \mathbf{\epsilon} \\
%\left . \mathbf{n} \cdot \mathbf{\sigma} = \mathbf{F_s} \right |_{bnd}
%\end{cases}
%\end{eqnarray}
%where, $\mathbf{\sigma}$ is the Cauchy stress tensor, $\mathbf{F_b}$ are the structural body forces per unit volume, $\mathbf{\epsilon}$ is the strain tensor, $\mathbf{U}$ is the displacement field, $\mathcal{C}$ is the fourth order stiffness tensor, $\mathbf{n}$ is the normal to the body surface and $\mathbf{F_s}$ are surface forces per unit area.
%Kirchoff model to relate stress and strains and kirchoff material in which C is constant. and hyperelastic materials
%The structural problem in system~(\ref{e:Structural_problem}) shows, in order, the equilibrium equation, the strain-displacement equation, featuring the geometrical nonlinearity, the constitutive equation for an elastic material and the boundary condition featuring external surface forces.
%
\par
In its FE discretized form the governing equation is: 
%The problem in system~(\ref{e:Structural_problem}) can be discretized an cast in form: 
%
% ---------------------------------
%       Solid mechanics discrete
% ---------------------------------
\begin{equation} \label{e:Structural_problem_discr}  
\mathcal{S} (\mathbf{u_s}) = \mathbf{f_{s}} - \mathbf{f_{int}} (\mathbf{u_s}) - \mathbf{f_{rig}} (\mathbf{u_s}) = \mathbf{0} 
\end{equation}
%------------------------------
where, $\mathbf{u_s}$, $\mathbf{f_{s}}$, $\mathbf{f_{int}}$ and $\mathbf{f_{rig}}$ are, respectively, the nodal generalized displacements (measured from the unloaded initial configuration), external and internal nodal forces vector, and the contribution of rigid elements to the residual. 
Equation~(\ref{e:Structural_problem_discr}) is solved iteratively with a Newton-Raphson method:
%
% ---------------------------------
%       Solid mechanics discrete Newton
% ---------------------------------
\begin{equation} \label{e:Structural_problem_discr_newton}  
\mathbf{K} \; \Delta \mathbf{u_s} = - \mathcal{S} (\mathbf{u_s})
\end{equation}
%------------------------------
with the Jacobian $ \mathbf{K} = \frac{\partial \mathcal{S} (\mathbf{u_s})}{\partial \mathbf{u_s}}$ retaining the nonlinear contributions of both beam and rigid elements. 
%
%At FE level, the Jacobian $ \mathbf{K} = \frac{\partial \mathcal{S} (\mathbf{u_s})}{\partial \mathbf{u_s}}$ retains the two contributions:
%
% ---------------------------------
%       tangent matrix
% ---------------------------------
%\begin{equation} \label{e:tangent_matr}  
%\mathbf{K} = \mathbf{K_t}+ \mathbf{K_r}
%\end{equation}
%------------------------------
%where $\mathbf{K_t}$ is the nonlinear elastic tangent matrix and $\mathbf{K_r}$ is the nonlinear contribution to the Jacobian of rigid elements.
%
% Rigid elements contributions to equation~(\ref{e:Structural_problem_discr_newton}) are retained in matrix $\mathbf{K_r}$ and in the problem residual $ \mathcal{S} (\mathbf{u_s})$. 
A load-stepping strategy, i.e., a progressive application of the external loads, is used in equation~(\ref{e:Structural_problem_discr}) to facilitate convergence. 
%\inred{[occhio, in realta' anche le forze esterne quando sono derivate per gli spostamenti cambiano, pero' il contributo non ce lo mettiamo anche se si potrebbe - un po' incasinato. Pero, per l'approccio nostro in cui, all'interno di una iterazione FSI le forze vengono passate e poi con esse congelate si itera a convergenza della struttura, la trattazione e' giusta ]}.
%
\par
PyBeam is coded in C++ and its top-level functions are wrapped in Python using SWIG\cite{swigpaper}, to be accessible by external solvers. 
Moreover, like SU2 CFD solver~\cite{albring2016}, it  employs AD by means of CoDiPack library for sensitivites evaluation.
%
%Towards sensitivity evaluation, it has been developed to handle AD, by means of CoDiPack library, in a similar fashion as it was done for the CFD solver of SU2~\cite{albring2016}.
%\emph{PyBeam} is coded in C++ and features CoDiPack for AD-based sensitivities evaluation; its top-level functions are wrapped in python using SWIG\cite{swigpaper}, to be accessible by external solvers. 
%\emph{PyBeam} is coded in C++ and features CoDiPack for AD-based sensitivity evaluation. The top-level functions are wrapped in \emph{python} using SWIG\cite{swigpaper} with the purpose of showing the of the viability of the proposed modular coupling strategy. 
%
%------------------------------
\paragraph{CFD solver}
%------------------------------
%
%
Focus is on transonic flows around aerodynamic bodies governed by the compressible Navier-Stokes equations. 
For this purpose, the flow solver available in the open-source multiphysics suite SU2 is chosen. 
%For this purpose, we use the flow solver available in the open-source multiphysics suite SU2.
Following the work of Economon \textit{et al}~\cite{SU2_AIAAJ2016}., the governing equations formulated in conservative form including the energy equation can be written as: 
\begin{equation}
\mathscr{F}(\mathbf{w}) = \frac{\partial \mathbf{w}}{\partial t} + \nabla \cdot \mathbf{F}^c(\mathbf{w}) - \nabla \cdot  \mathbf {F}^v(\mathbf{w}) - \mathbf{Q}(\mathbf{w}) = \mathbf{0}
\label{eq:constitutive_flow}
\end{equation}
where $\mathbf{w} = (\rho, \rho \mathbf{v}, \rho E)$ is the vector of conservative variables, $\rho$ the flow density, $\mathbf{v}$ the flow velocity and $E$ the total energy per unit mass. 
$\mathbf{Q}(\mathbf{w})$ is a generic source term, $\mathbf{F}^c(\mathbf{w})$ and $\mathbf{F}^v(\mathbf{w})$ are, respectively, the convective and viscous fluxes, and can be written as
\begin{equation}
\mathbf{F}^c(\mathbf{w}) = \begin{pmatrix}
\rho \mathbf{v}\\ 
\rho \mathbf{v} \otimes \mathbf{v} + p\mathbf{I}\\
\rho E \mathbf{v} + p\mathbf{v}
\end{pmatrix}
\label{eq:convective}
\end{equation}
\begin{equation}
\mathbf{F}^v(\mathbf{w}) = \begin{pmatrix}
\boldsymbol{0}\\ 
\boldsymbol{\tau}\\
\boldsymbol{\tau} \cdot \mathbf{v} + \mu^* C_p \nabla T
\end{pmatrix}
\label{eq:viscous}
\end{equation}
where $C_p$ is the specific heat at constant pressure and $T$ is the temperature, calculated using the ideal gas model. The viscous stress tensor is:
\begin{equation}
\boldsymbol{\tau} = \mu_{tot} \left(\nabla \mathbf{v} + \nabla \mathbf{v}^T -\frac{2}{3}\mathbf{I}(\nabla \cdot \mathbf{v})\right)
\label{eq:viscous_stress_tensor}
\end{equation}
where, based on the Boussinesq hypothesis\cite{Wilcox1998}, the total viscosity $\mu_{tot}$ is modelled as a sum of a laminar component which satisfies Sutherland's law\cite{White1974} ($\mu_{lam}$) and a turbulent component ($\mu_{turb}$) which is obtained from the solution of a turbulence model. Finally,
\begin{equation}
\mu^{*} =  \frac{\mu_{lam}}{Pr_l} + \frac{\mu_{turb}}{Pr_t}
\label{eq:mu_star}
\end{equation}
where $Pr_l$ and $Pr_t$ are the laminar and turbulent Prandtl numbers, respectively.
In this investigation, when viscous flows are considered, $\mu_{turb}$ is calculated by means of the one-equation Spalart-Allmaras turbulence model~\cite{SpalartAllmaras}. 

\par
SU2 core is written in C++ and top-level functions are wrapped in Python using SWIG.
Both continuous and discrete adjoint capabilities are provided; in particular, discrete adjoint implementation features CoDiPack for AD-based sensitivities evaluation.
%
%------------------------------
\paragraph{Fluid mesh deformation solver}
%------------------------------
%
For problems involving moving boundaries it is important to account for the modification of the fluid domain. Among the several strategies proposed in the  literature~\cite{Donea_2004}, it has been decided to rely on a linear elastic volume deformation method, which performs well in case of large deformations. 
Such strategy is implemented in the SU2 dedicated mesh deformation solver; it supports AD for gradient evaluation and  its top level functions are wrapped in Python using SWIG.
\par
%
%. This is carried out by the SU2 dedicated mesh deformation solver.
In order to find the new position of the nodes in the fluid domain, the mesh deformation problem can be treated as a pseudo-elastic linear problem\cite{Dwight2009401},
%
% ---------------------------------
%       Linear elastic mesh deformation problem
% ---------------------------------
\begin{equation} \label{e:Mesh_problem}   
\mathscr{M}(\mathbf{z, u_{f}}) = \mathbf{K_{m}} \cdot \mathbf{z} -  \mathbf{\tilde{f}}(\mathbf{ u_{f}}) = 0
\end{equation}
%
% ---------------------------------
% 
where $\mathbf{K_{m}}$ is a fictitious stiffness matrix and $\mathbf{\tilde{f}}$ are fictitious forces which ensure the boundary displacements $\mathbf{u_f}$. 
\par
In problems involving moving mesh boundaries, equation~(\ref{eq:constitutive_flow}) needs to be rewritten with the inclusion of the domain grid points position $\mathbf{z}$, following the Arbitrary Lagrangian-Eulerian (ALE) formulation~\cite{palacios_stanford_2013,palacios_stanford_2014,DONEA1982689,Donea_2004}: 
\begin{equation}
\mathscr{F}(\mathbf{w,z}) = \frac{\partial \mathbf{w}}{\partial t} + \nabla \cdot \mathbf{F}^c(\mathbf{w,z}) - \nabla \cdot  \mathbf {F}^v(\mathbf{w,z}) - \mathbf{Q}(\mathbf{w}) = \mathbf{0}
\label{eq:constitutive_flow_ALE}
\end{equation}
%
%------------------------------
\paragraph{Interfacing method}\label{s:MLS}
%------------------------------
%
To transfer information between the non-conformal structural FEM and CFD grids, an in-house Moving Least Square algorithm is implemented~\cite{SDSUteam_5jour}, based on Radial Basis Functions (RBF)~\cite{Romanelli_2012} and \emph{ANN} library\cite{ANNlibrary}. 
Briefly, given $\mathbf{x_s} \in \mathbb{R}^{N_s}$, the position of the structural nodes and $\mathbf{x_f} \in \mathbb{R}^{N_f}$, the position of the fluid nodes on the moving boundary, it is possible to build a so-called spline matrix $\mathbf{H}_{MLS} = \mathbf{H}_{MLS}(\mathbf{x_s},\mathbf{x_f}) \in \mathbb{R}^{N_f \times N_s}$ such that:
%
% ---------------------------------
%       Splining
% ---------------------------------
\begin{equation} \label{e:MLS_dis}   
\mathbf{u_f} =  \mathbf{H}_{MLS} \cdot \mathbf{u_s},
\end{equation}
\begin{equation} \label{e:MLS_forces}   
\mathbf{f_s} =  \mathbf{H}_{MLS}^{T} \cdot \mathbf{f_f}
\end{equation}
% ---------------------------------
% 
where $N_f$ and $N_s$ are, respectively, the dimensions of the aerodynamic moving surface and structural degrees of freedom, while $\mathbf{u_f}, \; \mathbf{f_f} \in \mathbb{R}^{N_f}$ and $\mathbf{u_s}, \; \mathbf{f_s} \in \mathbb{R}^{N_s}$ are, respectively, the displacements/forces defined on the aerodynamic/structural mesh. 
Aerodynamic forces, i.e.,  $\mathbf{f_f} = \mathbf{f_f} (\mathbf{w,z})$, are provided by the fluid solver for the converged solution of equation~(\ref{eq:constitutive_flow_ALE}).
As already stated in the work of Quaranta \textit{et al}~\cite{Mantegazza2}, employing the transpose of the spline matrix in equation~\eqref{e:MLS_forces} ensures the energy conservation. 
\par
The spline tool has been already successfully applied to a large variety of challenging cases, such as wings with mobile surfaces, free-flying aircraft models~\cite{SDSUteam_6jour} and other cases in which interpolation of information between 1D (structural) and 3D (aerodynamic) topologies had to be performed~\cite{UC3Mteam_Scitech_NITRO}.  
%
%The tool has been already successfully applied to a variety of cases, including the transfer of combined rigid-elastic displacements and featuring mobile surfaces~\cite{SDSUteam_6jour} and the interpolation of information between 1D (structural) models and 3D (aerodynamic) ones~\cite{UC3Mteam_Scitech_NITRO}.  
%
The module, coded in C++ into an independent library, has also been wrapped in Python. 
%
%
%to expose top-level functions and objects \inred{or classes, or methods?}.
%
%------------------------------
\paragraph{Coupling method}
%------------------------------
%
A partitioned approach is employed for the FSI system solution, following a three-field formulation\cite{FARHAT20033,Barcelos20081813}. 
This approach, according to Maute \textit{et al.}~\cite{maute2003911}, is suitable for problems featuring large structural deformations.
%\inmag{[FRASE  BRUTTA] Moreover, by exploiting the modularity of the different disciplines, it can be advantageous for practical applications (especially industry-oriented ones) on realistic test cases.}
%Such approach perfectly reflects the modular nature of the framework, which is considered by the authors as a milestone for deployment of the strategy in realistic and industry-oriented scenarios. % applications (especially industry-oriented ones) on realistic test-cases
%Being based on the principle of modularity of the different sub-solvers, it can be advantageous for practical applications (especially industry-oriented ones) on realistic test-cases.
%For the modular nature of the framework, a
%
%Moreover, according to Maute \textit{et al.}~\cite{maute2003911}, choosing a three-field approach is convenient for problems featuring large structural deformation. 
%According to Maute \textit{et al.}~\cite{maute2003911}, an advanta such approach is convenient for 
%
\par
Recalling the three fields under investigation, namely, structural $\mathcal{S}$, fluid $\mathcal{F}$ and mesh $\mathcal{M}$, the whole FSI system $\mathcal{G}$ can be expressed as a function of the state variables $\mathbf{u_s}$, $\mathbf{w}$ and $\mathbf{z}$ which are, respectively, structural displacements, aerodynamic state variables and fluid mesh nodes displacements. Hence, following Sanchez~\cite{SanchezThesis}, the primal problem is:
%
% ---------------------------------
%       Governing equations
% ---------------------------------
\begin{equation} \label{e:FSI_problem}   
\mathcal{G}(\mathbf{u_s},\mathbf{w},\mathbf{z}) = \begin{cases}
\mathcal{S}(\mathbf{u_s},\mathbf{w},\mathbf{z}) = 0, \\
\mathcal{F}(\mathbf{w},\mathbf{z}) = 0, \\
\mathcal{M}(\mathbf{u_s},\mathbf{z}) = 0,
\end{cases}
\end{equation}
% ---------------------------------
% 
Due to the non matching structural and fluid interfaces, the above system is closed by means of interfacing equations~(\ref{e:MLS_dis},\ref{e:MLS_forces}).
%representing the interfacing.
%\inblue{riscrivere} Due to the nonconformity of structural and fluid meshes, the above set of equations needs to be completed by equations \ref{e:MLS_dis} and \ref{e:MLS_forces} to be closed.
%which are necessary due to effects of nonconformity of structural and fluid meshes
%, tacked by the above-described splining, are implicitly assumed relating  procedure have already been included. \inred{[]DOVE? Semmai ]}.
%
\par
Due to the nonlinear nature of the FSI problem, an iterative approach based on Newton method is sought. At each iteration, the corresponding linear system is solved using a Block-Gauss-Seidel (BGS) strategy,  which suits well the the selected partitioned approach. 
The approximated linear system reads: 
%
% ---------------------------------
%       Linear elastic mesh deformation problem
% ---------------------------------
%\begin{equation} \label{e:NR_procedure}   
%\begin{bmatrix} 
%\frac{\partial \mathcal{S}}{\partial \mathbf{u_s}} & 0 & 0 \\
%0 & \frac{\partial \mathcal{F}}{\partial \mathbf{w}} & 0 \\
%\frac{\partial \mathcal{M}}{\partial \mathbf{u_s}} & 0 & \frac{\partial %\mathcal{M}}{\partial \mathbf{z}}
%\end{bmatrix} 
%\begin{Bmatrix} \Delta \mathbf{u_s} \\ \Delta \mathbf{w} \\ \Delta \mathbf{z}
%\end{Bmatrix} = - 
%\begin{Bmatrix}  \mathcal{S}(\mathbf{u_s},\mathbf{w},\mathbf{z}) \\
%\mathcal{F}(\mathbf{w},\mathbf{z}) \\
%\mathcal{M}(\mathbf{u_s},\mathbf{z})
%\end{Bmatrix},
%\end{equation}
% ---------------------------------
% 
%
% ---------------------------------
%       Linear elastic mesh deformation problem
% ---------------------------------
\begin{equation} \label{e:NR_procedure}   
\begin{bmatrix} 
 \frac{\partial \mathcal{F}}{\partial \mathbf{w}}   & 0 & 0 \\
0 & \frac{\partial \mathcal{S}}{\partial \mathbf{u_s}} & 0 \\
0 & \frac{\partial \mathcal{M}}{\partial \mathbf{u_s}} & \frac{\partial \mathcal{M}}{\partial \mathbf{z}}
\end{bmatrix} 
\begin{Bmatrix} \Delta \mathbf{w} \\  \Delta \mathbf{u_s} \\ \Delta \mathbf{z}
\end{Bmatrix} = - 
\begin{Bmatrix}  \mathcal{F}(\mathbf{w},\mathbf{z}) \\
\mathcal{S}(\mathbf{u_s},\mathbf{w},\mathbf{z}) \\
\mathcal{M}(\mathbf{u_s},\mathbf{z})
\end{Bmatrix}
\end{equation}
% ---------------------------------
% 
in which the upper-right part of the Jacobian has been set to 0\cite{Barcelos20081813}. 
In the work of Degroote \textit{et al.}~\cite{Degroote_2009} slow convergence or divergence of the BGS approach in case of strong FSI interactions (e.g., strong geometrical nonlinearities) is observed. 
%It is noticed in the work of Degroote \textit{et al.}~\cite{Degroote_2009}, chances of slow convergence or divergence of the BGS approach in case of strong FSI interactions (e.g. strong geometrical nonlinearities). 
To ensure the stability of the method, a relaxation parameter $\alpha$ is applied to the boundary displacements~\cite{Irons_1969}:
%
%------------------------------------------%
% EQUATION    relaxation                    %
%------------------------------------------%
\begin{equation} \label{e:relaxation}
\mathbf{u_{f}}^* = \alpha \mathbf{u_{f}}^n + (1-\alpha) \mathbf{u_{f}}^{n-1}.
\end{equation}
%------------------------------------------%
%
where ${n}$ and ${n-1}$ are, respectively, the current and previous BGS subiterations.
\par
\begin{figure}[!htbp]
     \centering
     \vspace{5mm}
         \includegraphics[width=\linewidth]{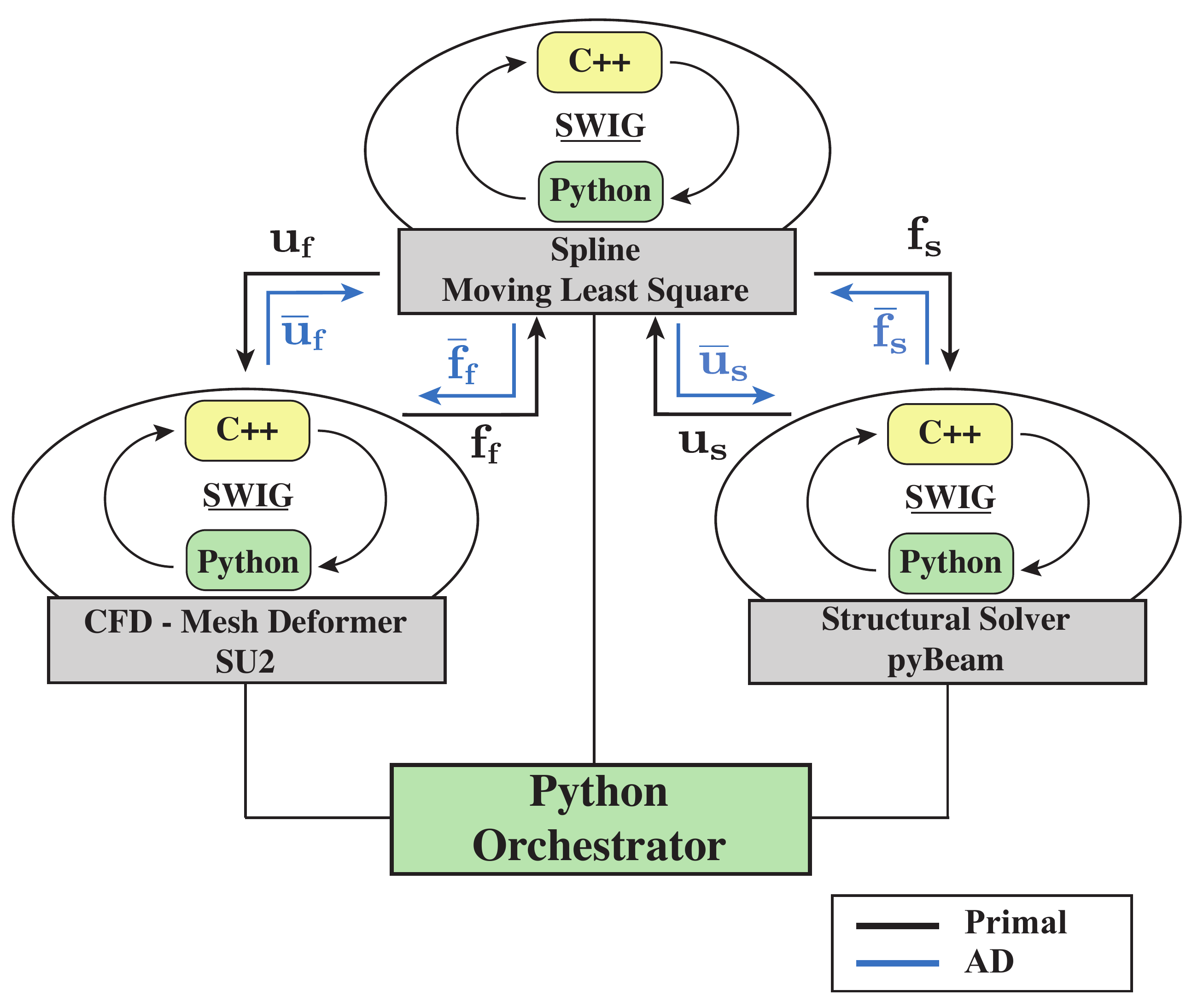}
         \caption{Framework layout for Primal and AD modes.}
         \label{f:Workflow_pybeam}
\end{figure}

 
Concerning the implementation, a Python orchestrator links the wrapped libraries and  allows the sequential solution of each discipline within a single FSI iteration. 
%
%The structural solver pyBeam has been developed for this work and operates as a C++ library, wrapped with python using SWIG\cite{swigpaper}. 
Structural displacements $\mathbf{u_s}$ are accessible from pyBeam module;  they are interpolated into the fluid boundary using equation~\eqref{e:MLS_dis} after the spline matrix has been assembled by the interface module.
The fluid boundary displacements are then transferred to the mesh solver in SU2 via its Application Programming Interface (API)\cite{Sanchez2016b}. 
A new value of the aerodynamic forces on the boundary is obtained after a CFD simulation in SU2 and interpolated back into the structural  model using equation~\eqref{e:MLS_forces}. 
The primal solver layout is shown in Figure~\ref{f:Workflow_pybeam}; 
its algorithm is resumed in Algorithm~\ref{a:Primal}.
%both the Primal and the reverse AD computational paths, and 
 % $\mathbf{w}$, $\mathbf{z}$, $\mathbf{u_{s}}$, $\mathbf{u_{f}}$, $\mathbf{f_{s}}$, $\mathbf{f_{f}}$ to 0
 %------------------------------------------% 
% algorithm   primal                    %
%------------------------------------------%
\begin{algorithm} \caption{Aerostructural primal solver} \label{a:Primal}
\setstretch{1.2}
\SetAlgoLined
Initialize $N =1, (\mathbf{w}$, $\mathbf{z}$, $\mathbf{u_{s}}$, $\mathbf{u_{f}}$, $\mathbf{f_{s}}$, $\mathbf{f_{f}}) = \mathbf{0}$\\
 \While{$N \leq N_{FSI}$}{
  Run CFD solver: $\mathbf{z} \rightarrow \mathbf{w}, \mathbf{f_{f}}, C_D, C_L$ \\
  \hspace{0.3cm} \textbf{while} $\norm{\mathbf{w}_{k_f} - \mathbf{w}_{k_f+1}} \leq \epsilon_f$ \textbf{do}: iterate $k_f$ \\
  %\hspace{0.1cm}\While{$\norm{\mathbf{w}_{k_f} - \mathbf{w}_{k_f+1}} \leq \epsilon_f$}{ \hspace{0.1cm} iterate $k_f$}\
  Transfer loads (Spline): $\mathbf{f_{f}} \rightarrow \mathbf{f_{s}}$ \\
  Run structural FEM solver: $\mathbf{f_{s}} \rightarrow \mathbf{u_{s}}$ \\
  \hspace{0.3cm} \textbf{while} $\norm{\mathbf{u_s}_{k_s} - \mathbf{u_s}_{k_s+1}} \leq \epsilon_s$ \textbf{do}: iterate $k_s$ \\
  %\While{$\norm{\mathbf{u_s}_{k_s} - \mathbf{u_s}_{k_s+1}} \leq \epsilon_s$}{iterate $k_s$} \
  Transfer displacements (Spline): $\mathbf{u_{s}} \rightarrow \mathbf{u_{f}}$ \\
  Run mesh deformation solver: $\mathbf{u_{f}} \rightarrow \mathbf{z}$ }
\end{algorithm}
%------------------------------------------%
%
%------------------------------------------%
% FIGURE    pyBeam Workflow                    %
%------------------------------------------%
%------------------------------------------%
%
%
%===================================================
\subsection{Coupled aerostructural adjoint method}
\label{sec:Adjoint_theory}
%====================================================
%
It has already been mentioned that each of the described modules features CoDiPack for adjoint AD-based sensitivities evaluation, being one of the contributions of the present study to show how different AD-based modules can be interfaced, by means of a high-level Python wrapper, for the purpose of coupled sensitivities evaluation.
%in fact, to demonstrate, within the context of the here considered aerostructural problem, how different AD-based modules can be interfaced for the purpose of coupled sensitivities evaluation. 
To extend the generality of the proposed approach, the interested reader will notice that, using the method described in the following for sensitivities calculation, it is not essential for each discipline module to feature AD, as far as they can provide the adequate input/output cross-dependency terms. 
\par
Following the work of Sanchez~\textit{et al.}~\cite{sanchez2018} let's define the Objective Function (OF) and DVs for the current optimization problem. The OF that will be considered in this effort is the aerodynamic drag: 
%
%------------------------------------------%
%  Drag                                    %
%------------------------------------------%
\begin{equation} \label{e:obj}
J = C_D(\mathbf{w,z})
\end{equation}
%------------------------------
%
%DVs are the parameters describing the undeflected shape of the wing (jig-shape).
Table~\ref{t:list} summarizes the set of state and design variables. 
%
%------------------------------------------%
% TABLE Variables                  %
%------------------------------------------%
\begin{table}[!t]
\caption{Complete set of state and design variables for aerostructural shape optimization}
\label{t:list}       
\centering
\begin{tabular}{ l l }
\hline\noalign{\smallskip}
 & \textbf{State variables} \\ 
\noalign{\smallskip}\hline\noalign{\smallskip}
$\mathbf{u_s}$ & Structural displacements \\
$\mathbf{w}$ & Flow conservative variables \\
$\mathbf{z}$ & Volume mesh displacements \\
$\mathbf{f_f}$ & Fluid loads \\
$\mathbf{f_s}$ & Structural loads \\
$\mathbf{u_f}$ & Displacements of wing surface due to deflection \\
$\mathbf{u_{tot}}$ & Cumulative displacements of wing surface\\% due to structure deflection and jig-shape change \\
\noalign{\smallskip}\hline\noalign{\smallskip}
& \textbf{Design variables} \\ 
\noalign{\smallskip}\hline\noalign{\smallskip}
$\mathbf{u_{F_{\alpha}}}$ & Variation of the jig-shape\\
\noalign{\smallskip}\hline
\end{tabular}
\end{table}
%------------------------------------------%
%
Notice that the displacement of the wing surface $\mathbf{u_{tot}}$ is expressed as the sum of the displacements due to the jig-shape redesign $\mathbf{u_{F_{\alpha}}}$ and the displacements due to the deflection (aerostructural coupling). 
Even though  both OF and DVs are relative to aerodynamics, the problem is still multidisciplinary due to the aerostructural coupling of the governing equations. 
The proposed optimization framework is flexible and general, and allows to freely choose OF, constraints and DVs for any of the considered disciplines, accordingly with the implementation of the discipline solvers.
\par
Let the complete set of equations of the primal problem, illustrated in section~\ref{sec:primal}, be rewritten in the form of \emph{fixed-point} iterators~\cite{Korivi1992AnIS,albring2016}:
%------------------------------------------%
% equation    fixed-point                   %
%------------------------------------------%
\begin{subequations} \label{e:FP}
\begin{equation} \label{e:FP_fluid} 
\mathrm{\mathbf{F}}(\mathbf{w,z}) - \mathbf{w} = 0
\end{equation}
\begin{equation}\label{e:FP_fluidforces}  
\mathrm{\mathbf{F_f}}(\mathbf{w,z}) - \mathbf{f_f} = 0
\end{equation}
\begin{equation} \label{e:FP_mesh} 
\mathrm{\mathbf{M}}(\mathbf{u_{tot}}) - \mathbf{z} = 0
\end{equation}
\begin{equation}  \label{e:FP_forces}
\mathbf{H}_{MLS}^{T} \cdot \mathbf{f_f} -  \mathbf{f_s} =0
\end{equation}
\begin{equation}  \label{e:FP_structure} 
\mathrm{\mathbf{S}}(\mathbf{u_s,f_s}) - \mathbf{u_s} = 0
\end{equation}
\begin{equation}  \label{e:FP_displacements} 
\mathbf{H}_{MLS} \cdot \mathbf{u_s} -  \mathbf{u_f}  =0
\end{equation}
\begin{equation}  \label{e:FP_displ_composition} 
\mathbf{u_{tot}} - \mathbf{u_f} - \mathbf{u_{F_{\alpha}}} =0
\end{equation}
\end{subequations}
%------------------------------------------%
%------------------------------------------%
% equation    example                   %
%------------------------------------------%
%\begin{empheq}[right={\empheqrbrace\ \forall t \in \mathbb{N}}]{align}
%                  f(x) &= ax^2 + bx + c\zsaveposy{top}
%  \raisebox{-.5\dimexpr\zposy{top}sp-\zposy{bot}sp}[0pt][0pt]{$
%    \hspace{15pt}\left.\rule{0pt}{\dimexpr\zposy{top}sp-\zposy{bot}sp}\right\}\ \forall i \in \mathbb{N}
%  $} \\
%         ax^2 + bx + c &= g(x)\zsaveposy{bot} \\
%                  h(x) &= dx^3 + ex^2 + fx + h \hspace{35pt} \\
%  dx^3 + ex^2 + fx + h &= i(x)
%\end{empheq}

%------------------------------
%

\par%
In system~(\ref{e:FP}) equation~(\ref{e:FP_fluid}) is the fixed-point version of equation~(\ref{eq:constitutive_flow_ALE}) and, together with fluid loads and objective function evaluation (equations~(\ref{e:FP_fluidforces},\ref{e:obj}) respectively) represents the core of the aerodynamic solver. 
Equation~(\ref{e:FP_mesh}) is the fixed-point form of mesh deformation problem  (equation~(\ref{e:Mesh_problem})), whereas equation~(\ref{e:FP_structure}) is the fixed-point version of the structural problem (equation~(\ref{e:Structural_problem_discr})). % evaluated by pyBeam FEM solver. 
Operators $\mathrm{\mathbf{F}}$, $\mathrm{\mathbf{F_f}}$, $\mathrm{\mathbf{M}}$ and $\mathrm{\mathbf{S}}$ are only defined at the solution of the primal problem. 
\par
Equations~(\ref{e:FP_fluid}-\ref{e:FP_mesh},\ref{e:obj}) are handled by the SU2 suite while equation~(\ref{e:FP_structure}) is handled by pyBeam. 
Cross-dependencies (displacements $\mathbf{u_s}$/$\mathbf{u_f}$ and forces $\mathbf{f_s}$/$\mathbf{f_f}$) are handled by the orchestrator at high-level after the spline matrix has been assembled. 
%
%%
%------------------------------------------%
% TABLE Variables                  %
%------------------------------------------%
%------------------------------------------%
%
\par
The optimization problem is formulated as:
%------------------------------------------%
% equation    minimization                   %
%------------------------------------------%
%\begin{equations} \label{e:minimization}
\begin{align}
\min_{\mathbf{u_{F_{\alpha}}}} \;\;\;& J(\mathbf{w,z}) \\
\mbox{subject to:} \;\;\;& \mathrm{\mathbf{F}}(\mathbf{w,z}) - \mathbf{w} = 0 \nonumber \\
 & \mathrm{\mathbf{F_f}}(\mathbf{w,z}) - \mathbf{f_f} = 0 \nonumber \\
 & \mathrm{\mathbf{M}}(\mathbf{u_{tot}}) - \mathbf{z} = 0 \nonumber \\
 & \mathbf{H}_{MLS}^{T} \cdot \mathbf{f_f} -  \mathbf{f_s} =0 \nonumber \\ 
  & \mathrm{\mathbf{S}}(\mathbf{u_s,f_s}) - \mathbf{u_s} = 0 \nonumber \\ 
  & \mathbf{H}_{MLS} \cdot \mathbf{u_s} -  \mathbf{u_f}  =0 \nonumber \\
  & \mathbf{u_{tot}} - \mathbf{u_f} - \mathbf{u_{F_{\alpha}}} =0 \nonumber
\end{align}
%\end{equations}
%------------------------------------------%
%
Such problem can be reformulated in the equivalent unconstrained optimization problem defined with the Lagrangian $\mathscr{L}$:
%------------------------------------------%
% equation    lagrangian                   %
%------------------------------------------%
\begin{equation} \label{e:lagrangian}
\begin{aligned}
\mathscr{L}(\mathbf{w, \bar{w}, z, \bar{z}, u_s,\bar{u}_s,u_f,\bar{u}_f,f_s,\bar{f}_s,f_f, \bar{f}_f, u_{tot},\bar{u}_{tot} })  = \\
J(\mathbf{w,z})  + \mathbf{\bar{w}}^T \left[  \mathrm{\mathbf{F}}(\mathbf{w,z}) - \mathbf{w}\right] + \mathbf{\bar{z}}^T \left[ \mathrm{\mathbf{M}}(\mathbf{u_{tot}}) - \mathbf{z}\right] +\\ 
 \mathbf{\bar{u}_{tot}}^T \left[  \mathbf{u_{tot}} - \mathbf{u_f} - \mathbf{u_{F_{\alpha}}} \right] +  \mathbf{\bar{u}_{f}}^T \left[  \mathbf{H}_{MLS} \cdot \mathbf{u_s} -  \mathbf{u_f} \right] + \\
 \mathbf{\bar{u}_{s}}^T \left[  \mathrm{\mathbf{S}}(\mathbf{u_s,f_s}) - \mathbf{u_s} \right]+  \mathbf{\bar{f}_{s}}^T \left[  \mathbf{H}_{MLS}^{T} \cdot \mathbf{f_f} -  \mathbf{f_s}  \right] + \\
 \mathbf{\bar{f}_{f}}^T \left[  \mathrm{\mathbf{F_f}}(\mathbf{w,z}) - \mathbf{f_f} \right]
\end{aligned}
\end{equation}
%------------------------------------------%
%
in which the Lagrangian multipliers $\mathbf{\bar{w}}$, $\mathbf{\bar{z}}$, $\mathbf{\bar{u}_{tot}}$, $\mathbf{\bar{u}_f}$, $\mathbf{\bar{u}_s}$, $\mathbf{\bar{f}_s}$ and $\mathbf{\bar{f}_f}$, corresponding to the adjoint of the state variables, are introduced. 
\par
Imposing the Karush-Kuhn-Tucker (KKT) conditions it is possible to: retrieve the state equations~(\ref{e:FP}) by differentiation of the Lagrangian with respect to the adjoint variables; obtain the set of adjoint equations differentiating the Lagrangian with respect to the state variables:
%------------------------------------------%
% equation    FP adjoint equations         %
%------------------------------------------%
\begin{subequations} \label{e:adjoint}
\begin{equation} \label{e:ad_w} 
\frac{\partial \mathscr{L}}{\partial \mathbf{w}} = \frac{\partial J}{\partial \mathbf{w}} + 
\mathbf{\bar{w}}^T \left[ \left. \frac{\partial  \mathrm{\mathbf{F}} }{\partial \mathbf{w}} \right|_{\mathbf{w^*}, \mathbf{z^*}} -1\right] + 
\mathbf{\bar{f}_{f}}^T \;  \left. \frac{\partial  \mathrm{\mathbf{F_f}} }{\partial \mathbf{w}} \right|_{\mathbf{w^*}, \mathbf{z^*}} = \mathbf{0}
\end{equation}
\begin{equation} \label{e:ad_z} 
\frac{\partial \mathscr{L}}{\partial \mathbf{z}} = \frac{\partial J}{\partial \mathbf{z}} + 
\mathbf{\bar{w}}^T \;  \left. \frac{\partial  \mathrm{\mathbf{F}} }{\partial \mathbf{z}} \right|_{\mathbf{w^*}, \mathbf{z^*}} + \mathbf{\bar{f}_{f}}^T \;  \left. \frac{\partial  \mathrm{\mathbf{F_f}} }{\partial \mathbf{z}} \right|_{\mathbf{w^*}, \mathbf{z^*}} +  \mathbf{\bar{z}}^T = \mathbf{0}
\end{equation}
\begin{equation} \label{e:ad_utot} 
\frac{\partial \mathscr{L}}{\partial \mathbf{u_{tot}}} = \frac{\partial J}{\partial \mathbf{u_{tot}}} + 
\mathbf{\bar{z}}^T \;  \left. \frac{\partial  \mathrm{\mathbf{M}} }{\partial \mathbf{u_{tot}}} \right|_{\mathbf{u_{tot}^*}} + \mathbf{\bar{u}_{tot}}^T = \mathbf{0}
\end{equation}
\begin{equation} \label{e:ad_uf} 
\frac{\partial \mathscr{L}}{\partial \mathbf{u_f}} = \frac{\partial J}{\partial \mathbf{u_f}} - \mathbf{\bar{u_{f}}}^T - \mathbf{\bar{u}_{tot}}^T = \mathbf{0}
\end{equation}
\begin{equation} \label{e:ad_us} 
\frac{\partial \mathscr{L}}{\partial \mathbf{u_s}} = \frac{\partial J}{\partial \mathbf{u_s}} + \mathbf{\bar{u}_s}^T \left[ \left. \frac{\partial  \mathrm{\mathbf{S}} }{\partial \mathbf{u_s}} \right|_{\mathbf{u_s^*}, \mathbf{f_s^*}} -1\right] + 
 \mathbf{\bar{u_{f}}}^T \;  \mathbf{H}_{MLS} = \mathbf{0}
\end{equation}
\begin{equation} \label{e:ad_fs} 
\frac{\partial \mathscr{L}}{\partial \mathbf{f_s}} = \frac{\partial J}{\partial \mathbf{f_s}} + \mathbf{\bar{u}_s}^T \left. \frac{\partial  \mathrm{\mathbf{S}} }{\partial \mathbf{f_s}} \right|_{\mathbf{u_s^*}, \mathbf{f_s^*}} - 
 \mathbf{\bar{f}_{s}}^T  = \mathbf{0}
\end{equation}
\begin{equation} \label{e:ad_ff} 
\frac{\partial \mathscr{L}}{\partial \mathbf{f_f}} = \frac{\partial J}{\partial \mathbf{f_f}} -\mathbf{\bar{f}_{f}}^T + \mathbf{\bar{f}_{s}}^T \;  \mathbf{H}_{MLS}^T = \mathbf{0},
\end{equation}
\end{subequations}
%------------------------------
%
and, finally, retrieve the optimality condition differentiating the Lagrangian with respect to the DVs. For a local minimum, it holds that:
%------------------------------------------%
% equation    sensitivity                   %
%------------------------------------------%
\begin{equation}\label{eq:gradient_dvs}  
\frac{\mathrm{d} J}{\mathrm{d} {\mathbf{u_{F_{\alpha}}}}} = 
\frac{\partial \mathscr{L}}{\partial  {\mathbf{u_{F_{\alpha}}}} } = 
\frac{\partial J}{\partial  {\mathbf{u_{F_{\alpha}}}}} - \mathbf{\bar{u}_{tot}}^T = \mathbf{0}
\end{equation}
%------------------------------------------%
%
%
The adjoint variables can be computed solving the system~(\ref{e:adjoint}), and are then used for gradient evaluation in equation~(\ref{eq:gradient_dvs}). 
The matrix-vector products in the form $\mathbf{\bar{y}}^T \left. \frac{\partial  \mathrm{\mathbf{A}}}{\partial \mathbf{x} }\right|_{\mathbf{x^*}}$ in equations~(\ref{e:adjoint}) are evaluated using the AD tool CoDiPack about the solution of the aerostructural primal problem (i.e., at $\mathbf{{w}^*}$, $\mathbf{{z}^*}$, $\mathbf{{u_{tot}}^*}$, $\mathbf{{u_f}^*}$, $\mathbf{{u_s}^*}$, $\mathbf{{f_s}^*}$ and $\mathbf{f_f^*}$). 
\par
To solve the general problem $\mathbf{\bar{x}} = \mathbf{\bar{y}}^T \left. \frac{\partial  \mathrm{\mathbf{A}}}{\partial \mathbf{x} }\right|_{\mathbf{x^*}}$, given the solution of the primal problem $\mathbf{x^*}$, the primal solver $\mathrm{\mathbf{A}}$ is advanced for one iteration which is recorded with CoDiPack. 
Once the solution is recorded, CoDiPack evaluates $\mathbf{\bar{x}}$ for a given value of $\mathbf{\bar{y}}$, provided in input by the user. 
This strategy is thought to avoid to store and operate directly on large-scale matrices such as $ \frac{\partial  \mathrm{\mathbf{F}} }{\partial \mathbf{z}}$ and $ \frac{\partial  \mathrm{\mathbf{F}} }{\partial \mathbf{w}}$ of equations~(\ref{e:adjoint}), whose complexity has already been pointed out by Maute \textit{et al.}\cite{maute2003911} and Barcelos \textit{et al.}\cite{Barcelos20081813}.
%Equations~(\ref{e:adjoint}) are evaluated around the solution of the aerostructural problem (i.e. $\mathbf{{w}^*}$, $\mathbf{{z}^*}$, $\mathbf{{u_{tot}}^*}$, $\mathbf{{u_f}^*}$, $\mathbf{{u_s}^*}$, $\mathbf{{f_s}^*}$ and $\mathbf{\bar{f_f}^*}$) using an iterative BGS procedure in a similar fashion as done for the primal problem.
%
\par
\begin{figure}[!htbp]
     \centering
     \vspace{5mm}
         \includegraphics[width=0.9\linewidth]{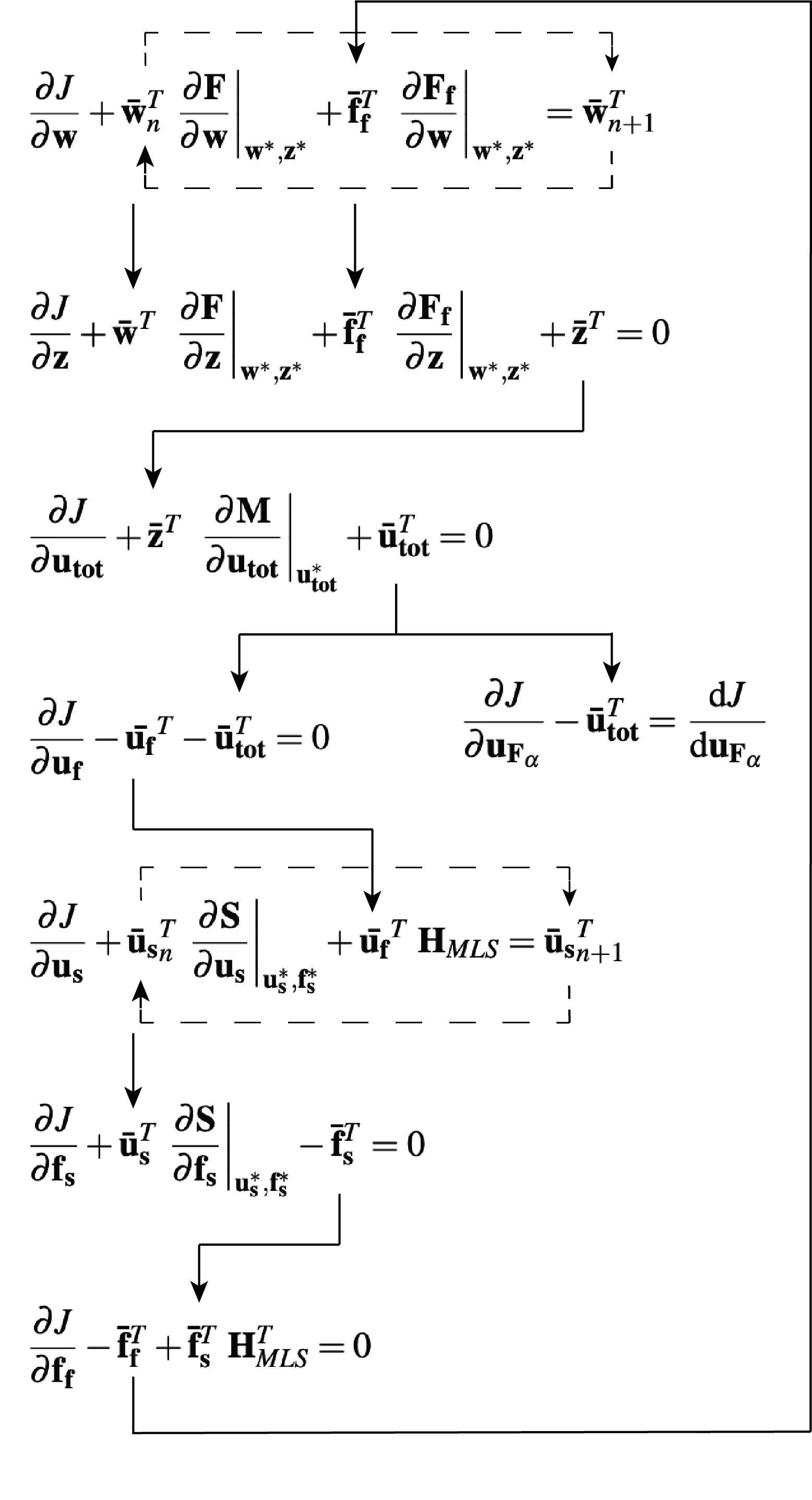}
         \caption{Reverse computational path of the adjoint FSI solver.}
         \label{f:Reverse}
\end{figure}

%
%
%\inblue{[ROCCO riscrivere]}
%The strategy for solving the adjoint system of equations needs to reflect the modular/partitioned approach and comply with practical constraints; hence, it is similar to the BGS staggered solution used in the primal problem (see Figure~\ref{f:Workflow_pybeam}) with the consequence that, even though the adjoint system of equations is linear, an iterative process is required. 
%
%One of the key point in solving  each field (or block fields) adjoint equation is how the matrix-vector products in the form $\mathbf{\bar{y}}^T \left. \frac{\partial  \mathrm{\mathbf{F}}}{\partial \mathbf{x} }\right|_{\mathbf{x^*}}$ in equations~(\ref{e:adjoint}) are handled. The matrices representing the 
%
%
%The field sensitivities with respect to the field state variables cannot be stored for typical  industrial problems. Moreover, modularity does not  allow for a simultaneous solution. 
%The procedure, repeated till convergence,
%The procedure, repeated till convergence, is similar to the BGS staggered solution used in the primal problem (see Figure~\ref{f:Workflow_pybeam}). Once the adjoint variables have been calculated, equation~(\ref{eq:gradient_dvs}) is used to evaluate the sensitivities. 
%Once the adjoint variables have been calculated, equation~(\ref{eq:gradient_dvs}) is used to evaluate the sensitivities. 
The reverse computational path for sensitivities calculation is summarized in Algorithm~\ref{a:Reverse} and Figure~\ref{f:Reverse}.
%ll matrix-vector products in the form $\mathbf{\bar{y}}^T \left. \frac{\partial  \mathrm{\mathbf{F}}}{\partial \mathbf{x} }\right|_{\mathbf{x^*}}$ in equations~(\ref{e:adjoint}) are evaluated using the AD tool \emph{CoDiPack}\cite{sagebaum2017high} around the solution of the aerostructural primal problem (i.e. $\mathbf{{w}^*}$, $\mathbf{{z}^*}$, $\mathbf{{u_{tot}}^*}$, $\mathbf{{u_f}^*}$, $\mathbf{{u_s}^*}$, $\mathbf{{f_s}^*}$ and $\mathbf{f_f^*}$). 
%\inblue{Qui sotto molto poco chiaro}.
%More in details, once the primal has converged, an extra iteration of the primal (in fixed form) is launched and all dependency paths registered by \emph{CoDiPack} in reverse mode. 
%
%For the generic problem $\mathbf{\bar{x}} = \mathbf{\bar{y}}^T \left. \frac{\partial  \mathrm{\mathbf{F}}}{\partial \mathbf{x} }\right|_{\mathbf{x^*}}$, given the solution of the general primal problem $\mathbf{x^*}$, the primal solver $ \mathrm{\mathbf{F}}$ is advanced for one iteration which is recorded with CoDiPack. For the recording procedure, $\mathbf{x}$ is set as input while $\mathbf{y}$ is the output. Once the solution is recorded, the adjoint problem is iteratively solved with AD.
%
%\inred{About the complexity of storing matrices such as $ \frac{\partial  \mathrm{\mathbf{F}} }{\partial \mathbf{z}}$ and $ \frac{\partial  \mathrm{\mathbf{F}} }{\partial \mathbf{w}}$ for adjoint applications both Maute \textit{et al.}\cite{maute2003911} and Barcelos \textit{et al.}\cite{Barcelos20081813}}.
%
Within the context of the current modular framework, AD is applied to every module and cross-term sensitivities are propagated backward to each discipline by the top-level orchestrator. 
The procedure, repeated till convergence, is conceptually similar to the BGS staggered solution used in the primal problem (see Figure~\ref{f:Workflow_pybeam}). 
%Once the adjoint variables have been calculated, equation~(\ref{eq:gradient_dvs}) is used to evaluate the sensitivities. 

 %------------------------------------------% 
% algorithm   primal                    %
%------------------------------------------%
\begin{algorithm} \caption{Aerostructural adjoint problem} \label{a:Reverse}
\setstretch{1.2}
\SetAlgoLined
Initialize $N =1, (\mathbf{\bar{w}},\; \mathbf{\bar{z}},\; \mathbf{\bar{u}_{tot}},\; \mathbf{\bar{u}_f},  \mathbf{\bar{u}_s}, \;\mathbf{\bar{f}_s},\; \mathbf{\bar{f}_f}) = \mathbf{0}$\\
 \While{$N \leq N_{ADJ}$}{
 Run fluid adjoint (eq.~(\ref{e:ad_w})): $ \mathbf{\bar{f}_f}\rightarrow \mathbf{\bar{w}}$ \\
 \hspace{0.3cm} \textbf{while} $\norm{\mathbf{\bar{w}}_{k_f} - \mathbf{\bar{w}}_{k_f+1}} \leq \epsilon_f$ \textbf{do}: iterate $k_f$ \\
 Evaluate $\mathbf{\bar{z}}$ (eq.~(\ref{e:ad_z})): $ \mathbf{\bar{f}_f}, \mathbf{\bar{w}} \rightarrow \mathbf{\bar{z}}$ \\
 Run mesh adjoint (eq.~(\ref{e:ad_utot})):  $ \mathbf{\bar{z}}\rightarrow \mathbf{\bar{u}_{tot}}$ \\
 Evaluate $\mathbf{\bar{u}_f}$ (eq.~(\ref{e:ad_uf})): $\mathbf{\bar{u}_{tot}} \rightarrow \mathbf{\bar{u}_f}$ \\
 Run struct. adjoint (eq.~(\ref{e:ad_us})): $\mathbf{\bar{u}_{f}} \rightarrow \mathbf{\bar{u}_s}$ \\
 \hspace{0.3cm} \textbf{while} $ \norm{\mathbf{\bar{u}_s}{}_{k_s} - \mathbf{\bar{u}_s}{}_{k_s+1}} \leq \epsilon_s $  \textbf{do}: iterate $k_s$ \\ 
 Evaluate $\mathbf{\bar{f}_s}$ (eq.~(\ref{e:ad_fs})): $ \mathbf{\bar{u}_s} \rightarrow \mathbf{\bar{f}_s}$ \\
 Evaluate $\mathbf{\bar{f}_f}$ (eq.~(\ref{e:ad_ff})): $\mathbf{\bar{f}_s}\rightarrow \mathbf{\bar{f}_f}$ \\
 }
\end{algorithm}
%------------------------------------------%

%------------------------------------------%
% FIGURE    adjoint reverse Workflow       %
%------------------------------------------%
%------------------------------------------%
%
%------------------------------
%
%------------------------------
\subsection{Aerostructural wing shape optimization}
\label{sec:optimization}
%------------------------------
%------------------------------
%This section summarizes the solution strategy of the proposed aerostructural shape optimization framework. 
To facilitate communication with the FSI orchestrator, the optimization tool is purely Python-based and wraps the modules used for OF, constraints, gradients and Jacobian evaluation.
The algorithm selected for the gradient-based optimization is the Sequential Least Square Quadratic Programming (SLSQP)~\cite{kraft1988software}, a gradient based algorithm that uses a Broyden Fletcher Goldfarb Shanno (BFGS)-based second order approximation of the objective function. 
\par
With respect to the previous section and for optimization purposes, the number of DVs is reduced by 
%means of a Free Form Deformation (FFD) technique. 
%relating displacement of wing surface 
%$\mathbf{u_{F_{\alpha}}}$ 
%As mentioned, OF is  the wing's $C_D$ and the chosen design variables are the aerodynamic grid surface nodes displacements with respect to the original jig-shape, as summarized in Table~(\ref{t:list}). \inred{[Occhio, hai usato diverse volte resume al posto di summerize]}
%
%\par
% 
%\inred{It's not the main reason to use FFD box}Although computational time is almost independent of the number of design variables for AD-based gradient evaluation methods, Kraft recommends only moderately large size problems for SLSQP~\cite{kraft1988software}.
%To perform the optimization it is decided to 
parametrizing the wing jig-shape with an FFD technique\cite{Samareh_2004}. %volume. 
Aerodynamic grid surface node positions are linked to the FFD Control Points (CPs)  with a trivariate interpolation based on Bezier's basis functions:
%
 %------------------------------------------% 
% algorithm   Bezier                    %
%------------------------------------------%
\begin{equation} \label{e:Bezier}
 %\mathbf{x}^{srf}
 \mathbf{x}_{\mathbf{F}_{\alpha}} = \sum_{i=0}^l \sum_{j=0}^m  \sum_{k=0}^n \; N_i(\mu) \;N_j(\nu) \; N_k(\xi) \;  \mathbf{x}^{CP}_{ijk}
\end{equation}
%------------------------------------------%
%
In equation~(\ref{e:Bezier}), $ {\mathbf{x}}_{{\mathbf{F}}_{\alpha}}$  is the coordinate vector of the generic node of the aerodynamic mesh lying on the wing surface, $\mathbf{x}^{CP}_{ijk}$ is the position of the CP, identified by indexes $i,j,k$; $N$ are Bernstein polynomials and $\mu$, $\nu$, $\xi$ are parametric coordinates evaluated with a point-inversion procedure\cite{UC3M_ASO1Pustina}. 
New DVs are then the FFD box CPs, and  the gradient of the OF with respect to them is easily evaluated applying the chain rule:
%by means of equations~(\ref{eq:gradient_dvs}-\ref{e:Bezier})
%
%----------
\begin{equation} \label{e:ChainRule}
\frac{\mathrm{d} J}{\mathrm{d} {\mathbf{u}^{CP}_{ijk}}  } = \frac{\mathrm{d} J} {\mathrm{d} {\mathbf{u_{F_{\alpha}}}}} 
\frac{\mathrm{d} {\mathbf{u_{F_{\alpha}}}}} {\mathrm{d} {\mathbf{u}^{CP}_{ijk}}  }
\end{equation}
%--------------------
%aerodynamic grid surface node displacements of equation~(\ref{eq:gradient_dvs}), i.e., $\frac{\mathrm{d} J}{\mathrm{d} {\mathbf{u_{F_{\alpha}}}}}$ are  projected to the CP displacements using equation~(\ref{e:Bezier}). 
%
FFD deformation strategies are largely used in literature\cite{Brezillon2012,Martins2004HighFidelityAD,kennedyScitech2014,KenwayMultipoint2014}, being independent of the grid topology and easy to use in automatic processes;
%and providing direct analytical relations between the two grids; 
they are well suited for cases in which the topology of the geometry is not expected to change (e.g., wings and fuselages)~\cite{Liem2017ExpectedDM}. 
Even though CPs do no have any direct engineering interpretation, strategical use of CP displacements can achieve consistent changes in the wing twist, chord and span~\cite{Hoogervorst2017WingAO} while ensuring the continuity of the surface. 
\par
An important reason to reduce the number of DVs is due to inherent limitations of the optimization algorithm:  Kraft recommends only moderately large size problems for SLSQP~\cite{kraft1988software}, although, for adjoint-based gradient evaluation methods, computational time is almost independent of the number of DVs.
\par
As a reasonable approximation, the interface matrix of equations~(\ref{e:MLS_dis},\ref{e:MLS_forces}) and the FFD box parametric coordinates of equation~(\ref{e:Bezier}) are evaluated for the initial jig-shape configuration and held constant throughout the whole optimization process, instead of being updated for each variation of the jig-shape. 
%Reason of that lies in the assumption that the effect of the change in position of the aerodynamic mesh nodes belonging to the wing surface due to redesign is negligible in building the interface matrix and performing the point-inversion.}
%------------------------------
\paragraph{Constraints}
%\label{sec:fem_solver}
%------------------------------
%
So far no mention to the optimization constraints has been made, to focus on the aerostructural coupling problem and relative sensitivities. 
%However, two kinds of constraints are considered during optimization. 
%
The lift coefficient $C_L$ at which the drag is  measured is prescribed and, hence, held constant throughout the optimization.
%at a single flight condition (single-point optimization), it is chosen to keep lift coefficient $C_L$ constant. %
%Being the objective function the drag \inmag{measured} at a single flight condition (single-point optimization), it is chosen to keep lift coefficient $C_L$ constant. %
In the proposed framework, fixed $C_L$ constraint is imposed by gradually changing the angle of attack during the iterative process~\cite{Martins2004HighFidelityAD}. 
With the above procedure, the constraint is accommodated internally in the aerostructural solver and is not treated at optimization level. 
This feature was originally present in the SU2 aerodynamic shape optimization tool\cite{albring2016}, and has been extended to the coupled aerostructural problem. 
\par
Geometrical constraints are imposed using SU2 module \emph{GEO}. 
Providing the topology of the aerodynamic body and exploiting the FFD box parametrization, SU2\_GEO can evaluate several kinds of constraints (e.g., wing curvature, volume and dihedral, airfoil chord, thickness, twist and LE radius) and their gradients with respect to the CP displacements, by means of FDs. 
%
%For the herein proposed optimization, the thickness ratios (t/c) measured at several control sections of the wing have been constrained to avoid reducing the internal volume and keep realistic shapes of the wing. 
%
%In this effort, the thickness ratio (t/c) measured at several control sections of the wing have been constrained to avoid reducing the internal volume and keep reasonable shapes of the wing. 
%
\par
As pointed out by Lyu \textit{et al.}\cite{Lyu2015AerodynamicSO} one of the weaknesses of single-point optimizations, without the use of an appropriate penalty, is the progressive thinning of wing leading edges.
This can be avoided performing a more costly multi-point optimization, as sharp leading edges would perform poorly in off design point.
In the proposed framework, instead, such issue is taken care of by manually setting to zero the OF sensitivities $\frac{\mathrm{d} J} {\mathrm{d} {\mathbf{u_{F_{\alpha}}}}}$ relative to grid points close to sharp edge regions. 
%
%
%As the resulting sharp leading edges perform poorly in off-design conditions, a more costly multi-point optimization would avoid such issue.
%
%For a single-point optimization, this noticeable drawback is avoided in the current framework manually setting to zero the OF sensitivities $\frac{\mathrm{d} J} {\mathrm{d} {\mathbf{u_{F_{\alpha}}}}}$ relative to grid points close to sharp edge regions. 
%\inred{[non capisco cosa signfiichi..viene misurata la sharpness e se sfora un certo valore la derivata rispetto a quel punto viene messa a a zero? Si, il valore lo sceglie lo user] Come viene misurata la sharpness? Boh}

%% file: Chapters/4_Testcases.tex
%------------------------------
\section{Aeroelastic test cases}
\label{sec:testcases}
%------------------------------
%
%In this section the aeroelastic test cases are presented, together with the main option used by the solvers for . 
%
\subsection{Test case based on ONERA M6}
\begin{figure*}[!htbp]
     \centering
     \vspace{5mm}
         \includegraphics[width=0.98\linewidth]{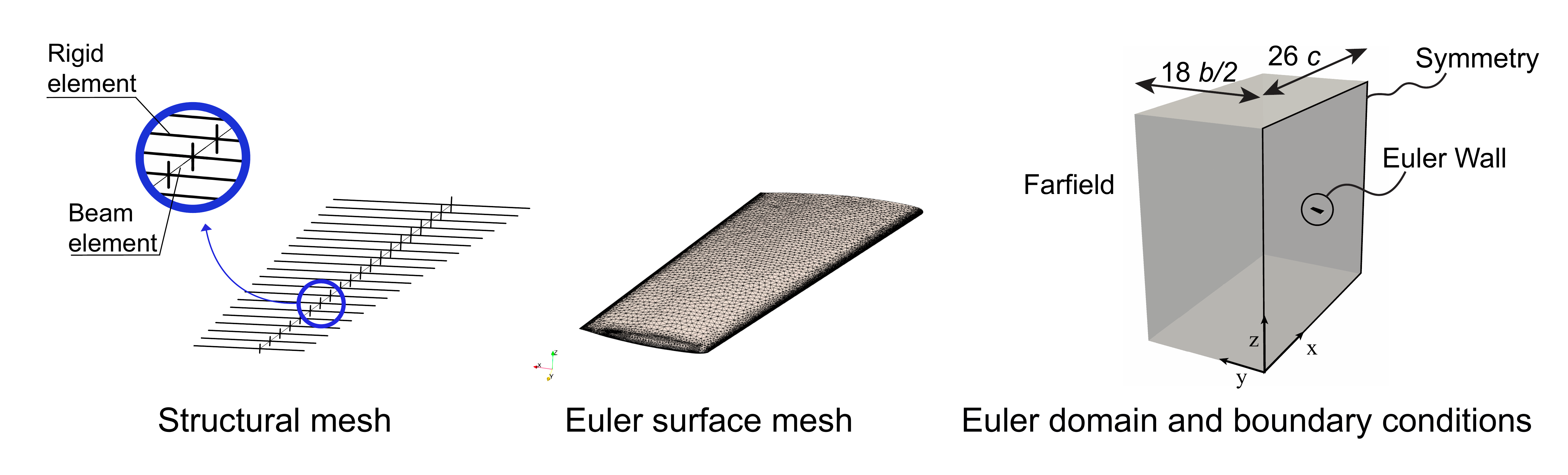}
         \caption{Meshes for the Euler-based ONERA M6 test case. Fluid domain dimensions are given as function of the wing root chord $c$ and semi-span $b/2$.  }
         \label{f:ONERAM6_meshes}
\end{figure*}

The first test case is based on ONERA M6 wing geometry. 
%The aerodynamic surface has been augmented with a synthetic structure similarly to what used in a previous effort by Bombardieri \textit{et al.}~\cite{UC3Mteam_Scitech_NITRO}. 
%In the original study, wing box properties (i.e. wing box cross section and material Young Modulus) were fine-tuned for the aeroelastic model to exhibit flutter in transonic regime; in the current effort these are changed for the purpose of the study, i.e., tuning the sought deflection of the wing in flying conditions.
%
A synthetic structure has been assembled based on the work of Bombardieri \textit{et al.}~\cite{UC3Mteam_Scitech_NITRO}, in which wing box properties (i.e. wing box cross section and material Young Modulus) were selected for the aeroelastic model to exhibit flutter in transonic regime; 
in the current effort such properties have been fine-tuned to obtain sought levels of wingtip deflection in flying conditions.
%
%for the purpose of the study, i.e., tuning the sought deflection of the wing in flying conditions.
%
The wing box, located at $25\%$ of the wing chord, is described by beam elements. Four rigid elements have been cross placed at several stations along the wing span to reproduce the position of the leading edge (LE), trailing edge (TE), upper and lower point positions of the current wing section (airfoil). 
%Such nodes are linked to the wing box with rigid elements to ensure, on the structural side, that the airfoil shape is kept during deformation. 
This solution has been found successful for a correct application of the spline algorithm in order to transfer information between solid and fluid boundary meshes. The structural model is clamped in correspondence of the wing root. 
Layout of the structural model is shown in Figure~\ref{f:ONERAM6_meshes}.
\par
Concerning the aerodynamic part of the problem, for this test case, flow has been model with Euler equations. The CFD mesh, inherited from efforts of Bombardieri \textit{et al.}~\cite{UC3Mteam_Scitech_NITRO,Bombardieri_uc3m_EUROGEN} is shown in Figure~\ref{f:ONERAM6_meshes}. Volume features 582,752 tetrahedral elements and wing surface is discretized  with 36,454 triangular elements. The computational domain is a box shape extending approximately for 13 root chords downstream, for 12 root chords upstream and for 9 semi-spans laterally. 
%Mesh size and its main parameters are summarized in Table~\ref{t:ONERA_CFD_mesh}. \inred{[Possibilita' di metterli in figura]}
%
%of approximately 26 root chord in the horizontal and vertical direction and 9 semi-spans in the span direction. 
%different meshes are used accordingly to the chosen flow model. 
%For Euler flow model, fluid domain is defined as a box with 582,752 tetrahedral elements, while the wing boundary features 36,454 triangular elements. 
%For this mesh, domain with the different boundaries and wing surface mesh are shown in Figure~\ref{f:ONERAM6_meshes}.  
%and 38,810 nodes. 
%The wing surface features 5,640 triangular elements and 2,871 nodes. 
%Wing surface discretization can be seen in Figure~\ref{f:ONERA_meshes_sens}. 
%
%
\par
Solution of the CFD equation has been performed with a 3 level Multi-Grid scheme together with  a 2$^{nd}$ order in space Jameson-Schmidt-Turkel (JST) scheme for the convective flux. Main CFD options for this test case are given in Table~\ref{t:Euler_CFD_setting}.
%Figure~\ref{f:ONERAM6_meshes} shows the structural and the CFD Euler mesh. 
%
\par
\begin{figure}[!htbp]
     \centering
     \vspace{5mm}
         \includegraphics[width=0.9\linewidth]{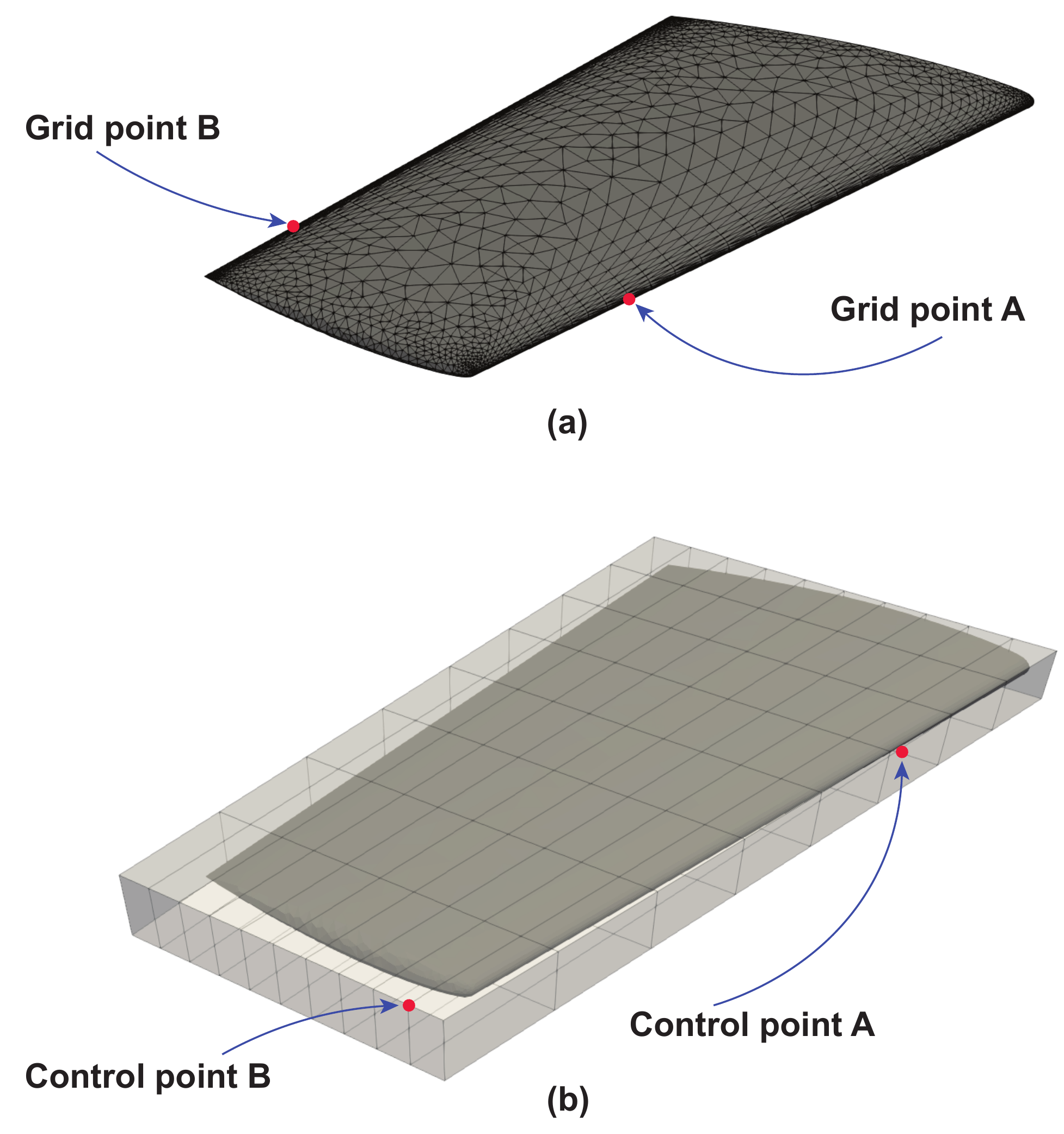}
         \caption{ONERA M6 coarse mesh wing surface with the grid points (a) and control points (b) considered for sensitivities validation.}
         \label{f:ONERA_meshes_sens}
\end{figure}

For the purpose of sensitivities validation, a coarser mesh, depicted in Figure~\ref{f:ONERA_meshes_sens}, is considered with 140,244 tetrahedral elements in the computational domain and 5,640 triangular elements on the wing surface.
%Flight condition is a steady flight at Mach number of $M = 0.8395$; in the optimization lift coefficient was set to  $C_L = 0.5$ being the angle of attack (AoA) of the undisturbed flow variable to accommodate such constraint.  For the validation campaign, either a fixed AoA or fixed  $C_L$ is considered.
For all applications concerning this test case, considered flight condition is a steady flight at $M = 0.8395$ and sea level ($\rho = 1.2250 \; kg /m^3$). 
%
%------------------------------------------%
% FIGURE    ONERA m6 different meshes                     %
%------------------------------------------%
%------------------------------------------%
%
%
%\inred{[Occhio qui , vediamo] A second mesh is used for RANS-based CFD application. }
%
%------------------------------------------%
%  ONERA M6 CASE    CFD                %
%------------------------------------------%
%\begin{table}[!t]
%\caption{ONERA M6 test case: Euler CFD mesh details.}
%\label{t:ONERA_CFD_mesh}       
%\centering
%\begin{tabular}{ l  l   }
%\hline\noalign{\smallskip}
%\textbf{Quantity} & \textbf{Value} \\
%Volume elements  & 582,752\\
%Surface elements  & 36,454\\
%Nr. of root chord upstream  & \sim 13 \\
%Nr. of root chord downstream  & \sim 12 \\
%Nr. of root chord in vertical direction  & \sim 13 \\
%Nr. of semi-spans in span direction  & \sim 9 \\
%\hline\noalign{\smallskip}
%\end{tabular}
%\end{table}
%--------------END  TABLE -----------------%

%------------------------------------------%
%  ONERA M6 CASE    CFD                %
%------------------------------------------%
\begin{table}[!t]
\caption{Numerical options for Euler-based CFD problems.}
\label{t:Euler_CFD_setting}       
\centering
\begin{tabular}{ l  l   }
\hline\noalign{\smallskip}
\textbf{Parameter} & \textbf{Value} \\
\noalign{\smallskip}\hline\noalign{\smallskip}
Multi-Grid levels nr.  & 3\\
Convective flow num. method  & JST\\
Pseudo-time num. method  & Euler implicit\\
Linear solver  & FGMRES\\
Linear solver precond. & LU\_SGS\\
\hline\noalign{\smallskip}
\end{tabular}
\end{table}
%--------------END  TABLE -----------------%
%
%
%------------------------------------------%
% FIGURE    ONERA M^ COARSE                     %
%------------------------------------------%
%------------------------------------------%

%% file: Chapters/4b_Testcases.tex
%
%====================================================
\subsection{Test cases based on the NASA CRM}
\begin{figure*}[!htbp]
     \centering
     \vspace{5mm}
         \includegraphics[width=0.98\linewidth]{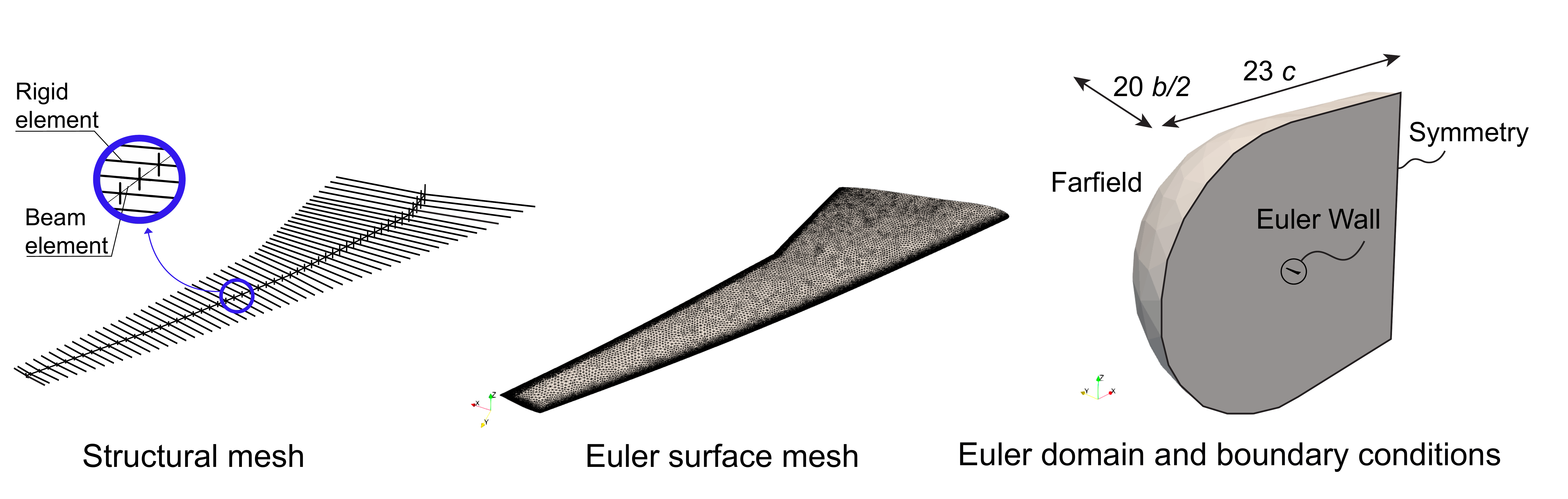}
         \caption{Meshes for the Euler-based QCRM test case. Fluid domain dimensions are given as function of the wing root chord $c$ and semi-span $b/2$.}
         \label{f:CRM_meshes}
\end{figure*}

%=====================================================
The second aeroelastic test case is based on the NASA CRM~\cite{Vassberg-2008} and will be referred to  as Quasi-CRM (QCRM). 
The QCRM structural beam model has been generated from the  global FEM (gFEM) model ``V15wingbox'' available in the NASA Common Research Model website repository~\cite{CRM_web}. 
%and used for linear flutter analysis. 
The gFEM has been converted to a beam-based FE model by means of a modal equivalence process\cite{vela2019aeroelastic} and is based on the CRM outer mold line at 1 \emph{g} load factor.
To exhibit the sought level of wingtip deflection in flying conditions and trigger geometric nonlinearities, the value of the synthetic Young Modulus has been fine-tuned. 
%later reduced for the structure to exhibit geometrical nonlinearities. 
%
The same strategy as for the previous test case, i.e., adding four rigid elements at several wing sections along the span, guarantees an appropriate interfacing between the structural and fluid meshes. 
%the Onera M6 test case is used to highlight the wing airfoils, by means of rigid elements. 
The model is simply supported at the section corresponding to the wing-fuselage intersection, and  symmetry constraint is applied to the far inboard section, see Figure~\ref{f:CRM_meshes}.
\par
With respect to the wing shape, the original NASA CRM, featuring a wing + body layout, has been modified removing the fuselage and extending the wing root till the symmetry plane, maintaining  the TE and LE sweep angles. 
The mesh used for the Euler case, built and validated in a previous effort~\cite{Luis019aeroelastic} features 1,529,927 tetrahedral elements while the wing boundary consists of 71,998 triangular elements. 
As depicted in Figure~\ref{f:CRM_meshes}, the computational domain has a bullet shape, extending approximately for 20 root chords downstream, for 21 root chords upstream and for 10 semi-spans laterally. 
%Mesh size and its main parameters are summarized in Table~\ref{t:QCRM_CFD_mesh}.
Numerical solution of the CFD equation has been performed with same options as for the previous test case (see  Table~\ref{t:Euler_CFD_setting}).
%
%------------------------------------------%
%  ONERA M6 CASE    CFD                %
%------------------------------------------%
%\begin{table}[!t]
%\caption{QCRM test case: Euler CFD mesh details.}
%\label{t:QCRM_CFD_mesh}       
%\centering
%\begin{tabular}{ l  l   }
%\hline\noalign{\smallskip}
%\textbf{Quantity} & \textbf{Value} \\
%Volume elements  & 1,529,927\\
%Surface elements  & 71,998\\
%Nr. of root chord upstream  & \sim 21 \\
%Nr. of root chord downstream  & \sim 20 \\
%Nr. of root chord in vertical direction  & \sim 21 \\
%Nr. of semi-spans in span direction  &  10 \\
%\hline\noalign{\smallskip}
%\end{tabular}
%\end{table}
%--------------END  TABLE -----------------%
%
\par
\begin{figure}[!htbp]
     \centering
     \vspace{5mm}
         \includegraphics[width=0.95\linewidth]{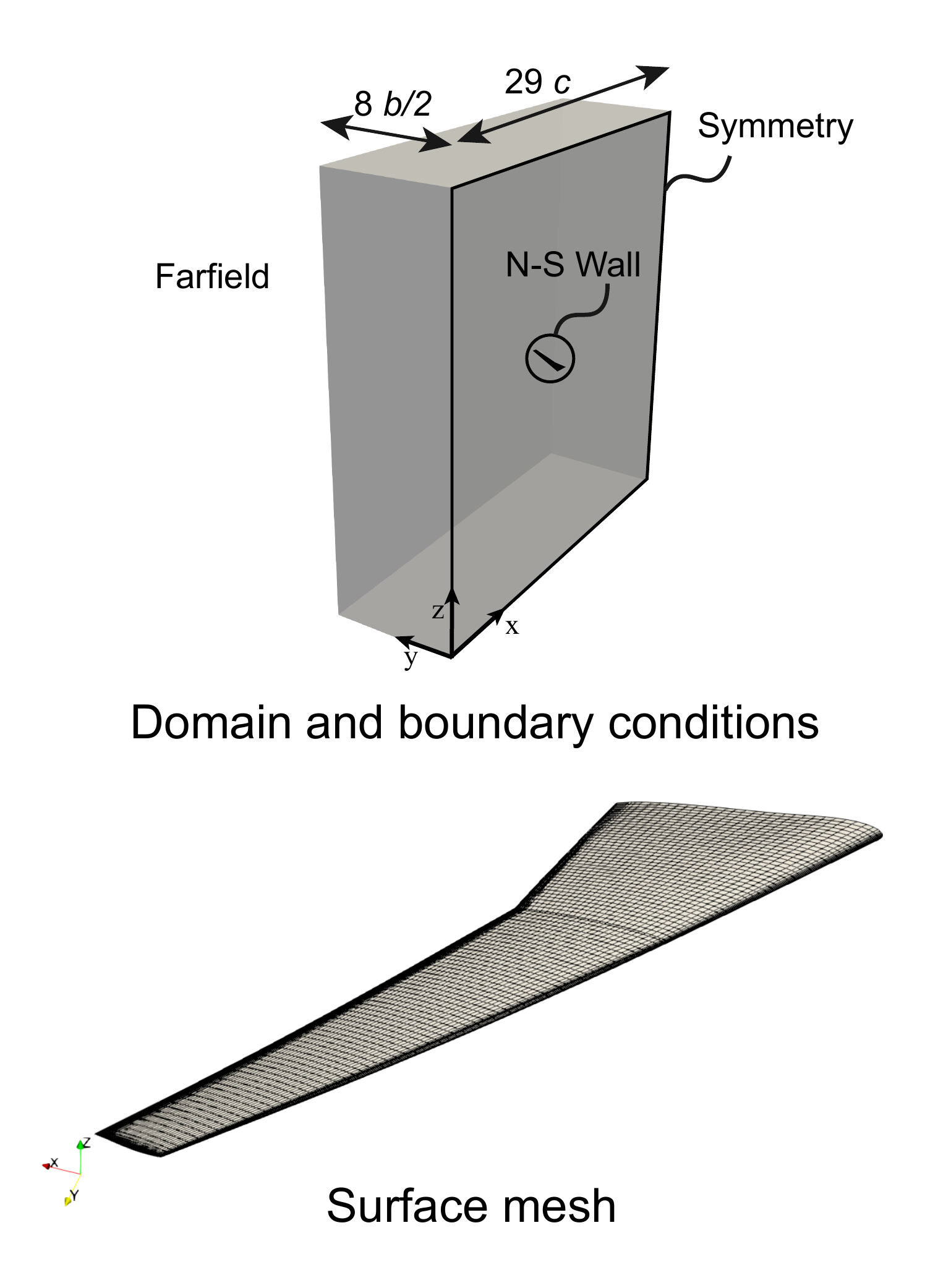}
         \caption{Aerodynamic mesh for the RANS-SA-based QCRM test case. Fluid domain dimensions are given as function of the wing root chord $c$ and semi-span $b/2$.}
         \label{f:CRM_meshes_RANS}
\end{figure}

For the RANS-SA simulation a different mesh is used which has been built and validated in a previous effort~\cite{Castro2020aerodynShapeOpt}. It consists of 1,549,052 hexahedral elements while the wing boundary features 10,669 elements. 
As depicted in Figure~\ref{f:CRM_meshes_RANS}, the computational domain is a box, extending approximately for 16 root chords downstream, for 12 root chords upstream and for 4 semi-spans laterally. Main CFD options for this test case are given in Table~\ref{t:RANS_CFD_setting}.
%
%\par
%
%Solution of the CFD equation has been performed with  a 2$^{nd}$ order in space Jameson-Schmidt-Turkel (JST) scheme for the convective flux and a scalar upwind scheme for the turbulent problem. Main CFD options for this test case are given in Table~\ref{t:RANS_CFD_setting}.
%
\par
For all applications concerning this test case, considered flight condition is a steady flight at $M = 0.85$ and sea level ($\rho = 1.2250 \; kg /m^3$). 
%
%Figure \ref~{f:CRM_meshes} summarizes the different meshes used for the QCRM aeroelastic model.
%
%SONO QUI. Rcorda di dire che la superficie viene estesa dentro 
%Two different meshes are employed for the aerodynamic domain, relative to the Euler and RANS-SA modeling of the equations. 
%For the Euler case, the mesh as the one created and validated in 
%Concerning the fluid part of the problem, flow is modeled with the Euler equations. 
%Aerodynamic mesh was built and validated in a previous effort\inred{thesis Jorge} removing the fuselage and extending the wing root till the symmetry plane, following the TE and LE sweep angles. %Like the structural model, the wing boundary is based on the CRM aerodynamic mold line at 1 \textit{g} load factor. 
%
 %Figure~\ref{f:CRM_meshes} shows the different meshes for the CRM model. 
%------------------------------------------%
% FIGURE    CRM different meshes                     %
%------------------------------------------%
%------------------------------------------%
%
\par
%
%
%Considered flight condition is a steady flight at  $M = 0.85$; in all simulations, lift coefficient is set to  $C_L = 0.5$ being the angle of attack (AoA) of the undisturbed flow variable to accommodate such constraint.  
%
%------------------------------------------%
% FIGURE    CRM different meshes                     %
%------------------------------------------%
%------------------------------------------%
%
%------------------------------------------%
%  ONERA M6 CASE    CFD                %
%------------------------------------------%
\begin{table}[!t]
\caption{Numerical options for the RANS-SA-based CFD problem.}
\label{t:RANS_CFD_setting}       
\centering
\begin{tabular}{ l  l   }
\hline\noalign{\smallskip}
\textbf{Parameter} & \textbf{Value} \\
\noalign{\smallskip}\hline\noalign{\smallskip}
Multi-Grid levels nr.  & 0\\
Convective flow num. method  & JST\\
%Convective turb. num. method  & Scalar upwind\\
%Turbulent flow num. method  & Scalar upwind\\
Pseudo-time num. method  & Euler implicit\\
Linear solver  & FGMRES\\
Linear solver precond. & ILU\\
\hline\noalign{\smallskip}
\end{tabular}
\end{table}
%--------------END  TABLE -----------------%

%% file: Chapters/5_Application.tex
%------------------------------
\section{Results}
\label{sec:results}
%------------------------------
%
In this section results of the validation campaign of the coupled aerostructural sensitivities are shown first. 
Thereafter,  wing shape optimizations carried out on the test cases are presented and results are discussed. 
%In particular, when showing optimization results, several subcases are considered to highlight, by means of physical considerations, the relevance of performing aerostructural optimization on the \emph{flying shape} (i.e., the wing in its deformed shape at aeroelastic equilibrium).
%
In particular, when showing optimization results, several subcases are discussed to highlight, by means of physical considerations, the relevance of performing wing optimization including the aerostructural coupling.
%
%optimizing on the actual shape in the flight condition (called flying-shape), determined by the aerostructural coupling. 
Since considered DVs are relative to the wing aerodynamic surface, such optimization will be referred to as AeroStructural Wing Shape Optimization (ASWSO).  
On the other hand, a less effective, though computationally less intensive, procedure would be performing the optimization of the wing surface considering a rigid configuration (i.e., without the inclusion of aerostructural coupling effects); such optimization will be referred to as Aerodynamic Wing Shape Optimization (AWSO).
%\st{To show the penalties with respect to an ASWSO, the optimal wing according to the AWSO will be kept free to deflect under aerodynamic loads of the target flight condition, and the drag compared to the optimal wing according to ASWSO.}\inred{lo diciamo dopo: inutile ripetersi.}
%In this section, validation of aerostructural sensitivities is shown first; later on, optimization resutls are proposed and discussed.
%
%------------------------------
\subsection{Sensitivities validation}
\label{sec:sens_validation}
%------------------------------
%
Extending the work of Bombardieri \textit{et al.}~\cite{Bombardieri_uc3m_EUROGEN} in which the calculation of selected AD-based cross sensitivities was validated for the proposed aeroelastic test case (i.e., the sensitivity of $C_D$ and $C_L$ with respect to the structure Young Modulus), in this effort further validation results are shown. 
Total derivative of the the drag coefficient with respect to the wing surface jig-shape parameters is considered for validation. 
Sensitivities are evaluated for the ONERA M6 aeroelastic test case using the coarse mesh version (see Section \ref{sec:testcases}) and Euler flow model. 
\par
First, sensitivities of $C_D$ with respect to the  vertical displacement for two selected wing surface nodes of the jig-shape (variable $\mathbf{u_{F_{\alpha}}}$ in Table~\ref{t:list}) are validated. Such nodes are depicted in Figure~\ref{f:ONERA_meshes_sens}(a) and are located outboard, in correspondence of the LE (grid point A) and of the TE (grid point B).
The AD-based sensitivities provided by the framework are compared to the ones evaluated by a central FD scheme. 
Results of such campaign are shown in Table~\ref{t:AoA_grid} for a fixed angle of attack ($AoA$ = 3.06) for two different synthetic Young Modulus E of the structure, whereas in Table~\ref{t:CL_grid} the same comparison is proposed for a fixed lift coefficient ($C_L$ = 0.22) and one value of E.
%
%
%First, AD-based sensitivities of the wing's $C_D$ with respect to the vertical displacement of the jig shape (variable $\mathbf{u_{F_{\alpha}}}$ in Table~\ref{t:list}) for two selected fluid boundary nodes are compared to the same values calculated using central scheme FD. 
%The two selected fluid boundary nodes can be seen in Figure~\ref{f:ONERA_meshes_sens}(a) and are located outboard, in correspondence of the LE (grid point A) and of the TE (grid point B). 
%
%Comparison is shown in Table~\ref{t:AoA_grid} for fixed angle of attack ($AoA$ = 3.06) for two different Young Modulus of the structure, while in Table~\ref{t:CL_grid} same comparison is proposed for fixed $C_L$ = 0.22 and one Young Modulus of the structure.
%
\par
For the calculation of the sensitivity with respect to the FFD CP displacements, the chain rule is applied % to sensitivity with respect to the fluid boundary jig shape displacements, 
as shown in equation~(\ref{e:ChainRule}) in Section~\ref{sec:optimization}. 
The FFD-box uses Bezier basis functions of order 10, 8, 1 chord-wise, span-wise and along the thickness, respectively. 
The relative CPs chosen for validation are shown in Figure~\ref{f:ONERA_meshes_sens}(b). 
%The FFD-box features 11 points chord-wise, 9 points span-wise and 2 in the thickness direction. %
CP A is located inboard, in correspondence of the compression side of the wing LE while CP B is located outboard, in correspondence of the suction side of the wing LE. 
Comparison between sensitivities predicted by the framework with AD  and by central FD is provided in Table~\ref{t:AoA_CP} for fixed $AoA$ = 3.06 and one value of E.
%In effort the of Bombardieri \textit{et al.}\inred{cite book chapter}, for the here discussed aerostructural framework, validation of coupled AD-based sensitivities calculation was proposed. 

%------------------------------------------%
% TABLE Sensitivity ONERA M6 CASE AoA                   %
%------------------------------------------%
\begin{table*}[!t]
\caption{Aerostructural sensitivities of $C_D$ with respect to vertical jig shape boundary displacements $\mathbf{u_{F_{\alpha_z}}}$ calculated using FD and AD for two values of the synthetic $E$, $AoA = 3.06$ and $M_{\infty} = 0.8395$.}
\label{t:AoA_grid}       
\centering
\begin{tabular}{ l l l l l  }
\hline\noalign{\smallskip}
 & E = 40 [GPa] & & E = 20 [GPa] & \\
\noalign{\smallskip}\hline\noalign{\smallskip}
Grid pt. A & Sens. & Relative error to FD & Sens. & Relative error to FD \\
FD & - 0.004372394124 & -- & 0.006691539871 & -- \\
AD & - 0.004366897122 & 0.1257\% & 0.006698868402 & 0.1094\% \\
\noalign{\smallskip}\hline\noalign{\smallskip}
Grid pt. B & Sens. & Relative error to FD & Sens. & Relative error to FD \\
FD & 0.011288968085 & -- & 0.001710656774 & -- \\
AD & 0.011265754131 & 0.2056\% & 0.001710362610 & 1.719e-04\% \\
\hline\noalign{\smallskip}
\end{tabular}
\end{table*}
%--------------END  TABLE -----------------%
%
%------------------------------------------%
% TABLE Sensitivity ONERA M6 CASE cl       %
%------------------------------------------%
\begin{table}[!t]
\caption{Aerostructural sensitivities of $C_D$ with respect to vertical jig-shape node displacements $\mathbf{u_{F_{\alpha_z}}}$ calculated using FD and AD. $C_L = 0.22$, $M_{\infty} = 0.8395$ and $E=40$ GPa.}
\label{t:CL_grid}       
\centering
\begin{tabular}{ l l l   }
\hline\noalign{\smallskip}
Grid pt. A & Sens. & Relative error to FD  \\
FD & 0.0080559423 & --  \\
AD & 0.0080589472 & 0.0373\%  \\
\noalign{\smallskip}\hline\noalign{\smallskip}
Grid pt. B & Sens. & Relative error to FD  \\
FD & 0.00162756262 & --  \\
AD & 0.00162795035 & 0.0238\%  \\
\hline\noalign{\smallskip}
\end{tabular}
\end{table}
%--------------END  TABLE -----------------%
%
%------------------------------------------%
% TABLE CP Sensitivity ONERA M6 CASE cl                   %
%------------------------------------------%
\begin{table}[!t]
\caption{Aerostructural sensitivities of $C_D$ with respect to vertical CP displacements $\mathbf{u_{F_{\alpha_z}}^{CP}}$ calculated using FD and AD.  $AoA = 3.06$, $M_{\infty} = 0.8395$ and $E=40$ GPa.}
\label{t:AoA_CP}       
\centering
\begin{tabular}{ l l l   }
\hline\noalign{\smallskip}
CP A & Sens. & Relative error to FD  \\
FD &  0.0043083629927 & --  \\
AD &  0.0043296403878 & 0.4914\%  \\
\noalign{\smallskip}\hline\noalign{\smallskip}
CP B & Sens. & Relative error to FD  \\
FD & 0.0120379174801 & --  \\
AD & 0.0120349705805 & 0.0244\%  \\
\hline\noalign{\smallskip}
\end{tabular}
\end{table}
%--------------END  TABLE -----------------%
%
%All above cases feature asymptotic Mach of 0.84.
Excellent agreement between sensitivities calculated by the two methods is found. It is, anyway, pointed out that, due to truncation errors, FD are not reliable~\cite{kenway2014} in detecting errors below $O(10^{-2})$. 
%\inred{Forse e' percentuale? No, valore assoluto (nd Rocco)}
%which is, anyway, beyond the purpose of this study.
%

%% file: Chapters/5b_Application.tex
%====================================================
\subsection{Euler-based optimization of the ONERA M6}
\label{sec:Euler_opt_ONERA}
%=====================================================
%
%Results of the optimization of the ONERA M6 aeroelastic test case are discussed in this section. 
This section discusses the results of the optimization of the ONERA M6 aeroelastic test case.
Optimization constraints are shown in Table~\ref{t:ONERAM6_constr}. 
DVs are vertical positions (z direction with respect to reference system in Figure~\ref{f:ONERAM6_meshes}) of CPs of the FFD-box shown in Figure~\ref{f:ONERA_meshes_sens}(b), employing Bezier basis functions of order 10, 8, 1 chord-wise, span-wise and along the thickness, respectively. 
CPs on the symmetry plane are kept fixed as an effective way to prevent a change in shape of the relative airfoil. 
%This is a common requirement in  aerostructural and aerodynamic shape optimization since wing sections in correspondence of the fuselage are subjected to volume costraints (due to fuel storage) more than aerodynamic ones.\inred{[Non mi soffermerei sul discorso del perche', terrei solo la prima frase]}
%\inred{CPs on the symmetry plane are kept fixed as an effective way to prevent change in shape of that corresponding airfoil}.
%
%A synthetic Young Modulus equal to $25=GPa$ has been chosen to exhibit the sought level of wing-tip deflection and trigger geometrical nonlinearities.
The synthetic Young Modulus has been tuned for the structure to exhibit wing-tip deflection of approximately 13\% of the semi-span and trigger geometrical nonlinearities (see Figure~\ref{f:ONERAM6_opt_comp}).
% 25 GPa
%------------------------------------------%
% TABLE Variables                  %
%------------------------------------------%
\begin{table}[!t]
\caption{Set of constraints and total number of DVs used for the optimization of the ONERA M6 aeroelastic test case.}
\label{t:ONERAM6_constr}       
\centering
\begin{tabular}{ l l l }
\hline\noalign{\smallskip}
 \textbf{Aerodynamic constraints} &&\\ 
\noalign{\smallskip}\hline\noalign{\smallskip}
$C_L$ & = & 0.286 \\
\noalign{\smallskip}\hline\noalign{\smallskip}
\textbf{Geometric constraints} &&\\ 
\noalign{\smallskip}\hline\noalign{\smallskip}
%Fixed CPs on symmetry plane&&\\
%t/c (sec. at 0\% span) & = & 9.73\%\\
t/c (sec. at 16.4\% span) & $\geq$& 9.64\% \\
t/c (sec. at 32.8\% span) & $\geq$& 9.60 \% \\
t/c (sec. at 49.2\% span) & $\geq$& 9.58\% \\
t/c (sec. at 65.7\% span) & $\geq$& 9.49\% \\
\noalign{\smallskip}\hline\noalign{\smallskip}
\textbf{Number of DVs} &=& 198 \\ 
%Thickness (section at $y = 0.2 m $) $\geq 0.0.072 m$ \\
%Thickness (section at $y = 0.4 m $) $\geq 0.0.066 m$ \\
%Thickness (section at $y = 0.6 m $) $\geq 0.0.060 m$ \\
%Thickness (section at $y = 0.8 m $) $\geq 0.0.054 m$ \\
\noalign{\smallskip}\hline
\end{tabular}
\end{table}
%--------------END  TABLE -----------------%
%
%===================================================
\subsubsection{Aerodynamic wing shape optimization}
\begin{figure}[!htbp]
     \centering
    % \vspace{5mm}
         \includegraphics[width=0.9\linewidth]{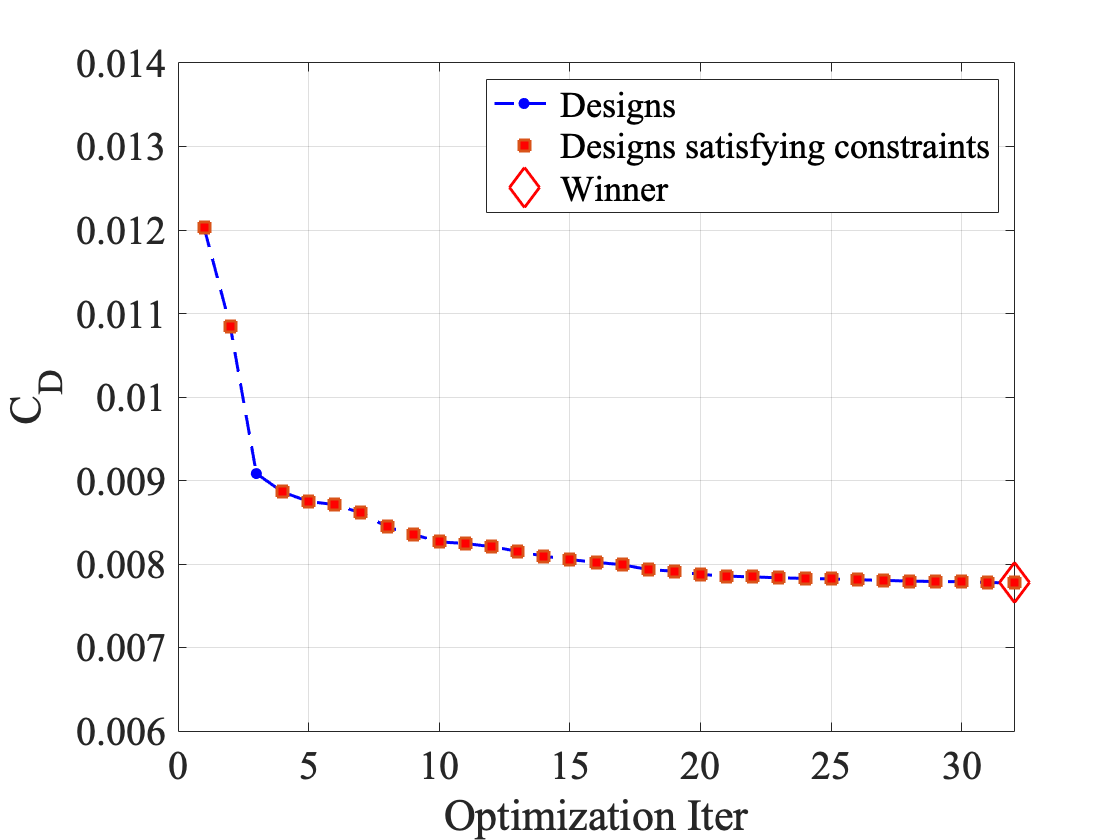}
         \caption{$C_D$ reduction for the Euler-based AWSO of the ONERA M6 wing.}
         \label{f:ONERA_Rigid_opt}
\end{figure}

\begin{figure}[!htbp]
     \centering
     \vspace{5mm}
         \includegraphics[width=\linewidth]{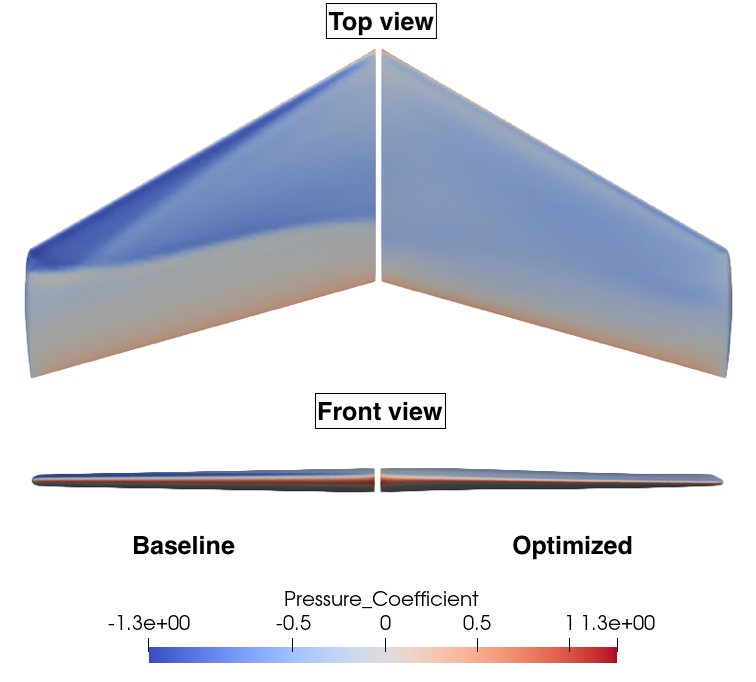}
         \caption{Euler-based AWSO of the ONERA M6 wing: $C_p$ distribution on the baseline and the optimized designs.}
         \label{f:ONERAM6_aerodWSO}
\end{figure}

%====================================================
%
First, an AWSO is run for the test case. Result of this optimization, in terms of $C_D$ vs optimization iterations is shown in Figure~\ref{f:ONERA_Rigid_opt}. 
For the optimal shape a $C_D$ reduction of 35.38\% with respect to the baseline configuration is achieved.
Figure~\ref{f:ONERAM6_aerodWSO} shows the $C_p$ distribution for the top and front view of the baseline (left) and optimal (right) designs. 
%while Fig\inred{to be added} shows, for selected sections along the wing span, the airfoil shape and $C_p$ distribution for the original and optimized designs for selected stations along the wing span.\inred{forse la mettiamo forse no}.
It can be noted how the optimal design doesn't feature the characteristic lambda shock of the original design.  
%------------------------------------------%
% FIGURE                       %
%------------------------------------------%
%------------------------------------------%
%
%------------------------------------------%
% FIGURE                        %
%------------------------------------------%
%------------------------------------------%
%
%===================================================
\subsubsection{Aerostructural wing shape optimization}
\begin{figure}[!htbp]
     \centering
     \vspace{5mm}
         \includegraphics[width=0.9\linewidth]{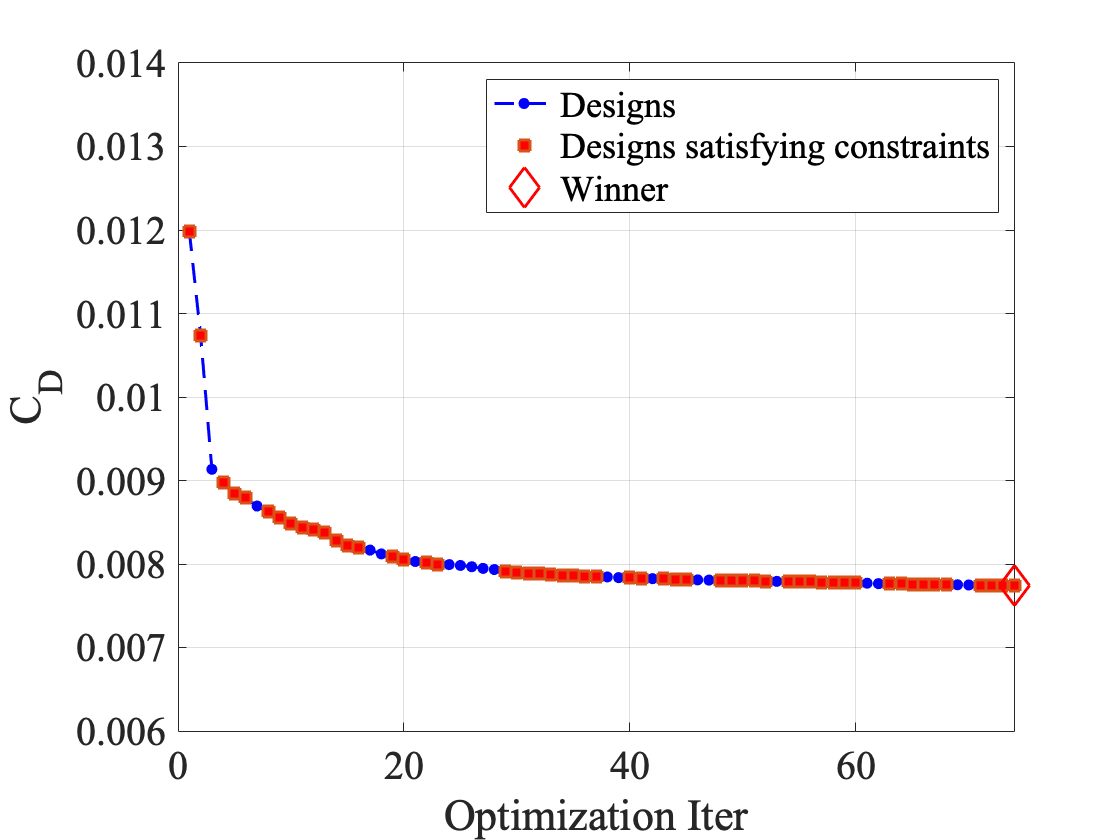}
         \caption{$C_D$ reduction for the Euler-based ASWSO of the ONERA M6 aeroelastic test case.}
         \label{f:ONERA_Flex_opt}
\end{figure}

%====================================================
%
An ASWSO is then run. $C_D$ evolution is shown for this case in Figure~\ref{f:ONERA_Flex_opt}. 
%
%------------------------------------------%
% FIGURE                       %
%------------------------------------------%
%------------------------------------------%
%
%For the optimal wing a reduction of $C_D$ of 35.39\% with respect to the baseline configuration (in its relative flying shape) is obtained.  
For the optimal wing a reduction of $C_D$ of 35.39\% is obtained with respect to the baseline configuration in its relative \emph{flying shape} (i.e., the wing in its deformed shape at aeroelastic equilibrium).
\par
\begin{figure}[!htbp]
     \centering
     %\vspace{5mm}
         \includegraphics[width=\linewidth]{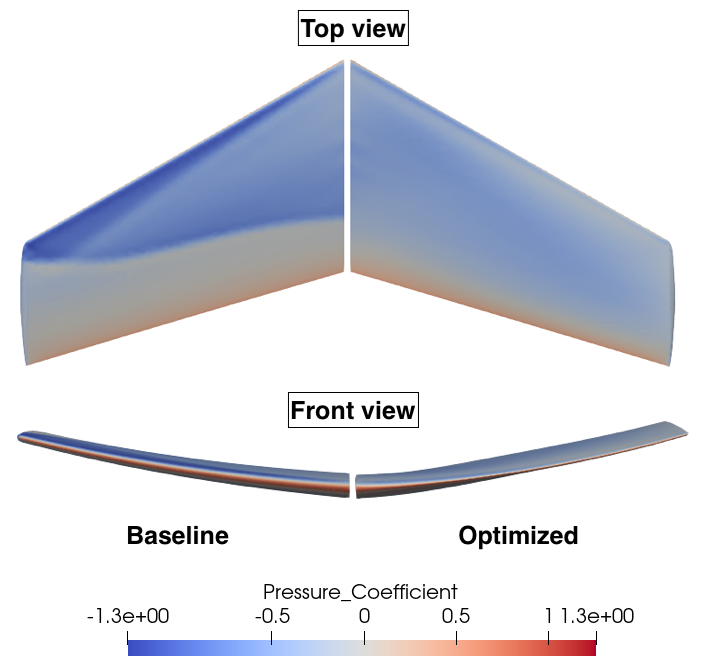}
         \caption{Euler-based ASWSO of the ONERA M6 wing: $C_p$ distribution on the baseline and the optimized designs at aeroelastic equilibrium.}
         \label{f:ONERAM6_aerosWSO}
\end{figure}

Figure~\ref{f:ONERAM6_aerosWSO} shows the $C_p$ distribution for the baseline (left) and optimal (right) designs. 
%
%------------------------------------------%
% FIGURE                        %
%------------------------------------------%
%------------------------------------------%
%
From the top view it is apparent how the original design at aeroelastic equilibrium features a similar shock wave pattern as the one observed in its undeflected condition.
%
%From the top view, it is apparent how the flying shape of the original design features a similar shock wave pattern as \inmag{in} its undeflected configuration.
%
Aerostructural optimal wing achieves a reduction of $C_D$ by alleviating the shock wave in its flying shape.
Front view allows to appreciate the maximum wing-tip deflection for both designs. 
%
%===================================================
\subsubsection{AWSO and ASWSO comparison}
%====================================================
%
%A comparison can be made between the different configurations discussed so far.
Relevance of pursuing an aerostructural optimization can be inferred from 
Table~\ref{t:ONERAM6_opt_comp}, which compares the $C_D$ for the flying shapes of the original design, the AWSO optimal design, and the ASWSO optimal design. 
%
%------------------------------------------%
% TABLE Variables                  %
%------------------------------------------%
\begin{table}[!t]
\caption{Comparison of $C_D$ between ASWSO, AWSO optima and the original configuration in flying shape, for the Euler-based ONERA M6 test case.}
\label{t:ONERAM6_opt_comp}       
\centering
\begin{tabular}{ l l r}
\hline\noalign{\smallskip}
 \textbf{Configuration} & $\mathbf{C_D}$ & Diff. \% \\ 
\noalign{\smallskip}\hline\noalign{\smallskip}
ASWSO optimum & 0.00775 & --\\
AWSO optimum &  0.00824 & 6.32\% \\
Original &  0.01199   &     35.39\%        \\
\noalign{\smallskip}\hline
\end{tabular}
\end{table}
%------------------------------------------%
%
%
It can be seen how the AWSO optimum in operation shows a value of the $C_D$ which is far from the ``real'' optimum evaluated by means of an ASWSO. 
%
%\inmag{[QUESTA FRASE HA PROBLEMI] This discrepancy can be interpreted as the result of optimizing for off-design conditions as consequence of not including flexibility of the structure}. 
%
%\inblue{[Alternative] 
This discrepancy can be explained by the fact that AWSO optimizes the wing around its rigid configuration which is, de-facto, an off-design point with respect to the flying shape in which the wing operates and which is naturally considered by an ASWSO. The more the wing is flexible, the more the rigid shape differs from the shape at aeroelastic equilibrium.
For this same reason, if large wing deflections are expected, geometrical nonlinearities should be considered in aerostructural optimization.
\par
\begin{figure}[!htbp]
     \centering
     \vspace{5mm}
         \includegraphics[width=0.98\linewidth]{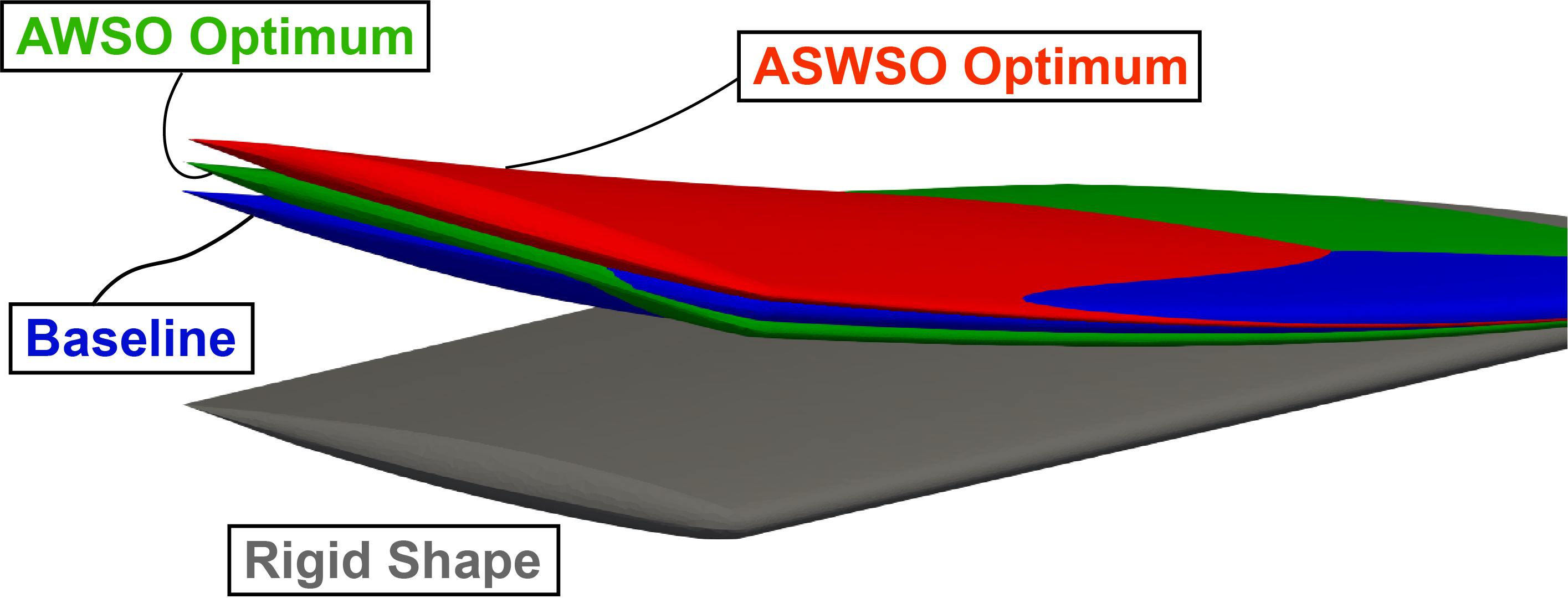}
         \caption{Comparison between ASWSO, AWSO optima and the original configuration flying shapes for the Euler-based ONERA M6 test case: detail of the wing tip.}
         \label{f:ONERAM6_opt_comp}
\end{figure}

Figure~\ref{f:ONERAM6_opt_comp} shows a detail of the tip deflection for the three configurations. 
%in their flying shape. 
It is interesting to notice how the wing optimized considering the aerostructural coupling displays a larger tip deflection than the other wings. % when in operation. 
\par
%
%It can be observed how, performing an aerodynamic optimization of a flexible wing means optimizing for an off-design point. This means, either 
%
%------------------------------------------%
% FIGURE                        %
%------------------------------------------%
%------------------------------------------%
%
\par
\begin{figure*}[!htbp]
     \centering
     \vspace{5mm}
         \includegraphics[width=\linewidth]{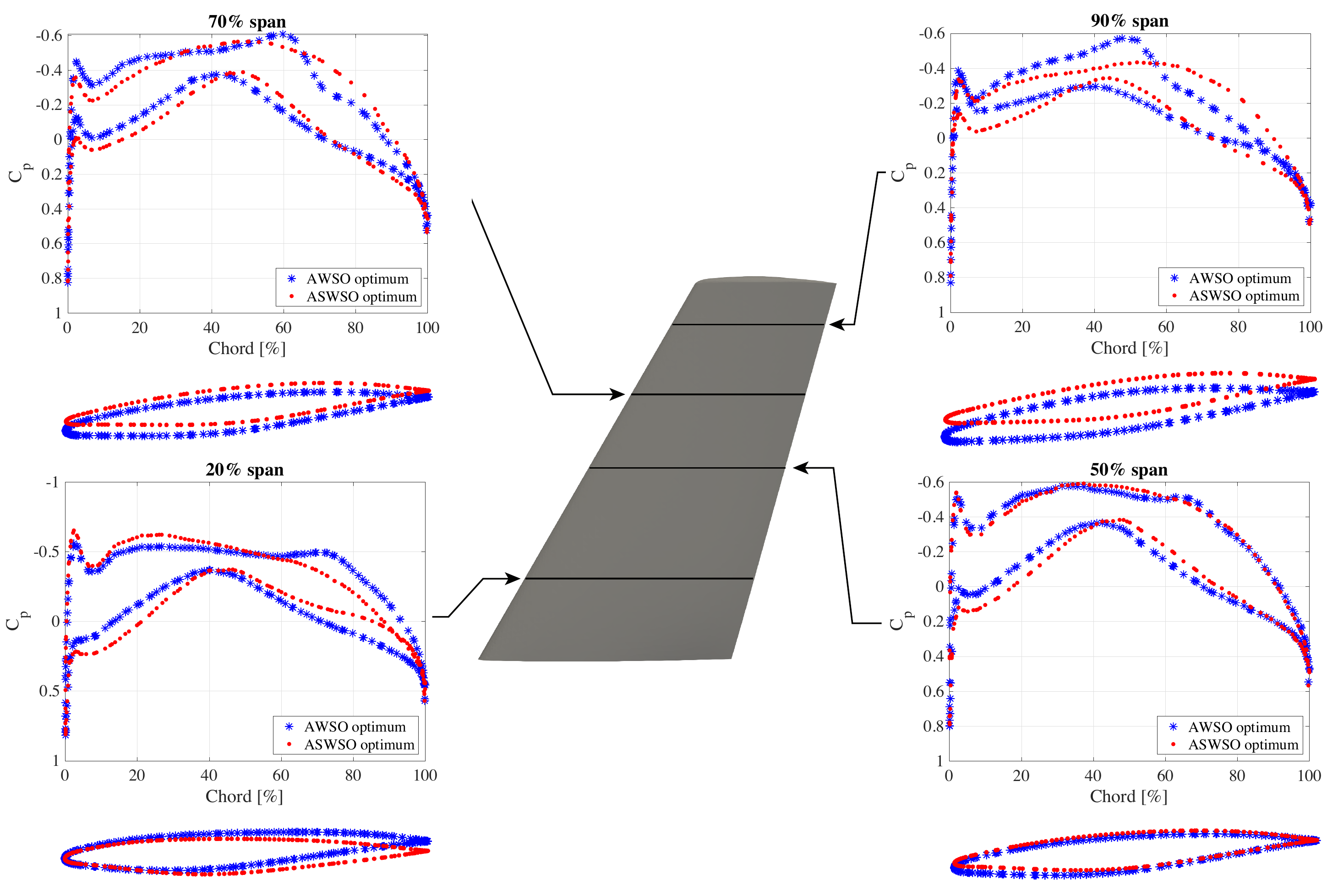}
         \caption{Flying shape comparison between AWSO and ASWSO optima for the Euler-based ONERA M6 test case. For selected stations along the wing span both $C_p$ distributions and airfoil shapes (with relative position in space) are shown.}
         \label{f:Cp_ONERA_comp}
\end{figure*}

For the flying shapes of the AWSO and ASWSO optima, Figure~\ref{f:Cp_ONERA_comp} shows, for selected sections along the wing span, the airfoil shapes (with their relative position in space) and the $C_p$ distribution.
%
%------------------------------------------%
% FIGURE                        %
%------------------------------------------%
%------------------------------------------%
%
Airfoils for both optimized configurations are different than the symmetric airfoils characteristic of the original ONERA M6 aerodynamic surface.
It can be noticed, close to the wing tip (sections at 70\% and 90\% span), the higher deflection of the ASWSO optimal wing. %optimum flying shape. 
%

%% file: Chapters/5c_Application.tex
%====================================================
\subsection{Euler-based optimization of the QCRM}
\label{sec:Euler_opt_QCRM}
\begin{figure}[!htbp]
     \centering
     \vspace{5mm}
         \includegraphics[width=0.9\linewidth]{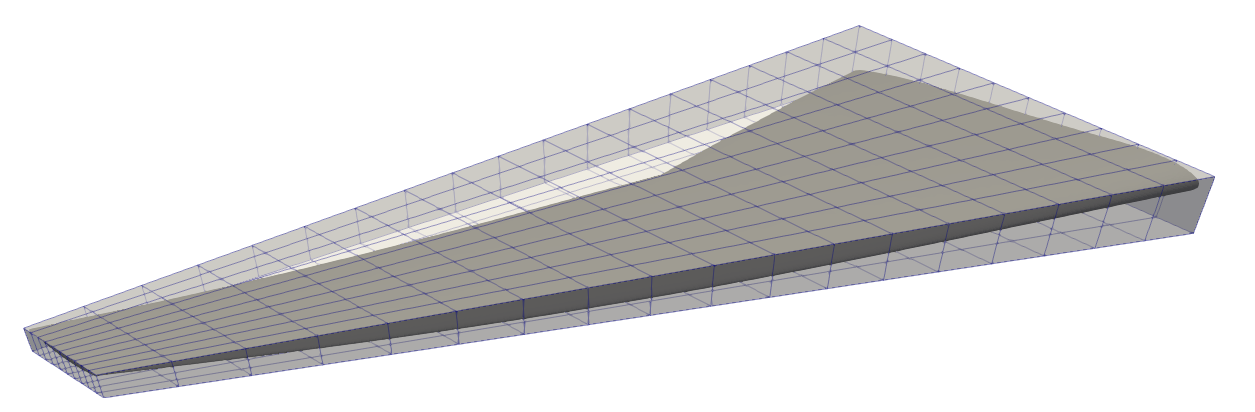}
         \caption{FFD box used for the Euler-based optimization of the QCRM aeroelastic model.}
         \label{f:CRM_FFD}
\end{figure}

%=====================================================
%
This section discusses the results of the Euler-based optimization performed on the QCRM aeroelastic test case. 
Optimization costraints are shown in Table~\ref{t:CRM_constr}. 
An FFD box is built based on Bezier functions of order 10, 18, 1 chord-wise, span-wise and along the thickness, respectively, as depicted in Figure~\ref{f:CRM_FFD}.
%
%------------------------------------------%
% FIGURE                        %
%------------------------------------------%
%------------------------------------------%
%
DVs are the vertical positions of all FFD-box CPs. 
CPs on the symmetry plane are kept fixed as an effective way to prevent changes in the shape of the relative airfoil.
The synthetic Young Modulus has been tuned for the structure to exhibit wing-tip deflection of approximately the 6\% of the semi-span (see Figure~\ref{f:CRM_opt_comp}).
%and trigger geometrical nonlinearities (see Figure~\ref{f:CRM_opt_comp}).
%590 GPa
%------------------------------------------%
% TABLE Variables                  %
%------------------------------------------%
\begin{table}[!t]
\caption{Set of constraints and total number of DVs used for the optimization of the Euler-based QCRM aeroelastic test case.}
\label{t:CRM_constr}       
\centering
\begin{tabular}{ l l l }
\hline\noalign{\smallskip}
 \textbf{Aerodynamic constraints} &&\\ 
\noalign{\smallskip}\hline\noalign{\smallskip}
$C_L$ & = & 0.5 \\
\noalign{\smallskip}\hline\noalign{\smallskip}
\textbf{Geometric constraints} &&\\ 
\noalign{\smallskip}\hline\noalign{\smallskip}
%Fixed CPs on symmetry plane&&\\
%t/c (sec. at 0\% span) & = & 26.0\%\\ %& = & 0.26020\\
t/c (sec. at 0.34\% span) & $\geq$&  15.6\% \\ %$\geq$&  0.15571 \\
t/c (sec. at 16.32\% span) & $\geq$& 12.5\% \\ %$\geq$& 0.12542 \\
t/c (sec. at 27.01\% span) & $\geq$& 11.2\% \\
t/c (sec. at 38.49\% span) & $\geq$& 10.4\% \\
t/c (sec. at 49.76\% span) & $\geq$& 10.0\% \\ %0.10003\\
t/c (sec. at 60.74\% span) & $\geq$& 9.8\% \\ %0.09756 \\
t/c (sec. at 71.89\% span) & $\geq$& 9.6\% \\ %0.09586 \\
t/c (sec. at 83.07\% span) & $\geq$& 9.5\% \\ %0.09536 \\
t/c (sec. at 94.14\% span) & $\geq$& 9.5\% \\ % 0.09487 \\
\noalign{\smallskip}\hline\noalign{\smallskip}
\textbf{Number of DVs} &=& 418\\ 
\noalign{\smallskip}\hline
\end{tabular}
\end{table}
%------------------------------------------%
%
%
%
%====================================================
\subsubsection{Aerodynamic wing shape optimization}
\begin{figure}[!htbp]
     \centering
     \vspace{5mm}
         \includegraphics[width=0.9\linewidth]{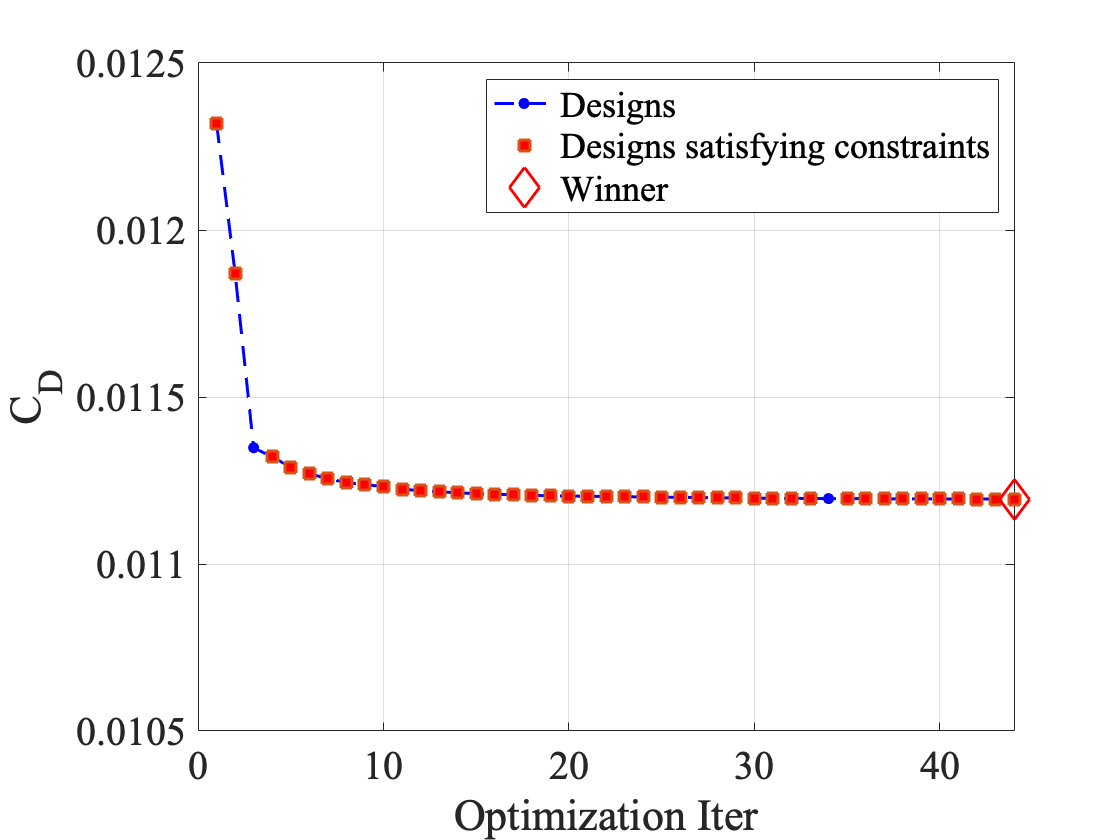}
         \caption{$C_D$ reduction for the Euler-based AWSO of the QCRM wing.} %In red designs that satisfy the imposed constraints.}
         \label{f:QCRM_Rigid_opt}
\end{figure}

\begin{figure}[!htbp]
     \centering
     \vspace{5mm}
         \includegraphics[width=\linewidth]{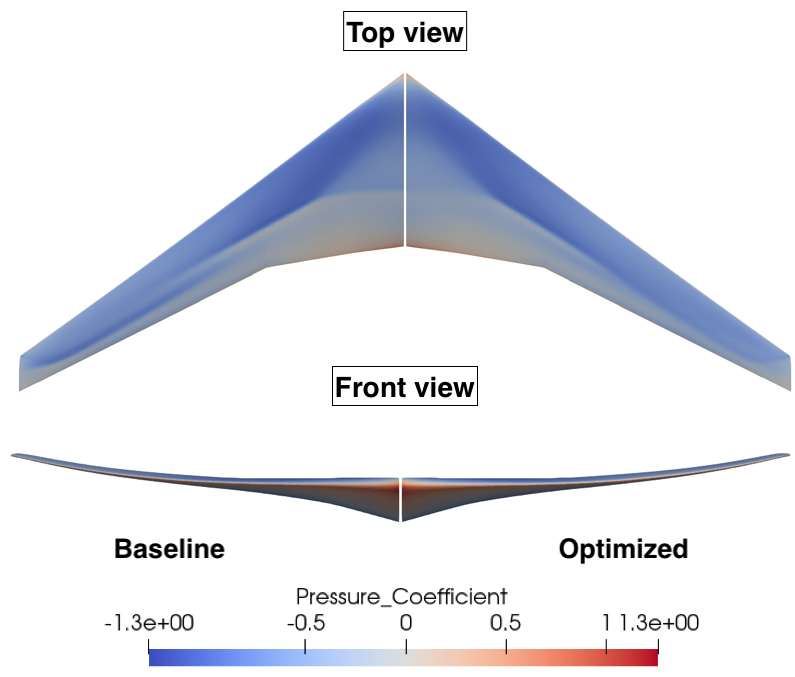}
         \caption{Euler-based AWSO of the QCRM wing: $C_p$ distribution on the baseline and the optimized designs.}
         \label{f:Cp_CRM_rig}
\end{figure}

%====================================================
%
Results of the AWSO process in terms of drag coefficient evolution are shown in Figure~\ref{f:QCRM_Rigid_opt}, from which a $C_D$ reduction of $9.13\%$ can be inferred. 
%\inblue{[COMMENTO SECONDO ME NON NECESSARIO] The smaller reduction of $C_D$, if compared with the ONERA M6 test case, can be explained by the fact that the CRM wing shape has been already optimized for transonic flight.} 
Figure~\ref{f:Cp_CRM_rig} shows the $C_p$ distribution for the top and front view of the baseline (left) and optimal (right) designs. 
On the optimal design, the shock redistribution and alleviation along the wing span are apparent. 
%------------------------------------------%
% FIGURE                       %
%------------------------------------------%
%------------------------------------------%
%
%
%------------------------------------------%
% FIGURE                        %
%------------------------------------------%
%------------------------------------------%
%
%====================================================
\subsubsection{Aerostructural wing shape optimization}
\begin{figure}[!htbp]
     \centering
     \vspace{5mm}
         \includegraphics[width=0.9\linewidth]{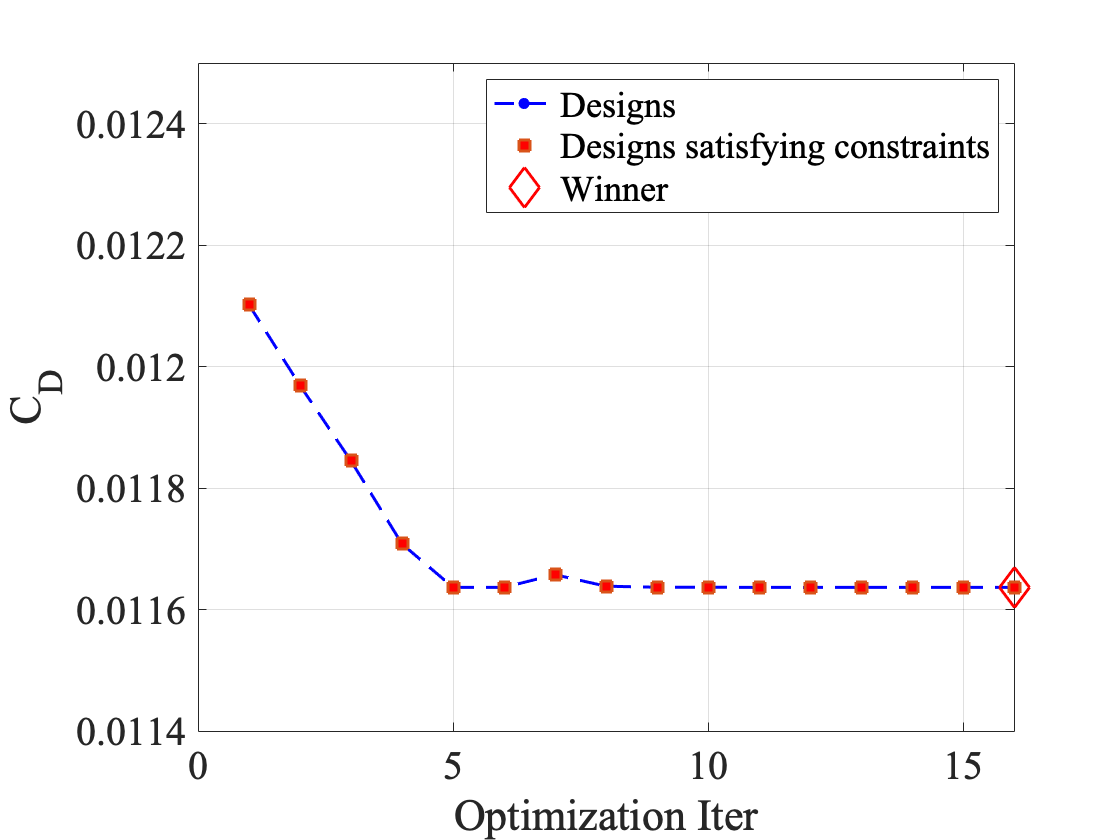}
         \caption{$C_D$ reduction for the Euler-based ASWSO of the QCRM wing.}
         \label{f:QCRM_Flexible_opt}
\end{figure}

\begin{figure}[!htbp]
     \centering
     \vspace{5mm}
         \includegraphics[width=\linewidth]{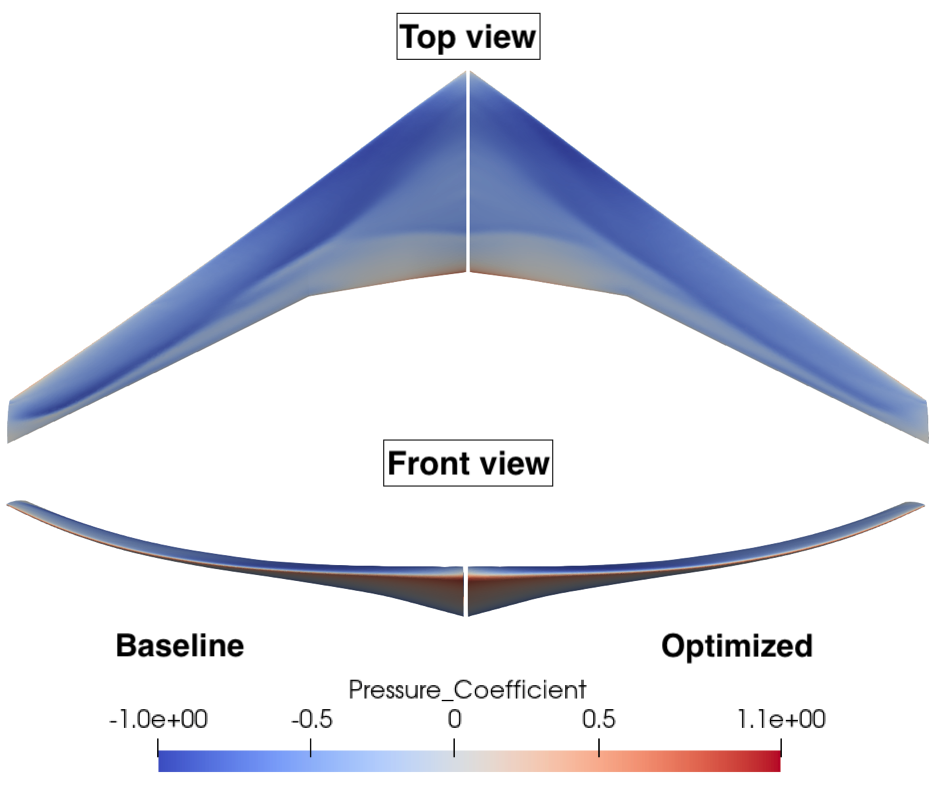}
         \caption{Euler-based ASWSO of the QCRM wing: $C_p$ distribution on the baseline and the optimized designs at aeroelastic equilibrium.}
         \label{f:Cp_CRM_flex}
\end{figure}

%====================================================
%
Figure~\ref{f:QCRM_Flexible_opt} depicts the drag coefficient evolution versus optimization iterations for an ASWSO on the aeroelastic test case. 
A $C_D$ reduction of 3.84\% with respect to the baseline configuration is obtained, which, even if representing a significant improvement, is smaller than the one observed for the AWSO case.
%with respect to the AWSO case, is a reduced (though still significant) drag reduction are achieved. 
It is also underlined how optimization parameters needed more tuning if compared to the AWSO case, witnessing an increased complexity of the problem due to the aerostructural coupling. 
It can be speculated that for such test case, opening the design space, in particular adding DVs relative to the structural domain (e.g., wing-box element sizes), may be needed for larger efficiency improvements.
%
%is needed if larger $C_D$ improvement were sought.
% 
%This may be due to the difficulty in resolving the imposed constraints for the aearostructural case, with respect to the aerodynamic one. 
%\inred{It can be inferred that better results can be achieved if introducing further DVs on the structural side of the problem}. 
%
Figure~\ref{f:Cp_CRM_flex} depicts the $C_p$ distribution on baseline (left) and optimal (right) designs. 
It can be noted, from the top view of the optimized configuration, the shock redistribution and alleviation close to the wing tip.
%
%------------------------------------------%
% FIGURE                       %
%------------------------------------------%
%------------------------------------------%
%
%
%------------------------------------------%
% FIGURE                        %
%------------------------------------------%
%------------------------------------------%
%
%===================================================
\subsubsection{AWSO and ASWSO comparison}
%====================================================
%
Results of the optimization campaign are summarized in Table~\ref{t:QCRM_opt_comp} where the $C_D$ for the flying shapes of the original design, the AWSO and the ASWSO optimal designs are shown.
%For the Euler-based QCRM test case, relevance of pursuing an aerostructural optimization can be inferred from Table~\ref{t:QCRM_opt_comp} which compares the $C_D$ for the flying shapes of the original design, the AWSO and the ASWSO optimal designs.
%
As already observed in the previous test case, the AWSO optimum in its flying shape has a considerably higher $C_D$  than the one of the ASWSO. 
%For this test case, as well, it can be seen how the AWSO optimum in operation presents a value of the OF which is far from the optimum evaluated by means of an ASWSO. 
%
However, this test case shows a relevant peculiarity: 
%that can lead to question the application of aerodynamic shape optimization on rigid wings:  
performances of the AWSO optimal wing are considerably poorer than the ones of the original (unoptimized) design at aeroelastic equilibrium. 
Hence, the computational costs of performing an aerodynamic shape optimization without considering the flexibility of the structure might not payback in more efficient wings when operating in their actual flying shape. 
%
%In this case it is also noticeable how the performances of the AWSO optimization flying shape are considerably poorer than the ones of the original design in operation. 
%
%This result highlights how performing aerodynamic shape optimizations of aircraft configurations that are expected to deflect considerably may be a conceptually wrong approach in the first place. 
%
%performing an AWSO and leads to worse performances than operating with the original design
%------------------------------------------%
% TABLE Variables                  %
%------------------------------------------%
\begin{table}[!t]
\caption{Comparison of $C_D$ between ASWSO, AWSO optima and the original configuration in flying shape, for the Euler-based QCRM test case.}
\label{t:QCRM_opt_comp}       
\centering
\begin{tabular}{ l l r}
\hline\noalign{\smallskip}
 \textbf{Configuration} & $\mathbf{C_D}$ & Diff. \% \\ 
\noalign{\smallskip}\hline\noalign{\smallskip}
ASWSO optimum & 0.01163 & --\\
AWSO optimum &  0.01243 & 6.87\% \\
Original &  0.01210   &     4.04\%   \\
\noalign{\smallskip}\hline
\end{tabular}
\end{table}
%------------------------------------------%
%
%
%------------------------------------------%
% FIGURE                        %
%------------------------------------------%
\begin{figure}[!htbp]
     \centering
     \vspace{5mm}
         \includegraphics[width=0.95\linewidth]{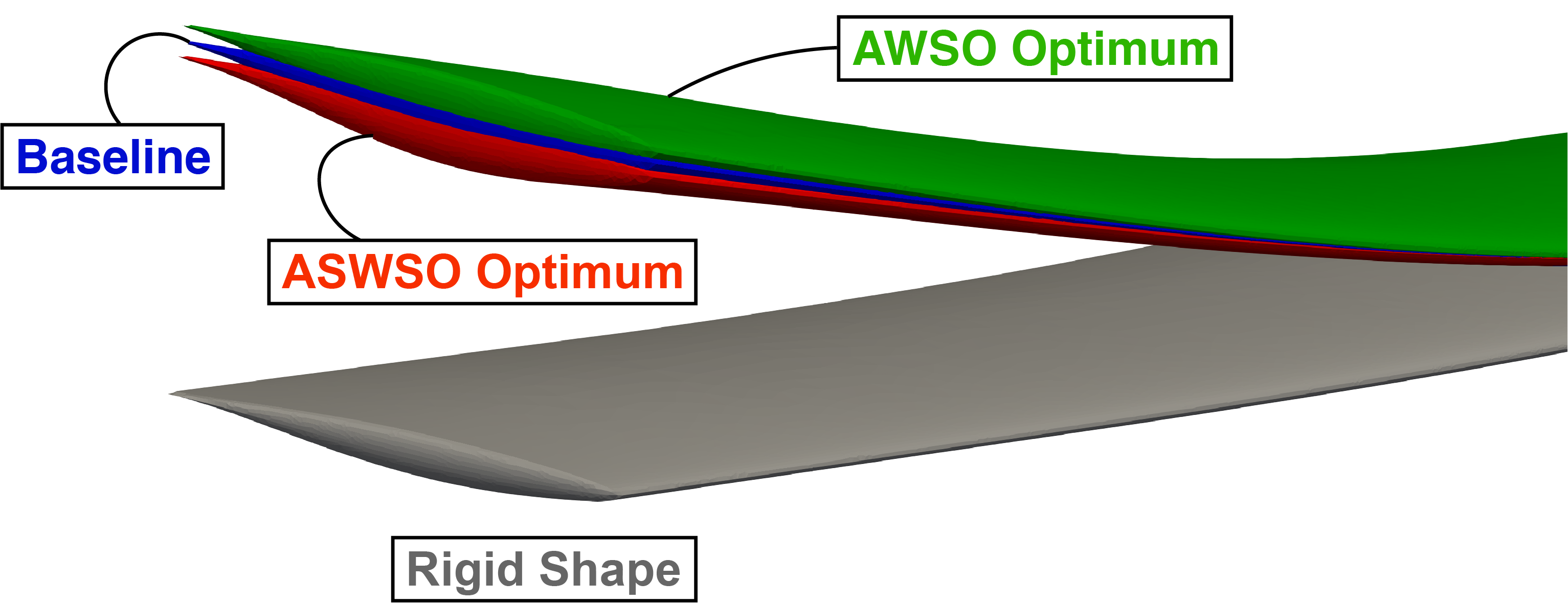}
         \caption{Comparison between ASWSO, AWSO optima and the original configuration flying shapes for the Euler-based QCRM test case: detail of wing tip.}
         \label{f:CRM_opt_comp}
\end{figure}
%------------------------------------------%
%
%
\par
\begin{figure*}[!htbp]
     \centering
     \vspace{5mm}
         \includegraphics[width=\linewidth]{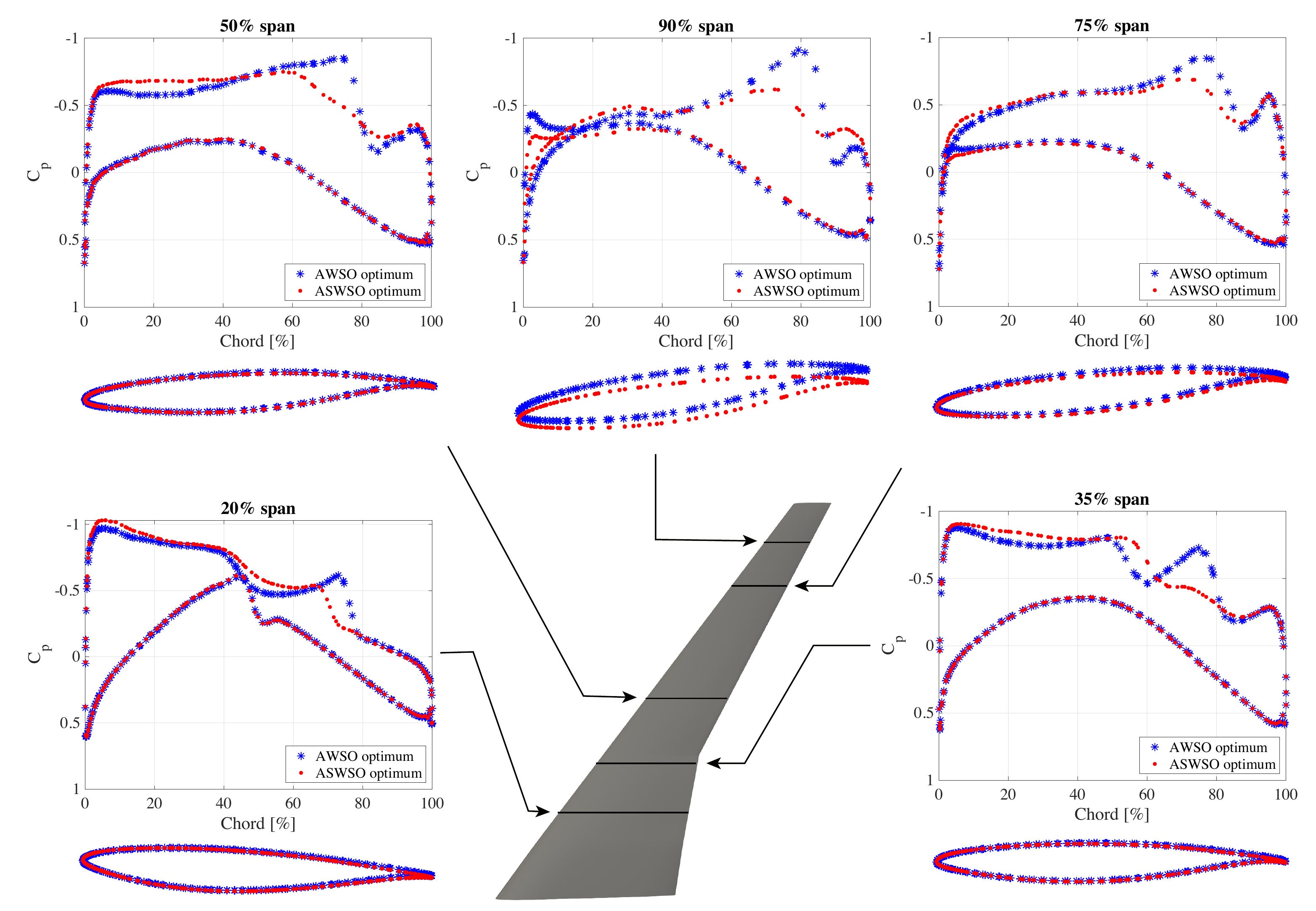}
         \caption{Comparison between AWSO and ASWSO optima in flight condition for the Euler-based QCRM M6 test case. $C_p$ distributions and airfoil shapes on selected stations.}
         \label{f:Cp_CRM_comp}
\end{figure*}

For the flying shapes of the AWSO and ASWSO optima, Figure~\ref{f:Cp_CRM_comp} shows, for selected sections along the wing span, the $C_p$ distribution and the airfoil shapes (with their relative position in space).
%
%------------------------------------------%
% FIGURE                        %
%------------------------------------------%
%------------------------------------------%
%
$C_p$ distribution highlights how shock-related gradients are generally much weaker  for the ASWSO optimum with respect to its AWSO counterpart, which has been, de-facto, optimized about a different operating condition. 
It can also be noted how ASWSO optimum has a smaller tip deflection compared to the AWSO one (as also shown in Figure~\ref{f:CRM_opt_comp}), showing hence an opposite trend with respect to the one seen for the ONERA M6 test case. 

%% file: Chapters/5d_Application.tex
%====================================================
\subsection{RANS-SA-based optimization of the QCRM}
\label{sec:RANS_opt_QCRM}
%=====================================================
One last optimization is performed for the QCRM aeroelastic test case considering a RANS-SA-based flow model for CFD: this counts as the highest-fidelity optimization performed within this effort. 
A Reynolds number of  $40$~millions is used in standard air conditions; Sutherland viscosity model is employed.
\par
With respect to the previous test case of Section~\ref{sec:Euler_opt_QCRM}, some changes have been performed to reduce the overall computational effort of the optimization:
the synthetic Young Modulus is 20\% larger and FFD box Bezier functions are of order 4, 9, 1 chord-wise, span-wise and along the thickness respectively. 
%
%the synthetic Young Modulus is 20\% larger and an FFD box is used with same shape as the one depicted in Figure~\ref{f:CRM_FFD}, employing Bezier functions of order 4, 9, 1 chord-wise, span-wise and along the thickness are respectively.
%
%, is used, with same shape as the one depicted in Figure~\ref{f:CRM_FFD}
%is used with respect to the structural model used for results of Section~\ref{sec:Euler_opt_QCRM}, while the symmetry constraint has been replaced by a fixed constraint. 
%
%To comply with the demanding computational time and resources necessary to run this case, settings have been specifically fine tuned to accelerate convergence of results. 
%Concerning the structural model, a 20\% higher Young Modulus is used with respect to the structural model used for results of Section~\ref{sec:Euler_opt_QCRM}, while the symmetry constraint has been replaced by a fixed constraint. 
%
%An FFD box with Bezier functions of order 4, 9, 1 chord-wise, span-wise and along the thickness, respectively, is used, with same shape as the one depicted in Figure~\ref{f:CRM_FFD}. 
DVs are the vertical positions of all FFD-box CPs, for a total number of 100. 
CPs on the symmetry plane are kept fixed as an effective way to prevent change in the shape of the relative airfoil. 

Aerodynamic and geometric constraints are the same as for the test case in  Section~\ref{sec:Euler_opt_QCRM}. 
To solve the fluid primal and relative adjoint problem, the non-frozen-turbulent approach is used~\cite{Lyu2015AerodynamicSO,Barcelos20081813}. 
\par
\begin{figure}[!htbp]
     \centering
     \vspace{5mm}
         \includegraphics[width=0.9\linewidth]{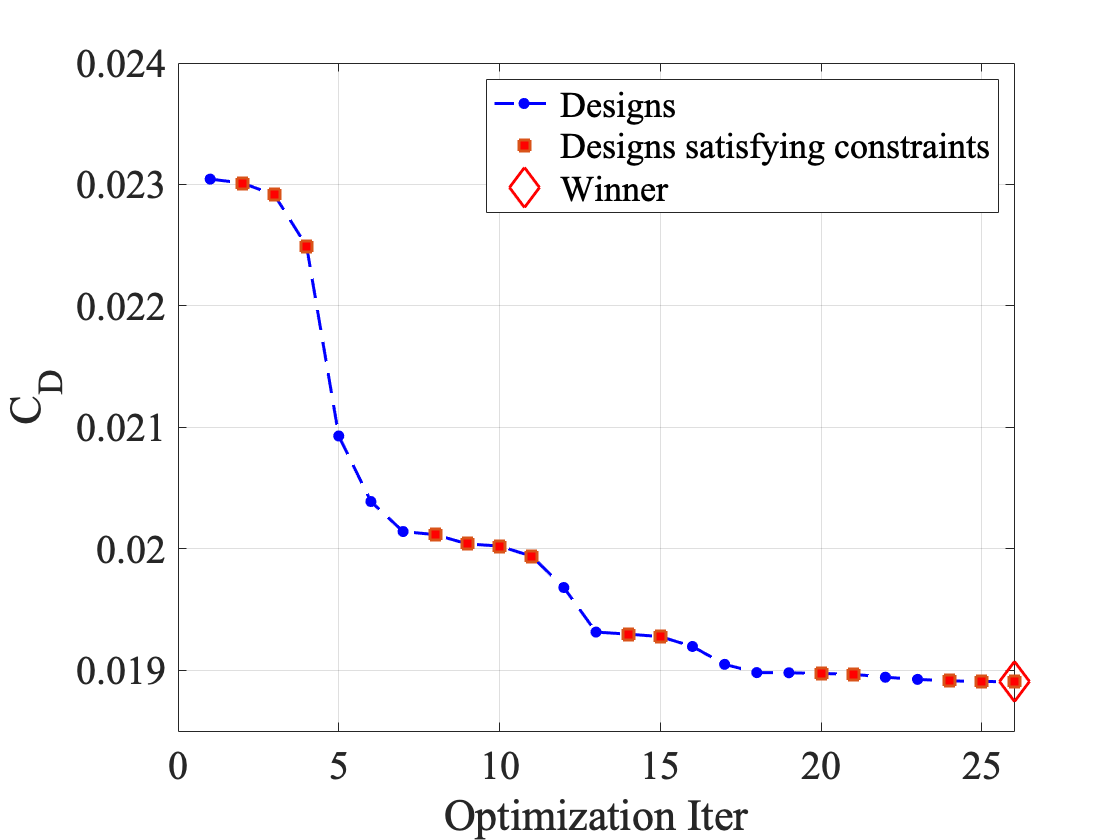}
         \caption{$C_D$ reduction for RANS-SA-based ASWSO of the QCRM wing.} %In red designs that satisfy the imposed constraints.}
         \label{f:RANS_QCRM_Flexible_opt}
\end{figure}
\begin{figure}[!htbp]
     \centering
     \vspace{5mm}
         \includegraphics[width=\linewidth]{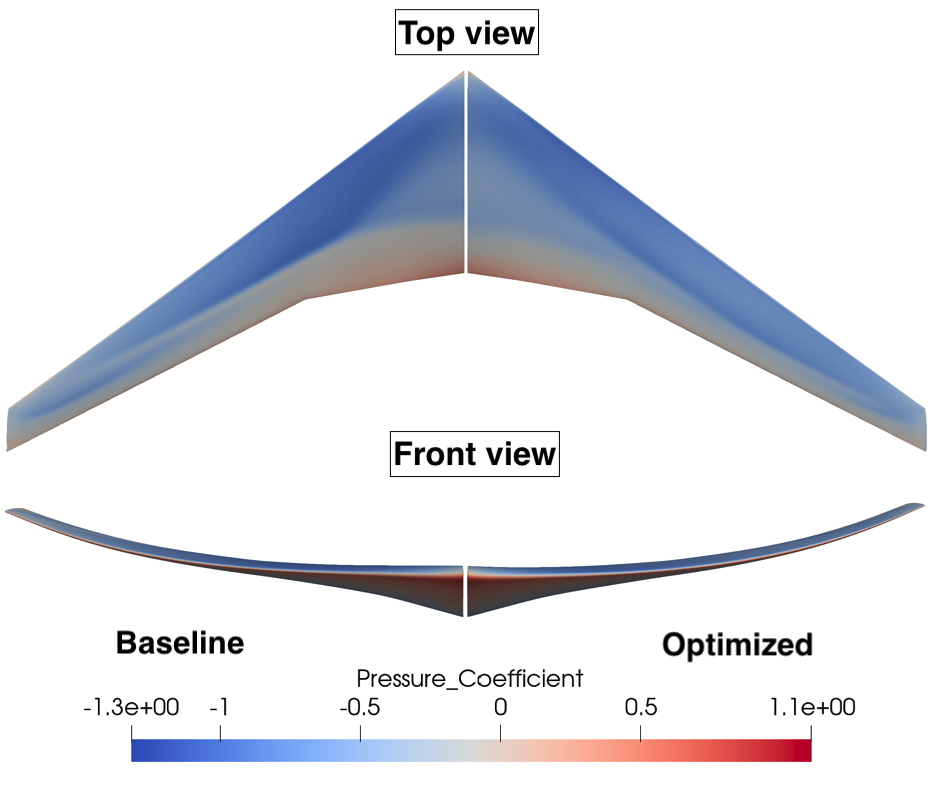}
         \caption{RANS-SA-based ASWSO of the QCRM wing: $C_p$ distribution on the baseline and the optimized designs at aeroelastic equilibrium.}
         \label{f:RANS_Cp_CRM_flex}
\end{figure}

Figure~\ref{f:RANS_QCRM_Flexible_opt} depicts the drag coefficient evolution versus optimization iterations for an ASWSO on test case. 
A $C_D$ reduction of 17.97\% with respect to the baseline configuration (in its relative flying shape) is obtained. 
%The optimization trend shown in figure highlights the higher complexity of the problem associated with the RANS-SA flow model.
%
%------------------------------------------%
% FIGURE                       %
%------------------------------------------%
%------------------------------------------%
%
Figure~\ref{f:RANS_Cp_CRM_flex} depicts the $C_p$ distribution on baseline (left) and optimal (right) designs.
It can be noted the shock alleviation in correspondence of the wing kink and its redistribution close to the wing tip. 
Moreover, the flying shape of the optimized configuration has a higher wing-tip deflection with respect to the baseline one, showing an opposite trend with respect to the one observed in the Euler case (Figure~\ref{f:CRM_opt_comp}).
%
%------------------------------------------%
% FIGURE                        %
%------------------------------------------%
%------------------------------------------%
%

%% file: Chapters/6_Conclusions.tex
%====================================================
\section{Conclusions and future works}
\label{sec:concl}
%====================================================
%
This work demonstrates a new methodology for high-fidelity aerostructural design and optimization of wings including aerodynamic and structural nonlinearities. %A gradient-based optimization method is presented
The proposed approach is modular: each one of the single discipline solvers is interfaced at high level through a Python wrapper to solve the static aeroelastic equilibrium (primal problem). 
Moreover, each solver has implemented its own adjoint capability employing algorithmic differentiation, hence, allowing the evaluation of coupled aerostructural sensitivities, to be used in gradient-based optimization. 
\par
For the fluid problem the solver is the CFD code SU2, whereas for the structural problem a nonlinear beam FEM solver embedding AD library CodiPack at native level has been ad-hoc developed to demonstrate the methodology. 
%in a similar fashion as already done in SU2. 
An interface/spline module provides loads/displacements transfer between the two solvers. 
\par
Capability of the method is demonstrated 
%Demonstration of the method is pursued by 
performing aerostructural wing shape optimization on test cases of potential industrial interest; namely, 
%of wings of industrial interest: i.e. 
aeroelastic test cases based on the ONERA M6 and CRM wings. 
Geometrical nonlinearities are always taken into account, and different levels of fidelity are employed at aerodynamic level (Euler and RANS-SA flow models). 
%, are used, whereascoupled with a geometrically nonlinear structural model. 
%
%
\par
Results of numerical optimization campaign show interesting trends, all highlighting the relevance of considering aerostructural coupling. Wings optimized neglecting such coupling, i.e., optimized not considering the deflection of the wing, perform relatively worse when considered in their actual flying shape configuration.
%performance are evaluated in their actual flying-shape. 
%
%relevant degradation in performances for wings optimized without including the aerostructural coupling 
%
%i.e., considering the actual flying shape, for wings optimized without including the aerostructural coupling. 
%(pure aerodynamic optimization). 
%when performing aerodynamic shape optimization 
%
For one test case  it is even found that optimization carried out neglecting the aerostructural coupling leads to wings with lower performances with respect to the ones of the initial non-optimized baseline, when both are considered in their relative flying shapes. 
Such result strongly warns against the practice of performing aerodynamic shape optimization without considering flexibility of the structure.
%, especially for relatively flexible wings. 
%
%that the initial baseline wing performs better than the one optimized not considering aerostructural coupling.  this warns against such optimization 
%For one test case the aerodynamic optimal flying shape performs worse not only with respect to the aerostructural optimum, but also with respect to the original design. 
%This result highlights how performing aerodynamic shape optimization without considering the flexibility of the structure may lead to worse designs than the original (unoptimized) one for highly flexible wings. 
%
%First, optimization of the Euler-based aeroelastic ONERA M6 test case is performed. Results are compared to the optimum of a classic aerodynamic shape optimization. It is noticed that optimizing the rigid configuration leads to worse performances if considering the relative flying shape with respect to the aerostructural optimum.
%
%
%\par
%
%Later on optimization of the Euler-based aeroelastic CRM test case is performed. For this case, the aerodynamic optimal flying shape performs worse not only with respect to the aerostructural optimum, but also with respect to the original design. This result highlights how performing aerodynamic shape optimization without considering the flexibility of the structure may lead to worse designs than the original (unoptimized) one for highly flexible wings. 
%
\par
The aerostructural optimization performed considering RANS-SA turbulence model showed the performances of the framework when modeling flow with higher level of fidelity. Thanks to the adoption of AD, non-frozen turbulence assumptions have been naturally employed (for both primal and adjoint solvers), and optimization of the wing shape delivered a noticeable drag coefficient reduction.
%
%higher flow modeling in rising the levle of fidelity %
%
%\inblue{Finally, an example of aerostructural shape optimization of the CRM test case is presented using a RANS-SA-based flow model. This counts as the highest level of fidelity used in this work. The optimization trend and results show the higher complexity of the problem associated with the RANS-SA flow model, if compared with the Euler-based case.}
%
%
\par
The optimization framework is released as open-source within the SU2 multiphysics suite in order to provide easy access to an aerostructural optimization tool (and relative primal aerostructural solver) to a potentially large audience. 
Thanks to the modular approach, users can easily experiment with different discipline modules, as far as they are provided with the adequate interfaces.
%
%One of the final goal of the modularity approach is, in fact, to allow users to experiment with different single discipline modules, as far as provided with the adequate interfaces.
%
\par
Concerning future works, the framework can be expanded, with little effort, to introduce structural objective functions and design variables, compatibly with the discipline solver capabilities. 
%With such augmented optimization process, design space will be enlarged and 
%for each of the considered disciplines. 
% 
%With such augmented capability 
%This will allow, in the future, to explore more optimization scenarios. 
%
%\inred{In terms of AD-level and RAM footprint during the registration process performed by CoDiPack, some fine-tuning may be needed to allow an efficient optimization process for computationally intensive cases featuring large meshes and higher-fidelity flow models (RANS or higher). Parliamone}
%Concerning performances, 
Moreover, revision of the AD operational scheme of the current framework may allow to achieve a reduced RAM footprint during the code registration process performed by CoDiPack, for computationally intensive cases featuring large meshes and higher-fidelity flow models (RANS or higher).
%
%
%intensive CFD meshes and higher-fidelity flow models (i.e. RANS).
%
%
%
%In order to tackle larger computational cases, closer to industrial needs, some fine-tuning at the AD-level needs to be performed to alleviate the RAM peak footprint  
%
%With reference to Revision of the AD operational scheme of the current framework will allow to achieve a reduced RAM footprint during the code registration process performed by CoDiPack, which is, by now, the biggest constraint for problems involving intensive CFD meshes and higher-fidelity flow models (i.e. RANS).

%% file: Chapters/7_Replication.tex
%====================================================
\section*{Replication of results}
%====================================================
%
The employed framework is currently on GitHub in the branch \textit{feature\_pyBeam\_ShapeDesignV2} of SU2 repository and will soon be available in the official release of the suite. 
\par
PyBeam organization on GitHub provides the complete set of test cases discussed above in the repository \textit{SAMO\_testcases}.

%% file: Chapters/1b_Statements.tex
%------------------------------------------%
\section*{Declaration}
%------------------------------------------%
%
%\paragraph{Funding} 
%Not applicable
%
%\par
%
\paragraph{Conflicts of interest/Competing interests}
The authors declare that they have no conflict of
interest.

\paragraph{Acknowledgments} 
Part of the simulations  were executed on the high performance cluster "Elwetritsch" at TU Kaiserslautern, which is part of the Alliance for High Performance Computing in Rhineland-Palatinate (AHRP). The authors would like to thank Dr. Beckett Y. Zhou and Guillermo Su\`arez of the Chair for Scientific Computing of TU Kaiserslautern for their assistance. % help in performing them.

%% file: main.bbl
\begin{thebibliography}{10}

\bibitem{HAFTKA-1977}
R.~T. Haftka.
\newblock Optimization of flexible wing structures subject to strength and
  induced drag constraints.
\newblock {\em AIAA Journal}, 15:1101--1106, 1977.

\bibitem{GROSSMAN_1986}
B.~Grossman, Z.~Gurdal, and R.~Haftka.
\newblock Integrated aerodynamic/structural design of a sailplane wing.
\newblock In {\em Aircraft Systems, Design and Technology Meeting}. American
  Institute of Aeronautics and Astronautics, oct 1986.

\bibitem{GROSSMAN_1989}
B.~Grossman, R.~Haftka, P.-J. Kao, D.~Polen, M.~Rais-Rohant, and
  J.~Sobieszczanski-Sobieski.
\newblock Integrated aerodynamic-structural design of a transport wing.
\newblock In {\em Aircraft Design and Operations Meeting}. American Institute
  of Aeronautics and Astronautics, jul 1989.

\bibitem{Martins2004HighFidelityAD}
Joaquim R. R.~A. Martins, J.~J. Alonso, and James~J. Reuther.
\newblock High-fidelity aerostructural design optimization of a supersonic
  business jet.
\newblock {\em Journal of Aircraft}, 41:523--530, 2004.

\bibitem{UC3M_ASO1Pustina}
L.~Pustina, R.~Cavallaro, and G.~Bernardini.
\newblock Nerone: An open-source based tool for aerodynamic transonic
  optimization of nonplanar wings.
\newblock {\em Aerotec. Missili Spaz.}, 98:85--104, 2019.

\bibitem{Peter_etal_2010}
Jacques~E.V. Peter and Richard~P. Dwight.
\newblock Numerical sensitivity analysis for aerodynamic optimization: A survey
  of approaches.
\newblock {\em Computers \& Fluids}, 39(3):373 -- 391, 2010.

\bibitem{Lyu2015AerodynamicSO}
Zhoujie Lyu, Gaetan K.~W. Kenway, and Joaquim R. R.~A. Martins.
\newblock Aerodynamic shape optimization investigations of the common research
  model wing benchmark.
\newblock {\em AIAA Journal}, 53:968--985, 2015.

\bibitem{maute_2001}
K.~Maute, M.~Nikbay, and C.~Farhat.
\newblock Coupled analytical sensitivity analysis and optimization of
  three-dimensional nonlinear aeroelastic systems.
\newblock {\em AIAA Journal}, 39(11):2051--2061, 2001.

\bibitem{martins2001ab}
J.R.R.A. Martins, J.J. Alonso, and J.J. Reuther.
\newblock Aero-structural wing design optimization using high-fidelity
  sensitivity analysis.
\newblock In {\em Proceedings - CEAS Conference on Multidisciplinary Aircraft
  Design Optimization}, Cologne, Germany,, 2001.

\bibitem{Martins_Review_Der2013}
Joaquim R. R.~A. Martins and John~T. Hwang.
\newblock Review and unification of methods for computing derivatives of
  multidisciplinary computational models.
\newblock {\em AIAA Journal}, 51(11):2582--2599, 2013.

\bibitem{maute2003911}
K.~Maute, M.~Nikbay, and C.~Farhat.
\newblock Sensitivity analysis and design optimization of three-dimensional
  non-linear aeroelastic systems by the adjoint method.
\newblock {\em International Journal for Numerical Methods in Engineering},
  56(6):911--933, 2003.

\bibitem{Barcelos20081813}
M.~Barcelos and K.~Maute.
\newblock Aeroelastic design optimization for laminar and turbulent flows.
\newblock {\em Computer Methods in Applied Mechanics and Engineering},
  197(19-20):1813--1832, 2008.

\bibitem{SpalartAllmaras}
P.~Spalart and S.~Allmaras.
\newblock A one-equation turbulence model for aerodynamic flows.

\bibitem{SanchezThesis}
R.~Sanchez.
\newblock {\em Coupled Adjoint-Based Sensitivities in Large-Displacement
  Fluid-Structure Interaction using Algorithmic Differentiation}.
\newblock PhD thesis, Imperial College London, 2018.

\bibitem{sagebaum2017high}
M.~Sagebaum, T.~Albring, and N.~R. Gauger.
\newblock High-performance derivative computations using codipack.
\newblock {\em arXiv preprint arXiv:1709.07229}, 2017.

\bibitem{Brezillon2012}
J.~Brezillon, A.~Ronzheimer, D.~Haar, M.~Abu-Zurayk, M.~Lummer, W.~Krüger, and
  F.~J. Natterer.
\newblock Development and application of multi-disciplinary optimization
  capabilities based on high-fidelity methods.
\newblock In {\em 53rd AIAA/ASME/ASCE/AHS/ASC Structures, Structural Dynamics
  and Materials Conference 2012}, 2012.

\bibitem{ghazlane2012}
Imane Ghazlane, Gerald Carrier, Antoine Dumont, and Jean antoine Desideri.
\newblock {\em Aerostructural Adjoint Method for Flexible Wing Optimization}.

\bibitem{Cambier2002}
L.~Cambier and M.~Gazaix.
\newblock Elsa: An efficient object-oriented solution to cfd complexity.
\newblock In {\em 40th AIAA Aerospace Sciences Meeting and Exhibit}, 2002.

\bibitem{kenway2014}
G.K.W. Kenway, G.J. Kennedy, and J.R.R.A. Martins.
\newblock Scalable parallel approach for high-fidelity steady-state aeroelastic
  analysis and adjoint derivative computations.
\newblock {\em AIAA Journal}, 52(5):935--951, 2014.

\bibitem{Weide2006}
Edwin van~der Weide, Georgi Kalitzin, Jorg Schluter, and Juan Alonso.
\newblock {\em Unsteady Turbomachinery Computations Using Massively Parallel
  Platforms}.

\bibitem{TACS2010}
Graeme Kennedy and Joaquim Martins.
\newblock {\em Parallel Solution Methods for Aerostructural Analysis and Design
  Optimization}.

\bibitem{Vassberg-2008}
J.~Vassberg, M.~Dehaan, M.~Rivers, and R.~Wahls.
\newblock Development of a common research model for applied cfd validation
  studies.
\newblock In {\em American Institute of Aeronautics and Astronautics 26th AIAA
  Applied Aerodynamics Conference - Honolulu, Hawaii}, 2008.

\bibitem{KenwayMultipoint2014}
Gaetan K.~W. Kenway and Joaquim R. R.~A. Martins.
\newblock Multipoint high-fidelity aerostructural optimization of a transport
  aircraft configuration.
\newblock {\em Journal of Aircraft}, 51(1):144--160, 2014.

\bibitem{Kennedy2014e}
Graeme~J. Kennedy, Gaetan~K. Kenway, and Joaquim R. R.~A. Martins.
\newblock A comparison of metallic, composite and nanocomposite optimal
  transonic transport wings.
\newblock Technical report, NASA, March 2014.
\newblock CR-2014-218185.

\bibitem{kennedyScitech2014}
Gaetan~K. Kenway, Graeme Kennedy, and Joaquim Martins.
\newblock {\em High Aspect Ratio Wing Design: Optimal Aerostructural Tradeoffs
  for the Next Generation of Materials}.

\bibitem{KenwayScitech2014}
Gaetan Kenway, Graeme Kennedy, and Joaquim Martins.
\newblock {\em Aerostructural optimization of the Common Research Model
  configuration}.

\bibitem{Brooks2018}
Timothy~R. Brooks, Gaetan K.~W. Kenway, and Joaquim R. R.~A. Martins.
\newblock Benchmark aerostructural models for the study of transonic aircraft
  wings.
\newblock {\em AIAA Journal}, 56(7):2840--2855, 2018.

\bibitem{Hoogervorst2017WingAO}
Jan~E.K. Hoogervorst and Ali Elham.
\newblock Wing aerostructural optimization using the individual discipline
  feasible architecture.
\newblock {\em Aerospace Science and Technology}, 65:90--99, 2017.

\bibitem{CramerIDF_1994}
Evin~J. Cramer, J.~E. Dennis, Jr., Paul~D. Frank, Robert~Michael Lewis, and
  Gregory~R. Shubin.
\newblock Problem formulation for multidisciplinary optimization.
\newblock {\em SIAM Journal on Optimization}, 4(4):754--776, 1994.

\bibitem{MartinsIDF}
Alp Dener, Jason~E. Hicken, Gaetan~K. Kenway, and Joaquim R. R.~A. Martins.
\newblock Enabling modular aerostructural optimization: Individual discipline
  feasible without the jacobians.

\bibitem{SU2_AIAAJ2016}
Thomas~D. Economon, Francisco Palacios, Sean~R. Copeland, Trent~W. Lukaczyk,
  and Juan~J. Alonso.
\newblock Su2: An open-source suite for multiphysics simulation and design.
\newblock {\em AIAA Journal}, Vol. 54(No. 3):pp. 828--846, 2016.

\bibitem{Nickbay2009}
Melike Nikbay, Levent Öncü, and Ahmet Aysan.
\newblock Multidisciplinary code coupling for analysis and optimization of
  aeroelastic systems.
\newblock {\em Journal of Aircraft}, 46(6):1938--1944, 2009.

\bibitem{Palacios2015}
F.~Palacios, T.~D. Economon, A.~D. Wendorff, and J.~J. Alonso.
\newblock Large-scale aircraft design using su2.
\newblock In {\em 53rd AIAA Aerospace Sciences Meeting}, 2015.

\bibitem{Pini2017}
M.~Pini, S.~Vitale, P.~Colonna, G.~Gori, A.~Guardone, T.~Economon, J.J. Alonso,
  and F.~Palacios.
\newblock Su2: the open-source software for non-ideal compressible flows.
\newblock {\em Journal of Physics: Conference Series}, 821(1):012013, 2017.

\bibitem{Gori2017}
G.~Gori, D.~Vimercati, and A.~Guardone.
\newblock Non-ideal compressible-fluid effects in oblique shock waves.
\newblock {\em Journal of Physics: Conference Series}, 821(1):012003, 2017.

\bibitem{Molina2017}
E.~S. Molina, C.~Spode, R.~G.~A. Da~Silva, D.~E. Manosalvas-Kjono,
  S.~Nimmagadda, T.~D. Economon, J.~J. Alonso, and M.~Righi.
\newblock Hybrid rans/les calculations in su2.
\newblock In {\em 23rd AIAA Computational Fluid Dynamics Conference, 2017},
  2017.

\bibitem{Zhou2017}
B.~Y. Zhou, T.~Albring, N.~R. Gauger, C.~Ilario, T.~Economon, and J.~J. Alonso.
\newblock Reduction of airframe noise components using a discrete adjoint
  approach.
\newblock {\em AIAA 2017-3658}, 2017.

\bibitem{albring2016}
T.~Albring, M.~Sagebaum, and N.R. Gauger.
\newblock Efficient aerodynamic design using the discrete adjoint method in
  su2.
\newblock In {\em 17th AIAA/ISSMO Multidisciplinary Analysis and Optimization
  Conference}, 2016.

\bibitem{sanchez2018}
R.~Sanchez, T.~Albring, R.~Palacios, N.~R. Gauger, T.~D. Economon, and J.~J.
  Alonso.
\newblock Coupled adjoint-based sensitivities in large-displacement
  fluid-structure interaction using algorithmic differentiation.
\newblock {\em International Journal for Numerical Methods in Engineering},
  113(7):1081--1107, 2018.

\bibitem{DAFoam2019}
Ping He, Charles~A. Mader, Joaquim R. R.~A. Martins, and Kevin~J. Maki.
\newblock Dafoam: An open-source adjoint framework for multidisciplinary design
  optimization with openfoam.
\newblock {\em AIAA Journal}, 58(3):1304--1319, 2020.

\bibitem{LevyBook}
R.~Levy and W.R. Spillers.
\newblock {\em Analysis of geometrically nonlinear structures}.
\newblock Number v. 1. Kluwer Academic Publishers, Dordrecht, Netherlands,
  2003.

\bibitem{Belytschko_book1}
T.~Belytschko, W.K. Liu, and B.~Moran.
\newblock {\em Nonlinear finite elements for continua and structures}.
\newblock Wiley, 2000.

\bibitem{swigpaper}
David~M. Beazley.
\newblock Swig: An easy to use tool for integrating scripting languages with c
  and c++.
\newblock In {\em Proceedings of the 4th Conference on USENIX Tcl/Tk Workshop,
  1996 - Volume 4}, TCLTK'96, pages 15--15, Berkeley, CA, USA, 1996. USENIX
  Association.

\bibitem{Wilcox1998}
D.~Wilcox.
\newblock {\em Turbulence Modeling for CFD}.
\newblock DCW Industries, Inc., 1998.

\bibitem{White1974}
F.~M. White.
\newblock {\em Viscous Fluid Flow}.
\newblock McGraw–Hill, New York, 1974.

\bibitem{Donea_2004}
Jean Donea, Antonio Huerta, J.-Ph. Ponthot, and A.~Rodríguez-Ferran.
\newblock {\em Arbitrary Lagrangian–Eulerian Methods}.
\newblock John Wiley \& Sons, Ltd, 2004.

\bibitem{Dwight2009401}
R.P. Dwight.
\newblock Robust mesh deformation using the linear elasticity equations.
\newblock pages 401--406, Ghent, 2009. Springer Berlin.

\bibitem{palacios_stanford_2013}
Francisco Palacios, Juan Alonso, Karthikeyan Duraisamy, Michael Colonno, Jason
  Hicken, Aniket Aranake, Alejandro Campos, Sean Copeland, Thomas Economon,
  Amrita Lonkar, Trent Lukaczyk, and Thomas Taylor.
\newblock Stanford {University} {Unstructured} ({SU}{\textasciicircum}2): {An}
  open-source integrated computational environment for multi-physics simulation
  and design.
\newblock In {\em 51st {AIAA} {Aerospace} {Sciences} {Meeting} including the
  {New} {Horizons} {Forum} and {Aerospace} {Exposition}}. American Institute of
  Aeronautics and Astronautics, 2013.

\bibitem{palacios_stanford_2014}
Francisco Palacios, Thomas~D. Economon, Aniket Aranake, Sean~R. Copeland,
  Amrita~K. Lonkar, Trent~W. Lukaczyk, David~E. Manosalvas, Kedar~R. Naik,
  Santiago Padron, Brendan Tracey, Anil Variyar, and Juan~J. Alonso.
\newblock Stanford {University} {Unstructured} ({SU}2): {Analysis} and {Design}
  {Technology} for {Turbulent} {Flows}.
\newblock In {\em 52nd {Aerospace} {Sciences} {Meeting}}. American Institute of
  Aeronautics and Astronautics, 2014.

\bibitem{DONEA1982689}
J.~Donea, S.~Giuliani, and J.P. Halleux.
\newblock An arbitrary lagrangian-eulerian finite element method for transient
  dynamic fluid-structure interactions.
\newblock {\em Computer Methods in Applied Mechanics and Engineering},
  33(1):689 -- 723, 1982.

\bibitem{SDSUteam_5jour}
Rauno Cavallaro, Andrea Iannelli, Luciano Demasi, and Alan~M{\'a}rquez
  Raz{\'o}n.
\newblock Phenomenology of nonlinear aeroelastic responses of highly deformable
  {Joined} {Wings}.
\newblock {\em Advances in Aircraft and Spacecraft Science}, 2(2):125--168,
  April 2015.

\bibitem{Romanelli_2012}
Michele Romanelli, Giulio;~Castellani, Paolo Mantegazza, and Sergio Ricci.
\newblock Coupled csd/cfd non-linear aeroelastic trim of free-flying flexible
  aircraft.
\newblock In {\em 53rd AIAA/ASME/ASCE/AHS/ASC Structures, Structural Dynamics
  and Materials Conference20th AIAA/ASME/AHS Adaptive Structures Conference14th
  AIAA - Honolulu, Hawaii.}, 2012.

\bibitem{ANNlibrary}
2010.

\bibitem{Mantegazza2}
Giuseppe Quaranta, Pierangelo Masarati, and Paolo Mantegazza.
\newblock A conservative mesh-free approach for fluid structure problems in
  coupled problems.
\newblock In {\em International Conference for Coupled Problems in Science and
  Engineering, Santorini, Greece}, pages 24--27, 23-29 May 2005.

\bibitem{SDSUteam_6jour}
Rauno Cavallaro, Rocco Bombardieri, Luciano Demasi, and Andrea Iannelli.
\newblock Prandtlplane {J}oined {W}ing: Body freedom flutter, limit cycle
  oscillation and freeplay studies.
\newblock {\em Journal of Fluids and Structures}, 59:57--84, November 2015.

\bibitem{UC3Mteam_Scitech_NITRO}
Rocco Bombardieri, Rauno Cavallaro, Jorge Luis Sáez~de Teresa, and Moti
  Karpel.
\newblock Nonlinear aeroelasticity: a cfd-based adaptive methodology for
  flutter prediction.
\newblock Number AIAA 2019-1866. AIAA Scitech 2019 Forum, San Diego,
  California, 7-11 January 2019.

\bibitem{FARHAT20033}
Charbel Farhat, Philippe Geuzaine, and Gregory Brown.
\newblock Application of a three-field nonlinear fluid–structure formulation
  to the prediction of the aeroelastic parameters of an f-16 fighter.
\newblock {\em Computers \& Fluids}, 32, 2003.

\bibitem{Degroote_2009}
Joris Degroote, Klaus-Jürgen Bathe, and Jan Vierendeels.
\newblock Performance of a new partitioned procedure versus a monolithic
  procedure in fluid–structure interaction.
\newblock {\em Computers \& Structures}, 87(11–12):793 -- 801, 2009.
\newblock Fifth MIT Conference on Computational Fluid and Solid Mechanics.

\bibitem{Irons_1969}
Bruce~M. Irons and Robert~C. Tuck.
\newblock A version of the aitken accelerator for computer iteration.
\newblock {\em International Journal for Numerical Methods in Engineering},
  1(3):275--277, 1969.

\bibitem{Sanchez2016b}
R.~Sanchez, H.~Kline, D.~Thomas, A.~Variyar, M.~Righi, T.D. Economon, J.J.
  Alonso, R.~Palacios, G.~Dimitriadis, and V.~Terrapon.
\newblock Assessment of the {F}luid-{S}tructure {I}nteraction {C}apabilities
  for {A}eronautical {A}pplications of the {O}pen-{S}ource {S}olver {SU2}.
\newblock In {\em VII {E}uropean {C}ongress on {C}omputational {M}ethods in
  {A}pplied {S}ciences and {E}ngineering ({ECCOMAS} 2016)}, Crete Island,
  Greece, 5-10 June, 2016.

\bibitem{Korivi1992AnIS}
Vamshi~Mohan Korivi, Arthur~C. Taylor, Perry~A. Newman, Gene Hou, and H.~E.
  Jones.
\newblock An incremental strategy for calculating consistent discrete cfd
  sensitivity derivatives.
\newblock 1992.

\bibitem{kraft1988software}
D.~Kraft.
\newblock {\em A Software Package for Sequential Quadratic Programming}.
\newblock Deutsche Forschungs- und Versuchsanstalt f{\"u}r Luft- und Raumfahrt
  K{\"o}ln: Forschungsbericht. Wiss. Berichtswesen d. DFVLR, 1988.

\bibitem{Samareh_2004}
J.~Samareh.
\newblock Aerodynamic shape optimization based on free-form deformation.
\newblock Number AIAA 2004-4630. 10th AIAA/ISSMO Multidisciplinary Analysis and
  Optimization Conference, August 20104.

\bibitem{Liem2017ExpectedDM}
Rhea~Patricia Liem, Joaquim R. R.~A. Martins, and Gaetan K.~W. Kenway.
\newblock Expected drag minimization for aerodynamic design optimization based
  on aircraft operational data.
\newblock {\em Aerospace Science and Technology}, 63:344--362, 2017.

\bibitem{Bombardieri_uc3m_EUROGEN}
R.~Bombardieri, R.~Sanchez, R.~Cavallaro, and N.~R. Gauger.
\newblock Towards an open-source framework for aero-structural design and
  optimization within the su2 suite.
\newblock In A.~Gaspar-Cunha, J.~Periaux, K.C. Giannakoglou, N.R. Gauger,
  D.~Quagliarella, and D.~Greiner, editors, {\em Advances in Evolutionary and
  Deterministic Methods for Design, Optimization and Control in Engineering and
  Sciences, \textit{in press}}, chapter~19. Springer, 2021.

\bibitem{CRM_web}
NASA.
\newblock Nasa common research model, 2008.

\bibitem{vela2019aeroelastic}
Javier Vela~Pe{\~n}a.
\newblock Aeroelastic calculations on an equivalent beam-based nasa crm.
\newblock {B.Sc.} thesis uc3m, 2019.

\bibitem{Luis019aeroelastic}
Jorge Luis S\`aez~de Teresa.
\newblock Towards correction of aerodynamic coefficients for aeroelastic
  calculations with open-source su2 software.
\newblock {M.Sc.} thesis uc3m, 2019.

\bibitem{Castro2020aerodynShapeOpt}
Enrique Castro~L\`opez.
\newblock Aerodynamic shape optimization of wings in transonic regime using
  discrete adjoint method.
\newblock {B.Sc.} thesis uc3m, 2020.

\end{thebibliography}
